\documentclass[%
reprint,
superscriptaddress,
nofootinbib,
amsmath,amssymb,
aps,
prl,
]{revtex4-1}

\usepackage{graphicx} 
\usepackage{dcolumn}
\usepackage{bm} 
\usepackage{braket}
\usepackage{dsfont}
\usepackage{color,colortbl}
\usepackage[table,xcdraw]{xcolor}
\usepackage{hyperref}
\usepackage{multirow}
\usepackage{subcaption}
\usepackage{mwe}
\usepackage{booktabs}
\usepackage{caption}
\usepackage{lipsum}
\usepackage{babel,blindtext}
\usepackage[toc,page]{appendix}

\begin{document}

\title{Fast Uncertainty Estimates in Deep Learning Interatomic Potentials} 

\author{Albert Zhu}
\affiliation{Harvard University}

\author{Simon Batzner$^{\dagger}$}
\affiliation{Harvard University}

\author{Albert Musaelian}
\affiliation{Harvard University}

\author{Boris Kozinsky$^{\dagger}$}
\affiliation{Harvard University}
\affiliation{Robert Bosch Research and Technology Center}

\def\thefootnote{$\dagger$}\footnotetext{Corresponding authors\\S.B., E-mail: \url{batzner@g.harvard.edu}\\B.K., E-mail: \url{bkoz@seas.harvard.edu}\\ }\def\thefootnote{\arabic{footnote}}

\begin{abstract}
\subsection{\\ Abstract}
Deep learning has emerged as a promising paradigm to give access to highly accurate predictions of molecular and materials properties. A common short-coming shared by current approaches, however, is that neural networks only give point estimates of their predictions and do not come with predictive uncertainties associated with these estimates. Existing uncertainty quantification efforts have primarily leveraged the standard deviation of predictions across an ensemble of independently trained neural networks. This incurs a large computational overhead in both training and prediction that often results in order-of-magnitude more expensive predictions. Here, we propose a method to estimate the predictive uncertainty based on a \emph{single} neural network without the need for an ensemble. This allows us to obtain uncertainty estimates with virtually no additional computational overhead over standard training and inference. We demonstrate that the quality of the uncertainty estimates matches those obtained from deep ensembles. We further examine the uncertainty estimates of our methods and deep ensembles across the configuration space of our test system and compare the uncertainties to the potential energy surface. Finally, we study the efficacy of the method in an active learning setting and find the results to match an ensemble-based strategy at order-of-magnitude reduced computational cost.
\end{abstract}
\maketitle

\section{Introduction}

Over the past decade, the construction of high-dimensional potential energy surfaces (PES) based on machine learning (ML) has become a promising avenue to enable linear-scaling and computationally efficient molecular simulations that retain the quantum chemical accuracy of their training data \cite{blank1995neural, handley2009optimal, behler2007generalized, gaporiginalpaper, thompson2015spectral, shapeev2016moment, schnet_jcp, sgdml, physnet_jctc, drautz2019atomic, christensen2020fchl, klicpera2020directional, nequip, allegro, dcf, gnnff, xie2021bayesian, xie2022uncertainty, deepmd, vandermause2020fly, vandermause2021active, anderson2019cormorant, kovacs2021linear}. A large variety of methods have been proposed to regress energies and forces obtained from  \textit{ab-initio} calculation as a function of atomic positions and chemical species, including kernel-based approaches \cite{gaporiginalpaper, vandermause2020fly, sgdml}, linear models \cite{thompson2015spectral, shapeev2016moment}, and neural networks (NNs) \cite{behler2007generalized, schnet_jcp, klicpera2020directional, nequip, allegro}. Among these, deep NNs in particular have shown remarkable accuracy and fast progress in their predictive accuracy \cite{schnet_neurips, nequip, qiao2021unite, klicpera2021gemnet, allegro}. The high accuracy of NN-based approaches, however, comes at a cost: common to all existing neural approaches is that they provide only point estimates of their predictions instead of the full predictive distribution. This differs from Bayesian methods such as Gaussian Processes, which inherently come with a measure of predictive uncertainty. Uncertainties have been shown to be of tremendous value in ML-driven molecular simulations \cite{vandermause2020fly, vandermause2021active, xie2021bayesian, xie2022uncertainty, johansson2022micron}. In particular, uncertainties have been used to bootstrap simulations without the need for a training set via an active learning loop \cite{vandermause2020fly}. In such an approach, the model's uncertainty is assessed at every integration step: if the uncertainty is low, the model's predictions are used. If instead the uncertainty exceeds a certain threshold, high-accuracy quantum mechanical calculations such as density functional theory (DFT) are invoked, the new data point is added to the data set, and the model is re-trained. Provided the uncertainty measure used is of high fidelity, such an approach can greatly enhance robustness, reliability, and ease-of-use of ML-driven atomistic simulations. In this work, we present a novel, computationally-inexpensive method to obtain uncertainty measures in deep learning interatomic potentials, evaluate its performance compared to existing approaches, and demonstrate that it produces order-of-magnitude faster uncertainty estimates. 

\subsection{Related Work}

Given the high impact reliable uncertainty estimates would have on the usefulness of NN-based machine learning interatomic potentials (MLIPs), great effort has gone into the development of techniques that enhance the point estimate predictions of NNs with a measure of predictive uncertainty \cite{behler2015constructing, schran2020committee, zhang2019active, tran2020methods, hirschfeld2020uncertainty, hu2022robust, soleimany2021evidential, wen2020uncertainty, janet2019quantitative, lakshminarayanan2017simple, gal2016dropout}. The most widely used approach among these is an \textit{ensemble} of NNs all trained on the same data that differ in their initial weights and perhaps other training or model hyperparameters. The mean of the predictions of all constituent networks is used as the ensemble's prediction and the standard deviation of the predictions is used as a measure of uncertainty. Intuitively, if a structure seen at test time falls within the input domain that the NNs are confident about, their predictions should agree. In contrast, if a test structure lies outside of the training distribution, the networks' predictions should differ, resulting in a higher standard deviation. This method has been widely used since the first generation of NN interatomic potentials \cite{behler2015constructing,  zhang2019active, schran2020committee}. Systematic analysis and improvements are desired in this direction, e.g., to avoid situations where all models share the same bias not captured in the training data or their functional form, resulting in models sharing confident but erroneous predictions. Recent work has explored the over-confidence of NN-based ensembles and the correlation between the test set error with the true predictive error \cite{Kahle_2022}. While larger ensembles can provide more statistics, the need to train and evaluate all constituent networks (often $N\geq10$) incurs significantly greater computational expenses, lowering inference speed and limiting practical applications of ensemble NN models for molecular dynamics or Monte Carlo sampling calculations. To alleviate this effect, a series of methods have been proposed \cite{janet2019quantitative, wen2020uncertainty}. However, these methods either still require multiple evaluations at inference time (and thereby only improve training cost), or have not been demonstrated to work in applications of molecular dynamics, where force uncertainty is the key objective.

\section{Methods}

\subsection{Ensembles of Neural Networks}

We investigate uncertainty quantification in Neural Equivariant Interatomic Potentials (NequIP), an E(3)-equivariant neural network for learning interatomic potentials that achieves state-of-the-art accuracy on a challenging and diverse set of molecules and materials at remarkable data efficiency in comparison to other MLIPs \cite{nequip}. To establish a baseline for our proposed uncertainty quantification approach, we train two sets of ensembles each consisting of ten NequIP neural networks: a ``traditional" ensemble consisting of networks differing solely in their weight initialization and the order in which mini-batches are sampled during training, and a ``diverse" ensemble consisting of three networks from the traditional ensemble and seven additional networks each with different hyperparameters (listed in SI table \ref{tab:ensemble-config}). To demonstrate the robustness of our methods and conclusions to the width/capacity of the networks, we train all networks with a hidden feature dimension of $f=32$ in one setting and $f=16$ in another setting. 

At run time, the force predictions of an ensemble, denoted $\bar{F}$, are calculated as the mean of the predictions of individual models in the ensemble, component-wise:
\begin{equation}
    \bar{F} = \left(\bar{F}_{x}, \bar{F}_{y}, \bar{F}_{z}
    \right)
\end{equation}
where $\bar{F}_{\alpha}$ denotes the mean of the $\alpha$-component of the predicted forces of all constituent models. To evaluate the model's fidelity, we calculate the per-atom root mean square error (RMSE), $\epsilon$, of the ensemble's predicted force as
\begin{equation}
    \epsilon = \sqrt{\frac{1}{3}\left((\bar{F}_{x} - F_x)^2 + (\bar{F}_{y} - F_y)^2 + (\bar{F}_{z} - F_z)^2\right)}
    \label{eqn:force-rmse-one-atom}
\end{equation}
and the force RMSE, $\bar{\epsilon}$, over all $N$ atoms in the test set as
\begin{equation}
    \bar{\epsilon} = \sqrt{\frac{1}{3N}\sum_{\alpha \in x, y, z} \left( \sum_{i=1}^N (\bar{F}_{i,\alpha}-F_{i,\alpha})^2\right)}
    \label{eqn:force-rmse}
\end{equation}
where $\bar{F}_{i, \alpha}$ and $F_{i, \alpha}$ denote the predicted and true 
$\alpha$-component of the force on atom $i$, respectively.

To obtain an uncertainty estimate, we calculate the standard deviation of a predicted force, $\sigma$, over the constituent networks. Because we are primarily interested in predictive uncertainties for molecular dynamics simulations, we investigate the uncertainty in the force components as opposed to the energies, since the forces determine the dynamics of the system. We thus calculate $\sigma$ as the square root of the mean of component-wise variances of the predicted forces:
\begin{equation}
    \sigma = \sqrt{\frac{1}{3J}\sum_{\alpha \in x, y, z}\left(\sum_j \left(\hat{F}_{j,\alpha} - \bar{F}_{\alpha}\right)^2\right)}
    \label{eqn:sigma}
\end{equation}
where $J$ is the number of constituent models (we use $J=10$) and $\hat{F}_{j,\alpha}$ denotes the $\alpha$-component of network $j$'s predicted force.

\subsection{Gaussian Mixture Model}
The aim of this work is to understand whether Gaussian mixture models (GMM), trained on a network's learned features, may provide a faster and more memory-efficent approach to uncertainty quantification in MLIPs. A GMM is a probabilistic model used in many applications -- including speaker verification \cite{reynolds2000speaker}, language identification \cite{torres2002language}, and computer vision \cite{huang2008new} -- due to its ability to represent a large class of data distributions \cite{reynolds2009gaussian}, which motivates us to investigate its capability of modeling a NequIP network's learned features. A GMM models a data distribution as a weighted sum of $M$ Gaussians,
\begin{equation}
    p(x|\theta) = \sum_{m=1}^M w_m\mathcal{N}(x|\mu_m, \Sigma_m),
\end{equation}
where $x$ is a $D$-dimensional, continuous-valued vector, $w_m$ is the weight of the $m$th Gaussian with the constraint $\sum_{m=1}^M w_m = 1$, and $\mathcal{N}(x|\mu_m, \Sigma_m)$ are the $D$-variate Gaussian densities,
\begin{equation}
    \mathcal{N}(x|\mu_m, \Sigma_m) = \frac{1}{(2\pi)^{D/2}|\Sigma_m|^{1/2}}e^{-\frac{1}{2}\left(x-\mu_m\right)^T\Sigma_m^{-1}\left(x-\mu_m\right)},
\end{equation}
with $D$-dimensional mean vector $\mu_m \in \mathbb{R}^D$ and $D \times D$-dimensional covariance matrix $\Sigma_m$ \cite{reynolds2009gaussian}. The parameters of the complete GMM are collected into $\theta$:
\begin{equation}
    \theta = \{w_m, \mu_m, \Sigma_m\}, \hspace{0.25cm} m\in [1,M].
\end{equation}

To construct the GMM, we first train and evaluate a NequIP model on the training set to access the per-atom final scalar features extracted from immediately before the linear projection down into the per-atom energy prediction. These final features are half the respective latent feature widths $f=16$ ($x \in \mathbb{R}^8$) and $f=32$ ($x \in \mathbb{R}^{16}$) used in our experiments. We then fit a GMM to model the distribution of these feature vectors as evaluated on the training set, denoted $X$, using the expectation maximization (EM) algorithm with each initial $\mu_m$ determined by k-means clustering \cite{reynolds2009gaussian, scikit-learn}. We fit the GMM using a full covariance matrix for each Gaussian, meaning that each $\Sigma_m$ is full rank and not shared between Gaussians \cite{reynolds2009gaussian}. We select the number of Gaussians using the Bayesian Information Criterion (BIC). To then estimate the uncertainty of a trained NequIP model on a test data point, we run a forward pass through NequIP to extract the final layer features for the atoms of that test structure. Subsequently, we evaluate the fitted GMM on the feature vector for each test atom $x$, obtaining a negative log-likelihood $\mathrm{NLL}({x|X})$:
\begin{equation}
    \mathrm{NLL}({x|X}) = -\log\left(\sum_{m = 1}^M w_m \mathcal{N}(x|\mu_m, \Sigma_m)\right). 
\end{equation}
A higher $\mathrm{NLL}({x|X})$ indicates higher uncertainty. Since the GMM is computationally light-weight, almost all computational burden lies in the evaluation of the NequIP features, which now occurs once instead of $J$ times in the ensembles.

\begin{figure}
    \includegraphics[scale=0.45, draft=false]{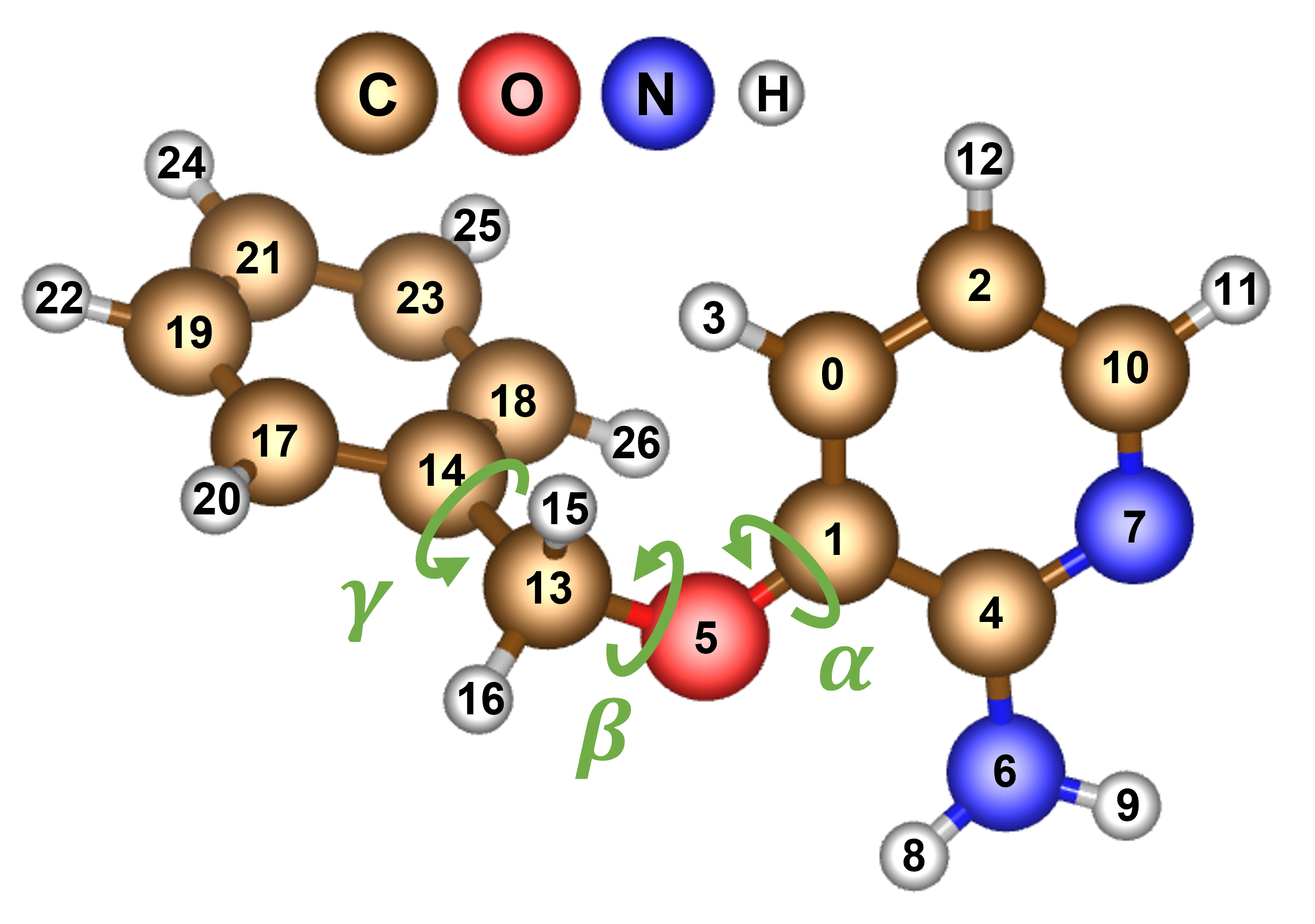}
    \caption{3D model of the 3BPA molecule with atomic indices and $\alpha, \beta,$ and $\gamma$ dihedral angles depicted.}
    \label{fig:bpa}
\end{figure}

\section{Results and Discussion}

\subsection{Data set}
We conduct our experiments on the 3-(benzyloxy)pyridin-2-amine (3BPA) transferability benchmark \cite{kovacs2021linear}. 3BPA (figure \ref{fig:bpa}) is a flexible drug-like molecule whose configurational diversity, which is largely determined by the three dihedral angles $\alpha$, $\beta$, and $\gamma$ and is explored more fully at higher temperatures, making it a challenging test case for MLIPs.

\begin{figure*}
    \includegraphics[width=\textwidth, draft=false]{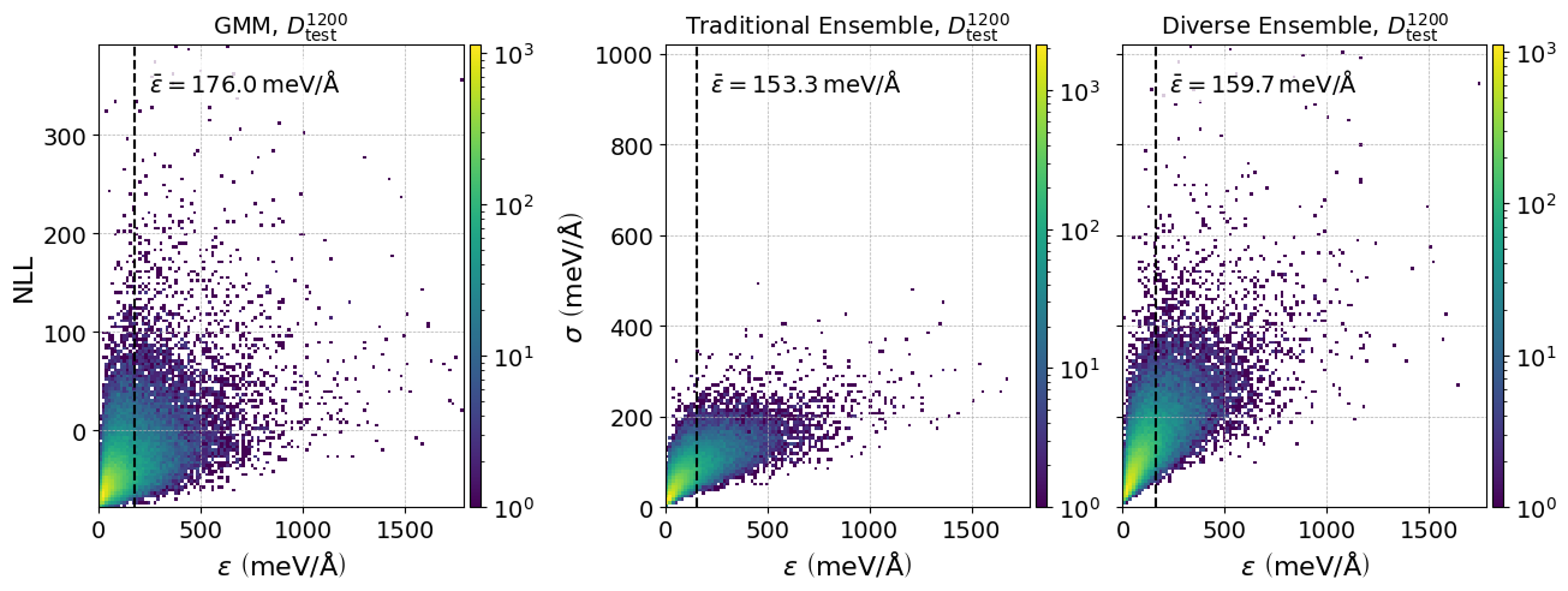}
    \caption{Plots of the uncertainty metric ($\sigma$ for the ensembles and NLL for the GMM) vs. $\epsilon$ for models with hidden feature dimension $f=32$, trained on $D_{\mathrm{train,100}}^{300}$ and evaluated on all atoms of all configurations in $D_{\mathrm{test}}^{1200}$. Each point $(\epsilon_i,U_i)$ in a given plot represents the model's force RMSE and predictive uncertainty, respectively, on a single atom $i$. The color bar represents the number of points within each bin. The vertical dashed line in each plot marks the average force RMSE $\bar{\epsilon}$ over all atoms (see Eq. \ref{eqn:force-rmse}).}
    \label{fig:uncertainty-vs-error-all-atoms-1200K}
\end{figure*}

In a first setting, we use three data sets of 3BPA structures sampled at three temperatures: 300K, 600K, and 1200K. We train on structures sampled at 300K (once with 50 structures, $D_{\mathrm{train, 50}}^{300}$, once with 100 structures, $D_{\mathrm{train, 100}}^{300}$). We set aside a pool of additional training data at 300K to select from ($D_{\mathrm{pool}}^{300}$). Finally, we have three test sets of structures at each temperature, i.e. $D_{\mathrm{test}}^{300}$, $D_{\mathrm{test}}^{600}$, and $D_{\mathrm{test}}^{1200}$, along with three additional data sets ($D_{\beta=120^{\circ}}$, $D_{\beta=150^{\circ}}$, and $D_{\beta=180^{\circ}}$), each consisting of structures with a fixed $\beta$ dihedral angle and uniformly sampled $\alpha$ and $\gamma$ angles.

In a second setting, we combine all data from the three temperatures and similarly split the combined data into training sets (of size 50, $D_{\mathrm{train, 50}}^{\mathrm{mixed}}$, and size 100, $D_{\mathrm{train, 100}}^{\mathrm{mixed}}$), a pool of additional training data to sample from ($D_{\mathrm{pool}}^{\mathrm{mixed}}$), and a test set ($D_{\mathrm{test}}^{\mathrm{mixed}}$).

\subsection{Uncertainty Quantification}

\begin{figure*}
\centering
     \begin{subfigure}[b]{\textwidth}
         \centering
         \includegraphics[width=0.8\textwidth]{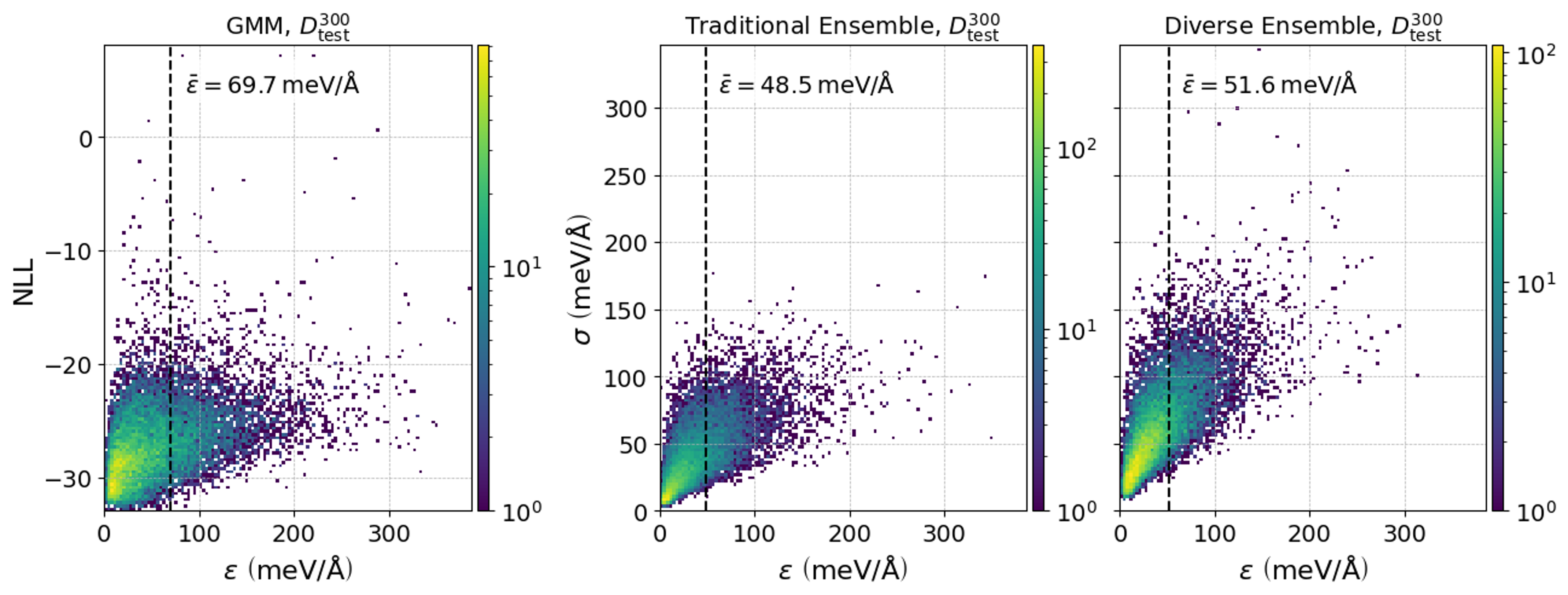}
     \end{subfigure}
     \hfill
     \begin{subfigure}[b]{\textwidth}
         \centering
         \includegraphics[width=0.8\textwidth]{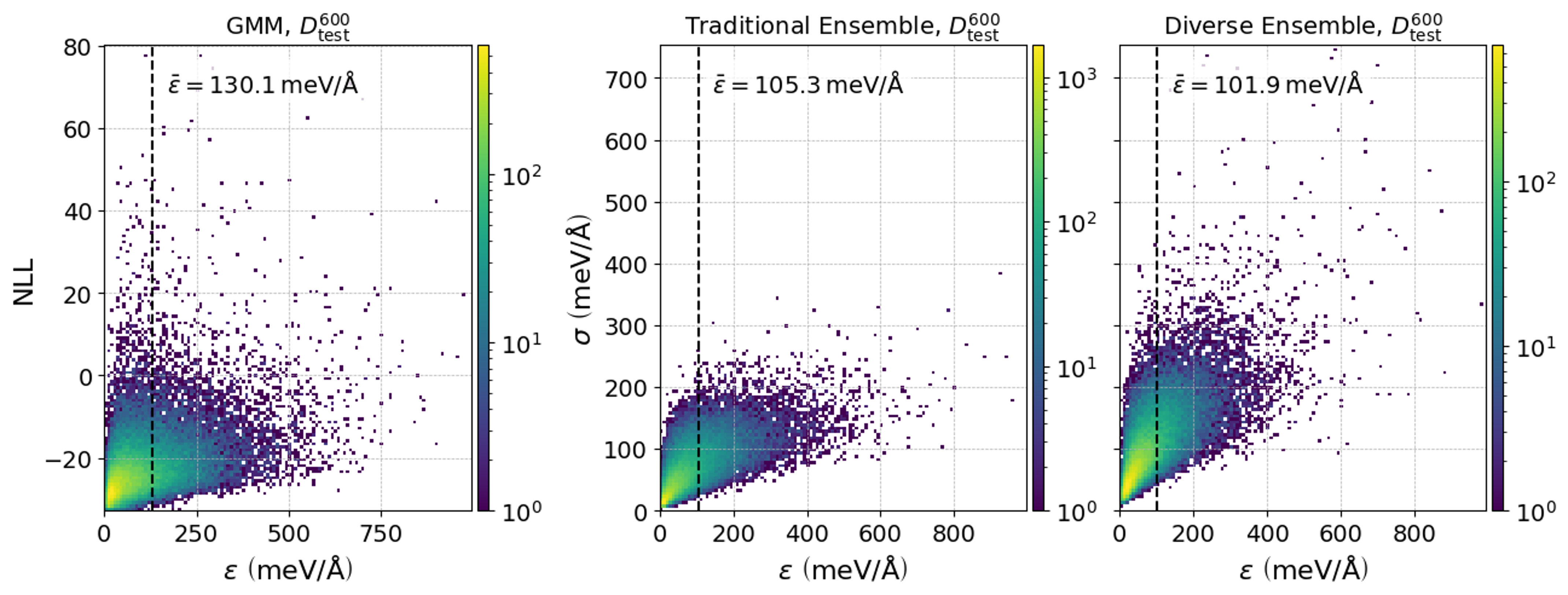}
     \end{subfigure}
     \hfill
    \begin{subfigure}[b]{\textwidth}
         \centering
         \includegraphics[width=0.8\textwidth]{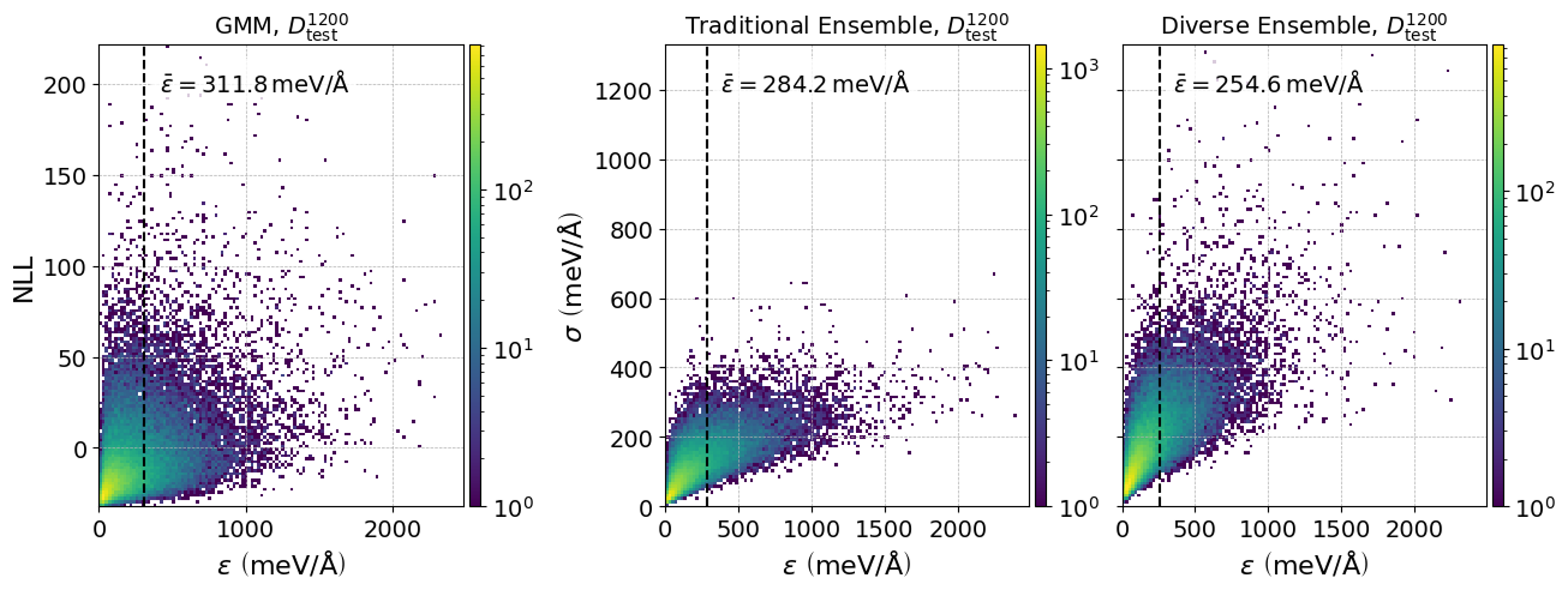}
     \end{subfigure}
     \hfill
     \begin{subfigure}[b]{\textwidth}
         \centering
         \includegraphics[width=0.8\textwidth]{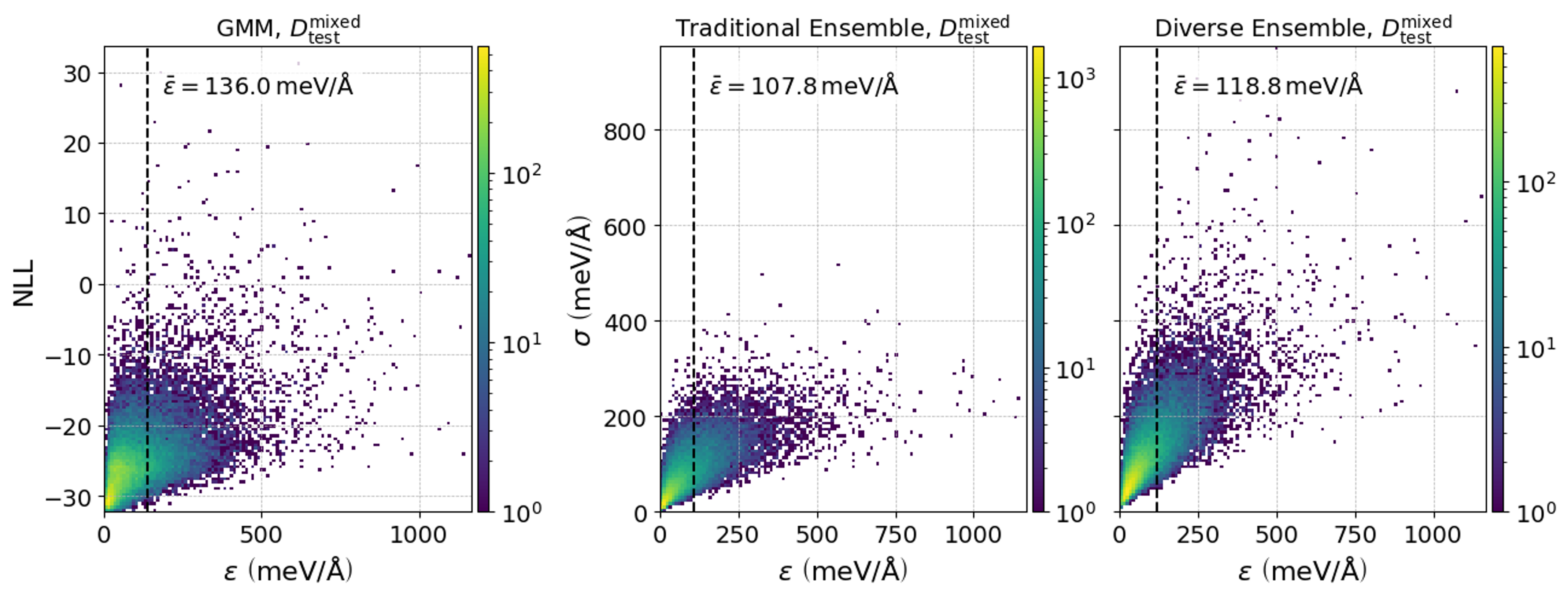}
    \end{subfigure}
    \caption{Uncertainty vs. $\epsilon$ for models with hidden feature dimension $f=16$ and trained on $D_{\mathrm{train,50}}^{300}$ for the single-temperature test sets and $D_{\mathrm{train,50}}^{\mathrm{mixed}}$ for the mixed-temperature test set.}
    \label{fig:uncertainty-vs-error-n50-f16-si}
\end{figure*}

\begin{figure*}
\centering
     \begin{subfigure}[b]{\textwidth}
         \centering
         \includegraphics[width=0.8\textwidth]{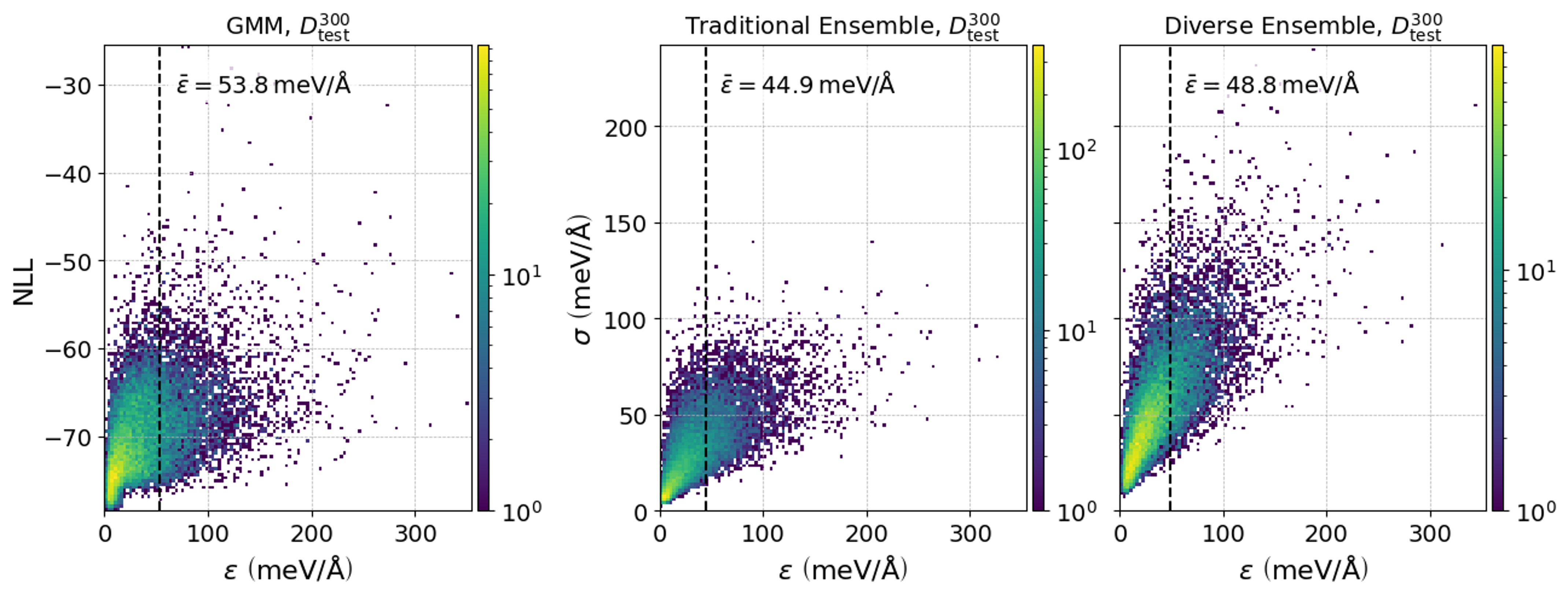}
     \end{subfigure}
     \hfill
     \begin{subfigure}[b]{\textwidth}
         \centering
         \includegraphics[width=0.8\textwidth]{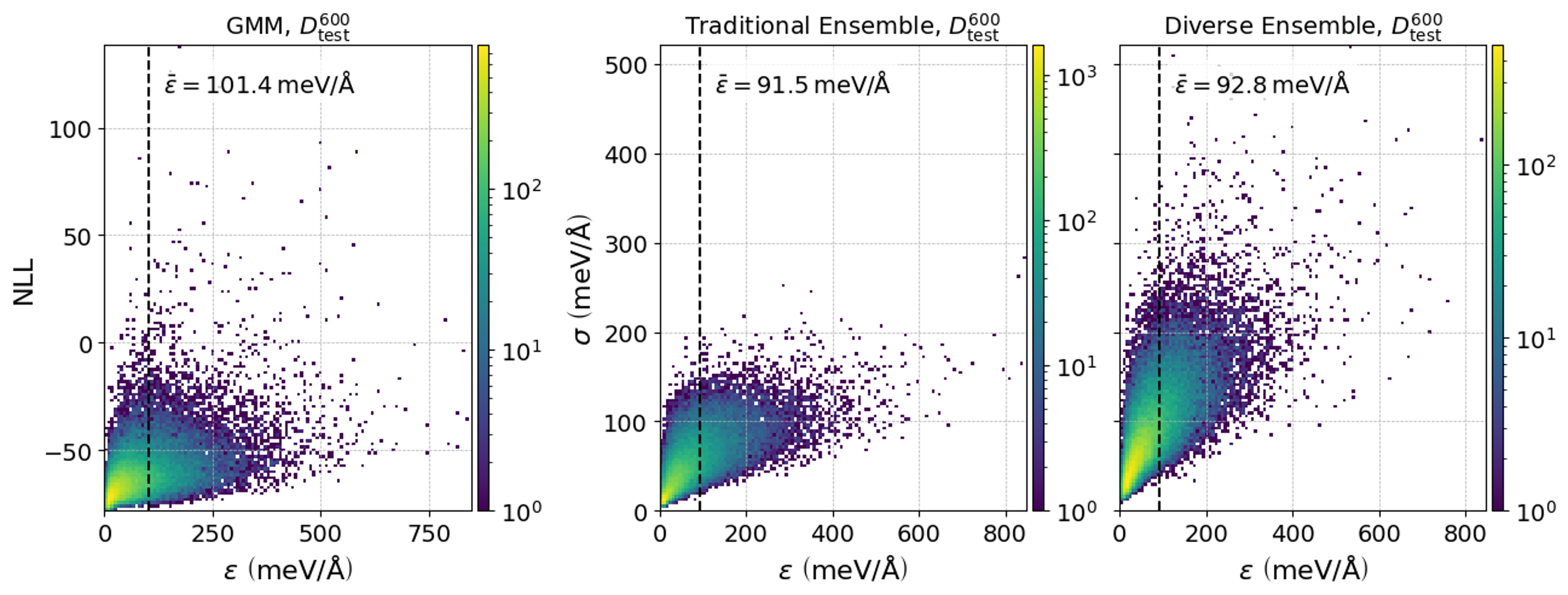}
     \end{subfigure}
     \hfill
    \begin{subfigure}[b]{\textwidth}
         \centering
         \includegraphics[width=0.8\textwidth]{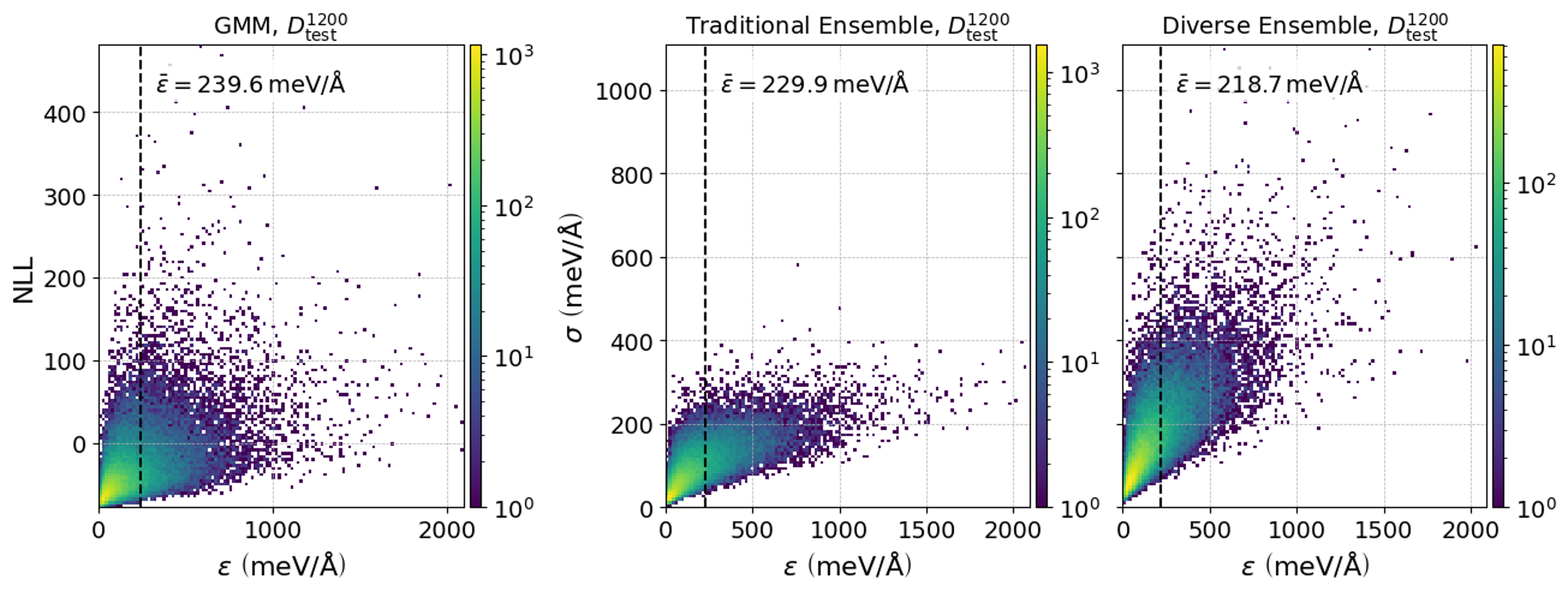}
     \end{subfigure}
     \hfill
     \begin{subfigure}[b]{\textwidth}
         \centering
         \includegraphics[width=0.8\textwidth]{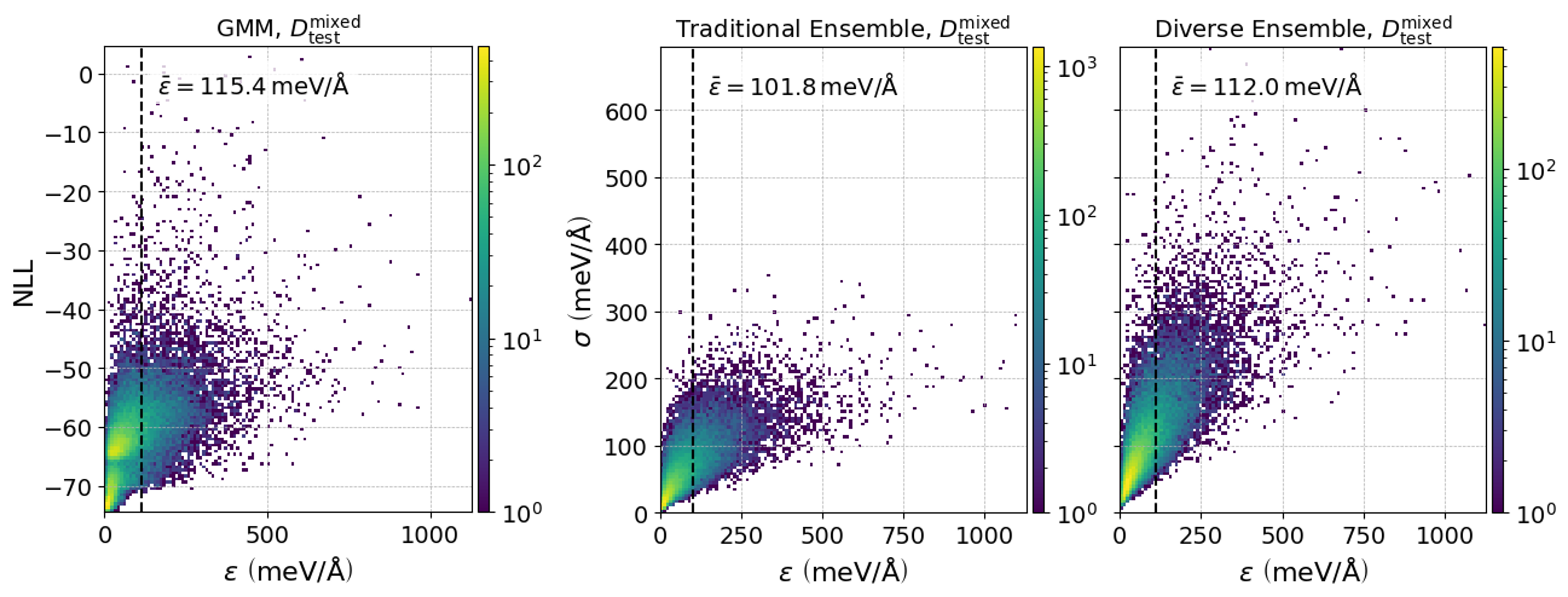}
    \end{subfigure}
    \caption{Uncertainty vs. $\epsilon$ for models with hidden feature dimension $f=32$ and trained on $D_{\mathrm{train,50}}^{300}$ for the single-temperature test sets and $D_{\mathrm{train,50}}^{\mathrm{mixed}}$ for the mixed-temperature test set.}
    \label{fig:uncertainty-vs-error-n50-f32-si}
\end{figure*}

\begin{figure*}
\centering
     \begin{subfigure}[b]{\textwidth}
         \centering
         \includegraphics[width=0.8\textwidth]{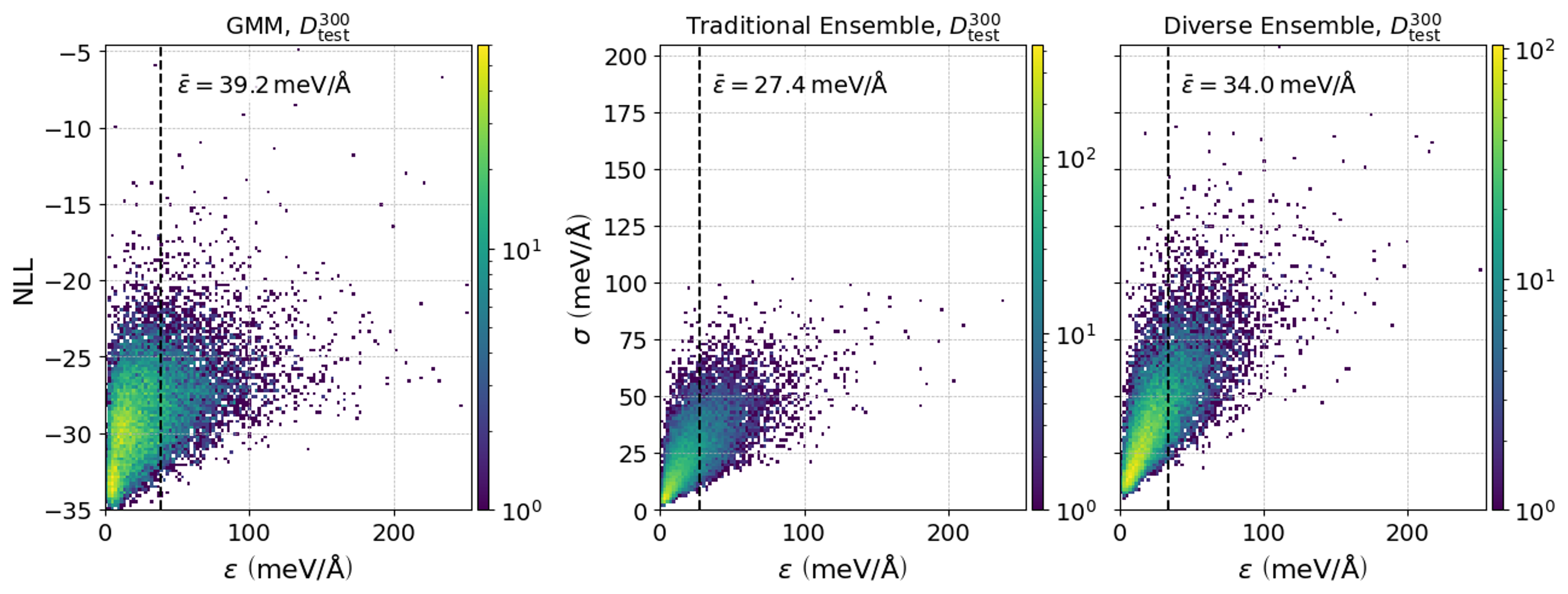}
     \end{subfigure}
     \hfill
     \begin{subfigure}[b]{\textwidth}
         \centering
         \includegraphics[width=0.8\textwidth]{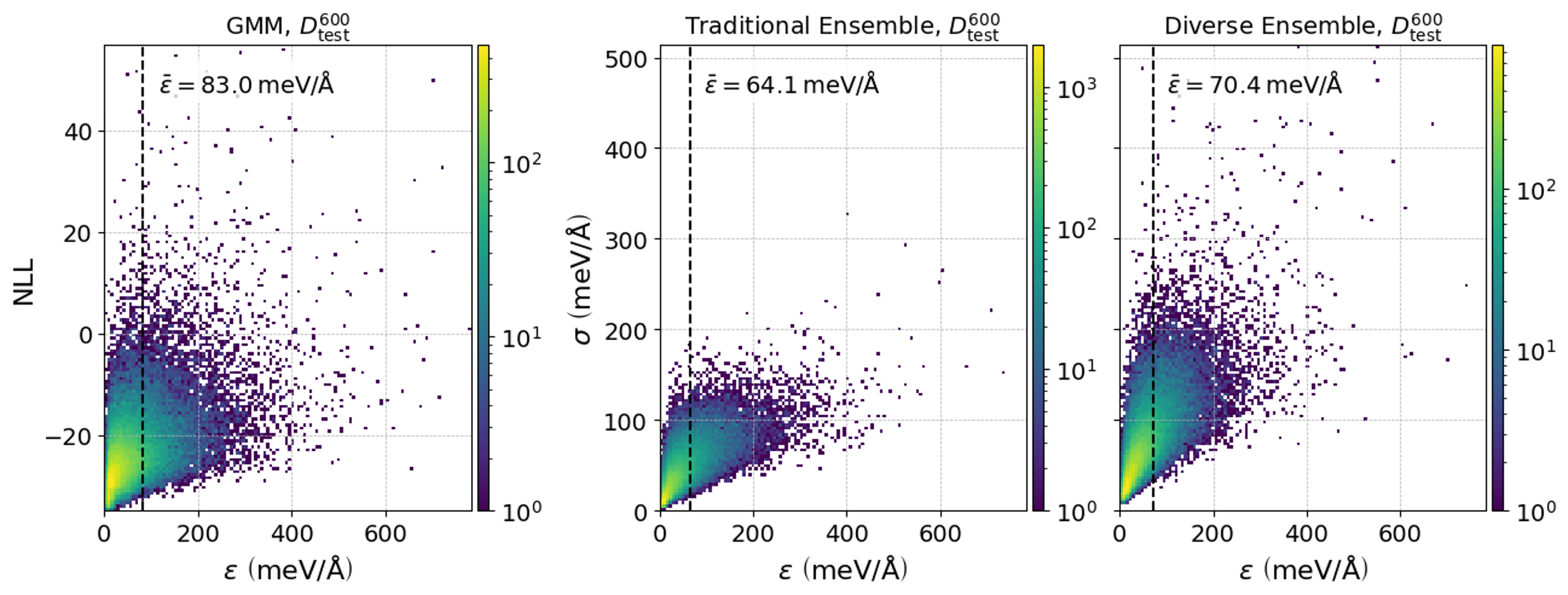}
     \end{subfigure}
     \hfill
    \begin{subfigure}[b]{\textwidth}
         \centering
         \includegraphics[width=0.8\textwidth]{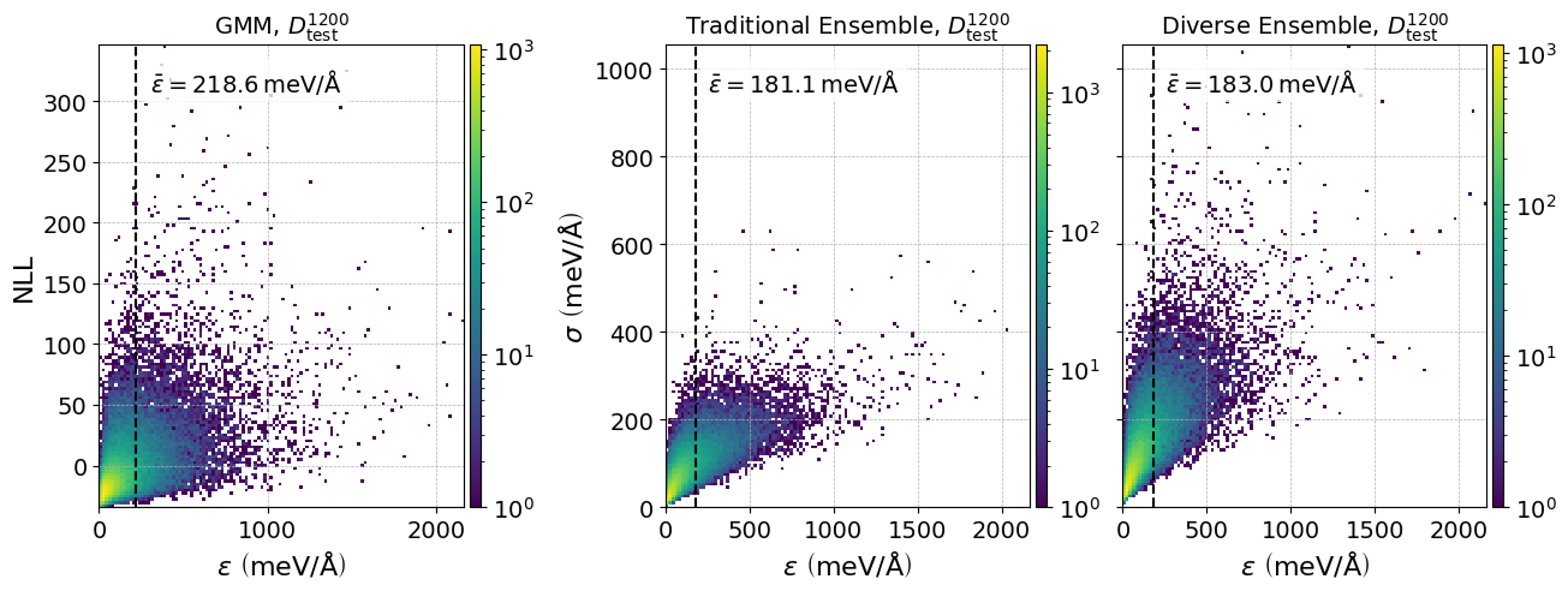}
     \end{subfigure}
     \hfill
     \begin{subfigure}[b]{\textwidth}
         \centering
         \includegraphics[width=0.8\textwidth]{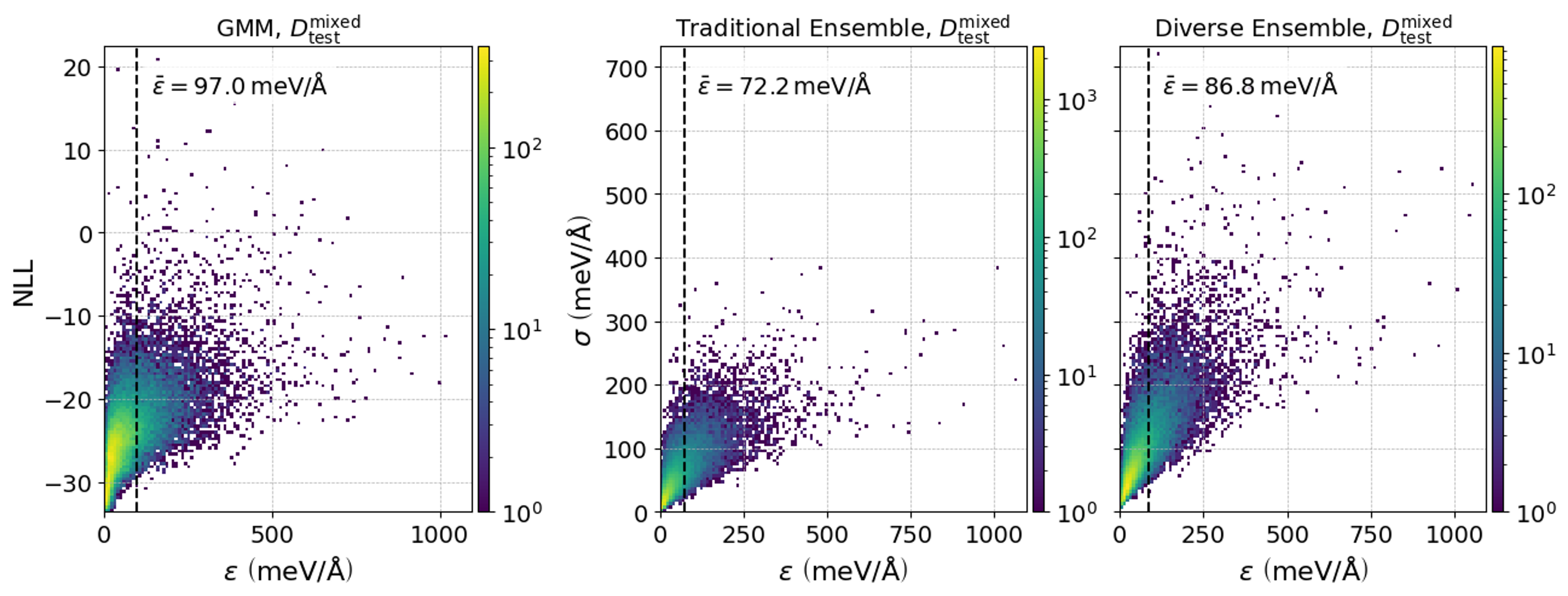}
    \end{subfigure}
    \caption{Uncertainty vs. $\epsilon$ for models with hidden feature dimension $f=16$ and trained on $D_{\mathrm{train,100}}^{300}$ for the single-temperature test sets and $D_{\mathrm{train,100}}^{\mathrm{mixed}}$ for the mixed-temperature test set.}
    \label{fig:uncertainty-vs-error-n100-f16-si}
\end{figure*}

\begin{figure*}
\centering
     \begin{subfigure}[b]{\textwidth}
         \centering
         \includegraphics[width=0.8\textwidth]{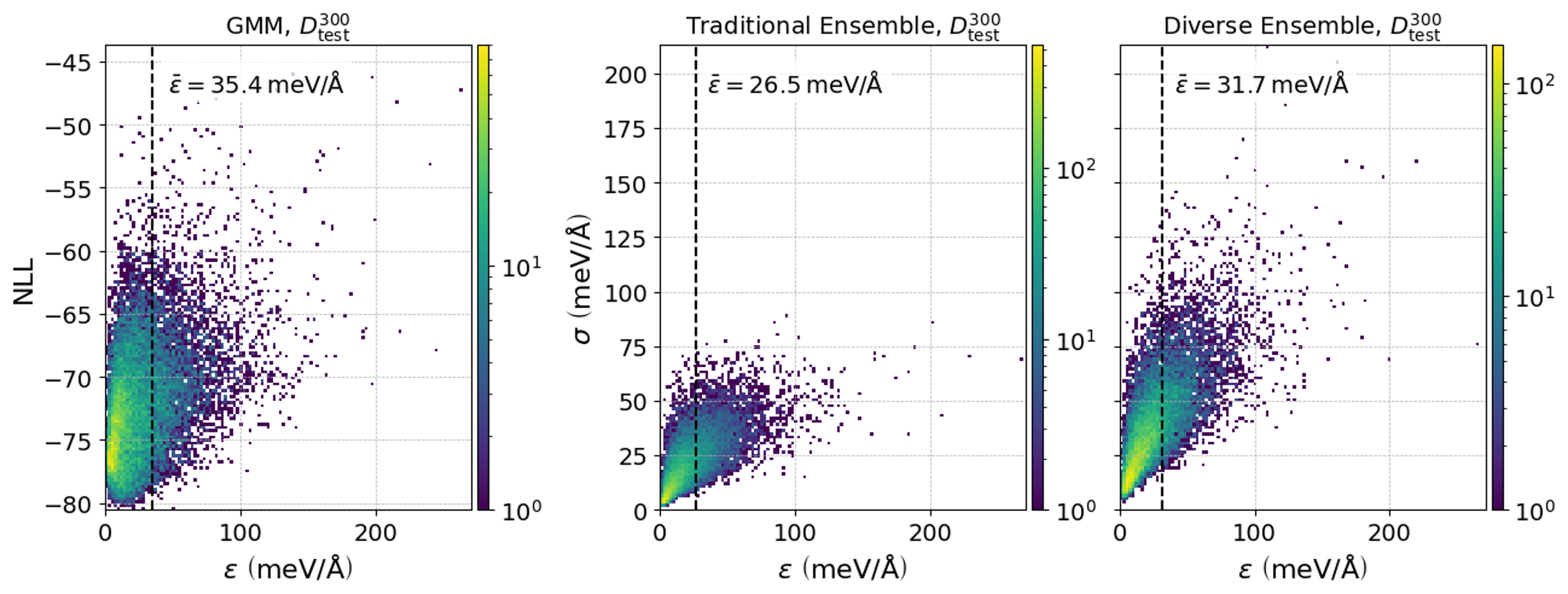}
     \end{subfigure}
     \hfill
     \begin{subfigure}[b]{\textwidth}
         \centering
         \includegraphics[width=0.8\textwidth]{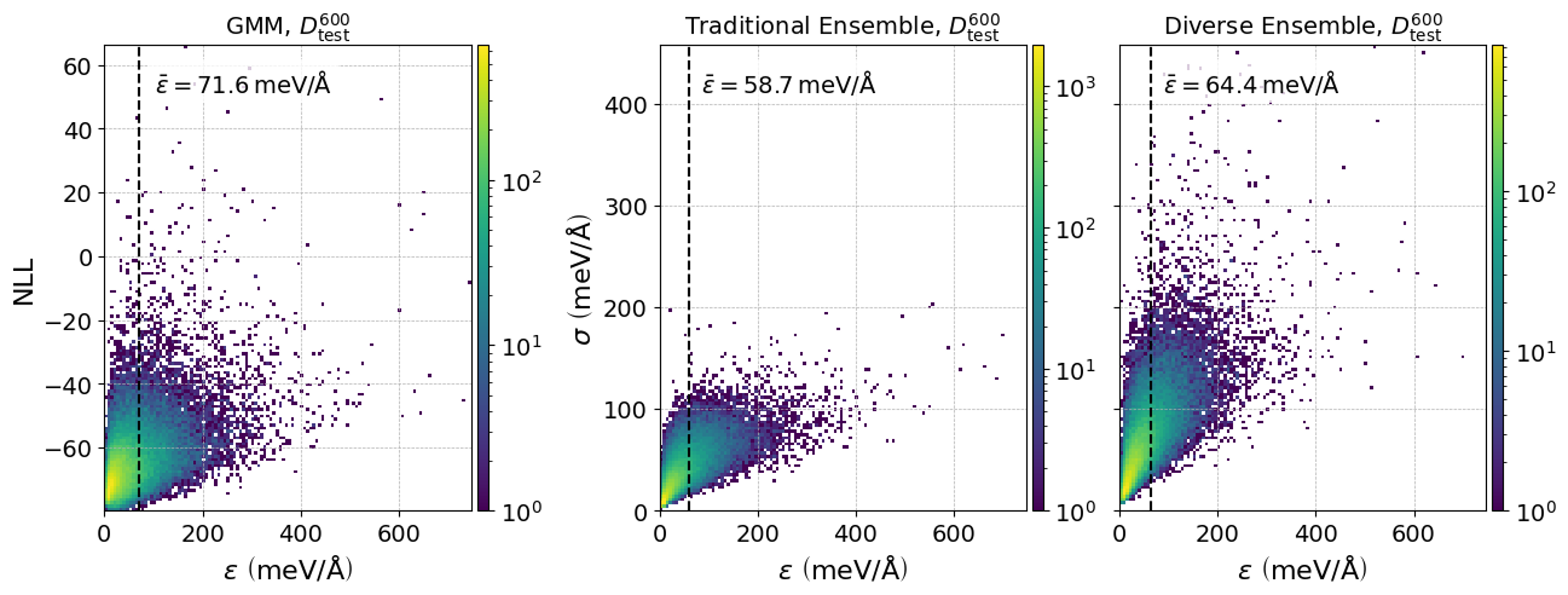}
     \end{subfigure}
     \hfill
    \begin{subfigure}[b]{\textwidth}
         \centering
         \includegraphics[width=0.8\textwidth]{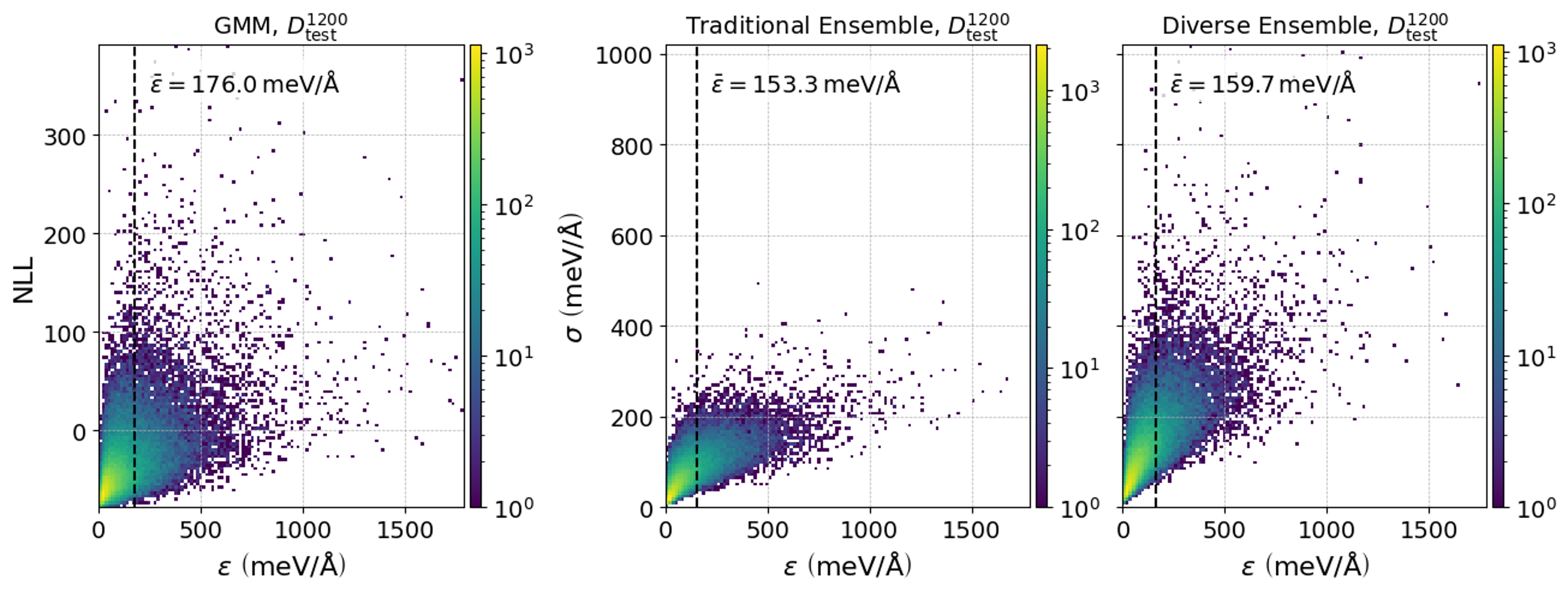}
     \end{subfigure}
     \hfill
     \begin{subfigure}[b]{\textwidth}
         \centering
         \includegraphics[width=0.8\textwidth]{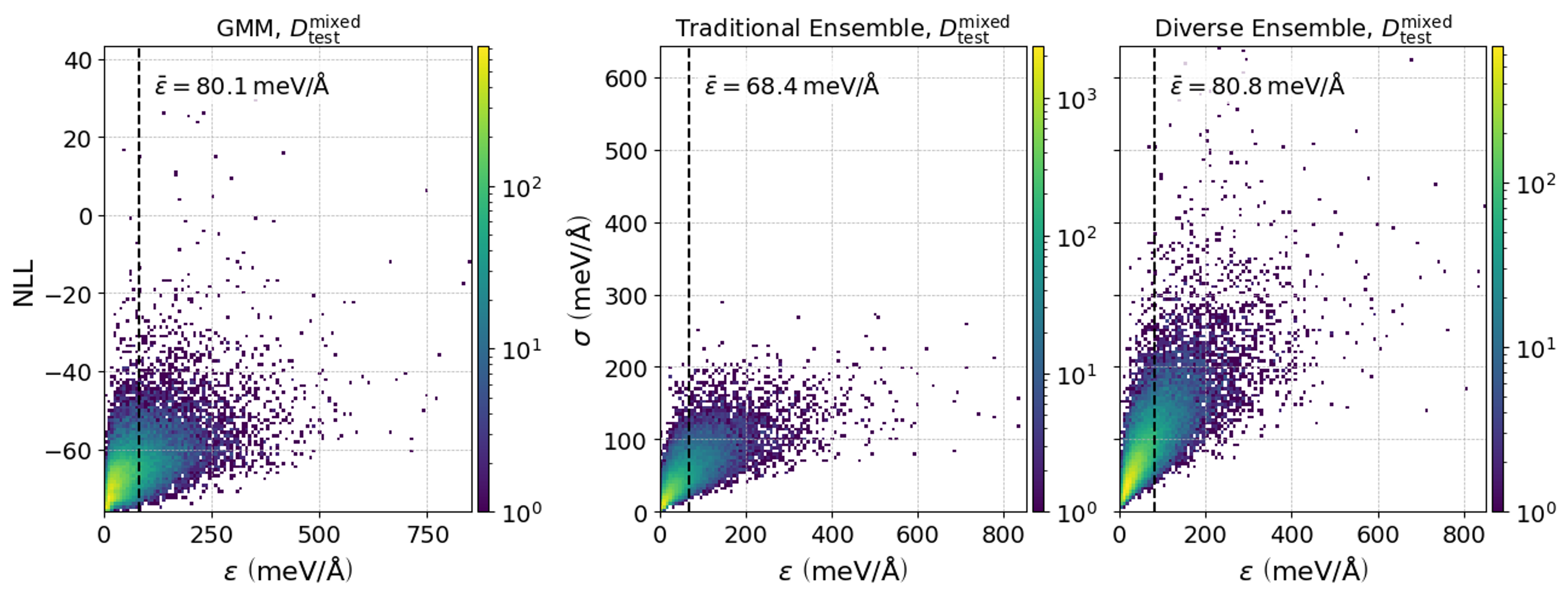}
    \end{subfigure}
    \caption{Uncertainty vs. $\epsilon$ for models with hidden feature dimension $f=32$ and trained on $D_{\mathrm{train,100}}^{300}$ for the single-temperature test sets and $D_{\mathrm{train,100}}^{\mathrm{mixed}}$ for the mixed-temperature test set. Third row of plots is presented in the manuscript.}
    \label{fig:uncertainty-vs-error-n100-f32-si}
\end{figure*}

In figure \ref{fig:uncertainty-vs-error-all-atoms-1200K}, we plot the uncertainty estimates of the ensembles and the GMM against the measured per-atom RMSE $\epsilon$ for models with hidden feature dimension $f = 32$, trained on $D_{\mathrm{train,100}}^{300}$ and evaluated on $D_{\mathrm{test}}^{1200}$. Plots for the evaluations on $D_{\mathrm{test}}^{300}$ and $D_{\mathrm{test}}^{600}$ and for the mixed-temperature setting show similar results (SI figure \ref{fig:uncertainty-vs-error-n100-f32-si}). Likewise, plots for models with hidden feature dimension $f=16$ and models trained on $D_{\mathrm{train,50}}^{300}$ and $D_{\mathrm{train,50}}^{\mathrm{mixed}}$ generally demonstrate the same results (SI figures \ref{fig:uncertainty-vs-error-n50-f16-si}, \ref{fig:uncertainty-vs-error-n50-f32-si}, \ref{fig:uncertainty-vs-error-n100-f16-si}).

We observe in figure \ref{fig:uncertainty-vs-error-all-atoms-1200K} that the traditional ensemble achieves the lowest $\bar{\epsilon}$  while the single model used for fitting the GMM has the highest $\bar{\epsilon}$. This result is expected as the average ensemble prediction has been observed to be more accurate than the prediction of a single model \cite{dietterich2000ensemble}. The traditional ensemble generally performs better than the diverse ensemble since the diverse ensemble contains simpler models on average. The distribution of $\sigma$ in the diverse ensemble is generally shifted towards higher values than the traditional ensemble, likely due to the differences in network architecture resulting in more diverse predictions. As expected, SI figure \ref{fig:uncertainty-vs-error-n100-f32-si} shows that the distribution of $\epsilon$ shifts towards higher errors as the temperature of the test set increases from 300K to 1200K. More notably, the distribution of $\sigma$ of all methods shifts towards larger positive values with increasing temperature, demonstrating the ability of all methods to detect high-energy, out-of-distribution configurations. Overall, we observe a modest correlation between the uncertainty metric and $\epsilon$ for each of the approaches, with the GMM's correlation similar to those of the ensembles. Most importantly, a GMM evaluated on a \emph{single network} has similar predictive power of the uncertainty as an ensemble of ten networks, providing a way to reduce the computational cost of uncertainty quantification by an order of magnitude while maintaining the state-of-the-art performance of ensemble-based uncertainties.\\

To further quantify the quality of these uncertainty metrics, we establish certain criteria that a good uncertainty metric should meet. Since uncertainty estimates are often used to identify high-error structures for which to invoke first principles calculations, we set some uncertainty cutoff $U_{\mathrm{cutoff}}$ for capturing such structures. In particular, we classify all configurations with uncertainty $U > U_{\mathrm{cutoff}}$ as ``high-error" and all configurations with uncertainty $U \leq U_{\mathrm{cutoff}}$ as ``low error." A good uncertainty metric should simultaneously classify a large proportion of configurations with $\epsilon > \epsilon_{\mathrm{cutoff}}$ as high-error to avoid missing configurations with high true $\epsilon$ while classifying a small proportion of configurations with true $\epsilon \leq \epsilon_{\mathrm{cutoff}}$ as high-error configurations to avoid redundant calls to DFT calculations. In other words, a good uncertainty metric should achieve a high true positive rate (TPR) and a high positive predictive value (PPV) \cite{Kahle_2022}:
\begin{equation}
    \mathrm{TPR = \frac{TP}{TP + FN}}
\end{equation}
\begin{equation}
    \mathrm{PPV = \frac{TP}{TP + FP}}
\end{equation}
where TP (true positives) is the number of configurations with $U > U_{\mathrm{cutoff}}$ and $\epsilon > \epsilon_{\mathrm{cutoff}}$, FN (false negatives) is the number of configurations with $U \leq U_{\mathrm{cutoff}}$ and $\epsilon > \epsilon_{\mathrm{cutoff}}$, and FP (false positives) is the number of configurations with $U > U_{\mathrm{cutoff}}$ and $\epsilon \leq \epsilon_{\mathrm{cutoff}}$. Thus, to quantify the quality of each method's uncertainty estimates, we calculate their TPR and PPV (as done in \cite{Kahle_2022}) for a range of $U_{\mathrm{cutoff}}$ and $\epsilon_{\mathrm{cutoff}}$. 

\begin{figure*}
    \includegraphics[width=0.7\textwidth, draft=false]{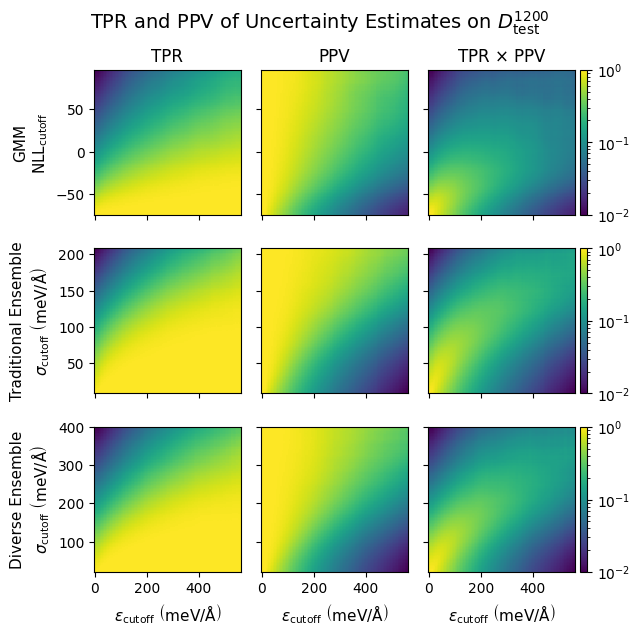}
    \caption{TPR and PPV profiles of uncertainty estimates on all atoms in $D_{\mathrm{test}}^{1200}$.}
    \label{fig:tpr-ppv-all-atoms}
\end{figure*}

\begin{figure*}
\centering
     \begin{subfigure}[b]{0.49\textwidth}
         \centering
         \includegraphics[width=\textwidth]{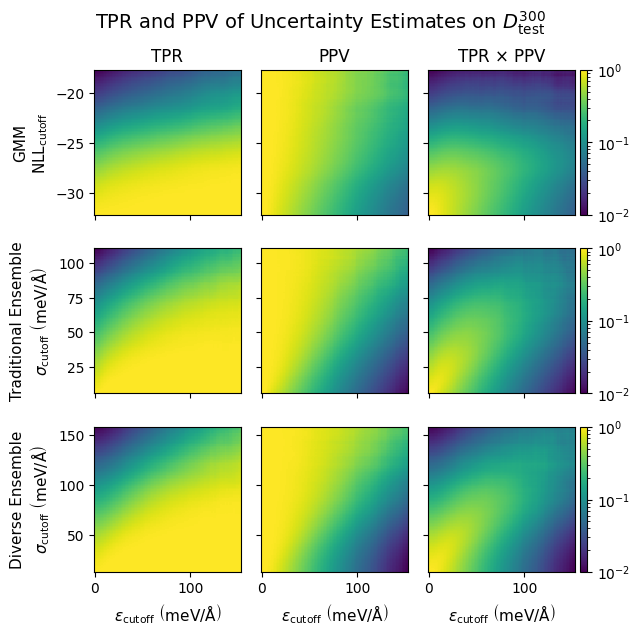}
     \end{subfigure}
     \hfill
     \begin{subfigure}[b]{0.49\textwidth}
         \centering
         \includegraphics[width=\textwidth]{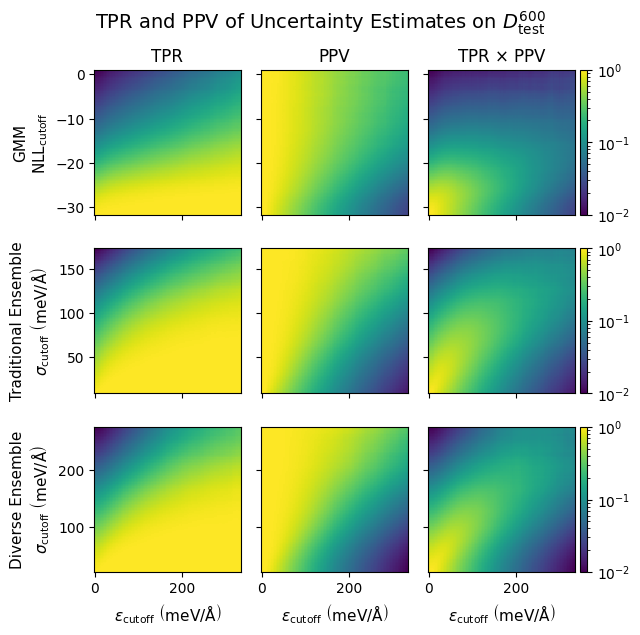}
     \end{subfigure}
     \hfill
     \begin{subfigure}[b]{0.49\textwidth}
         \centering
         \includegraphics[width=\textwidth]{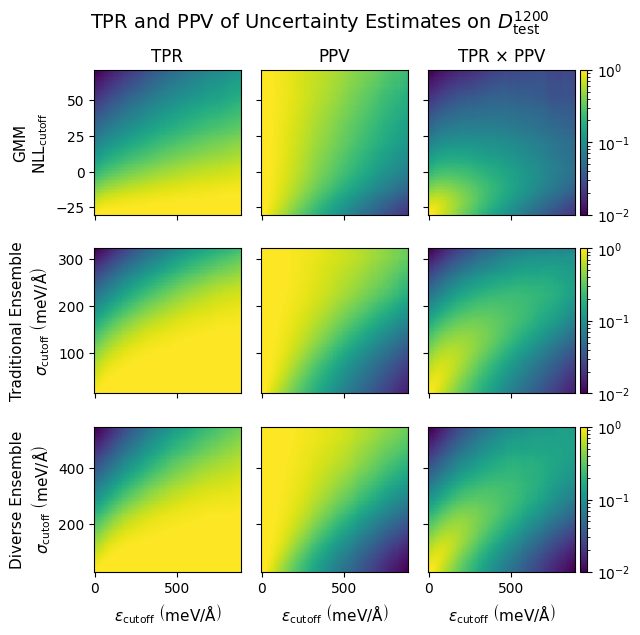}
     \end{subfigure}
     \hfill
     \begin{subfigure}[b]{0.49\textwidth}
         \centering
         \includegraphics[width=\textwidth]{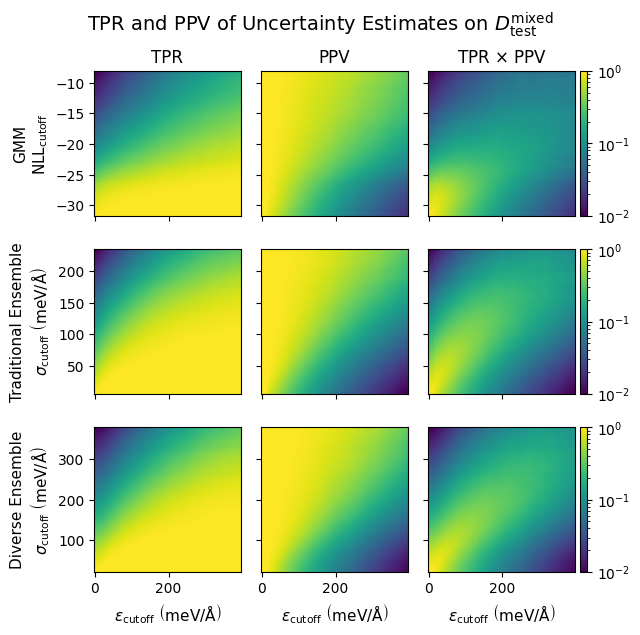}
    \end{subfigure}
    \caption{TPR and PPV profiles of uncertainty estimates on all atoms for models with hidden feature dimension $f=16$ and trained on $D_{\mathrm{train, 50}}^{300}$ for the single-temperature test sets and $D_{\mathrm{train,50}}^{\mathrm{mixed}}$ for the mixed-temperature test set.}
    \label{fig:tpr-ppv-n50-f16-si}
\end{figure*}

\begin{figure*}
\centering
     \begin{subfigure}[b]{0.49\textwidth}
         \centering
         \includegraphics[width=\textwidth]{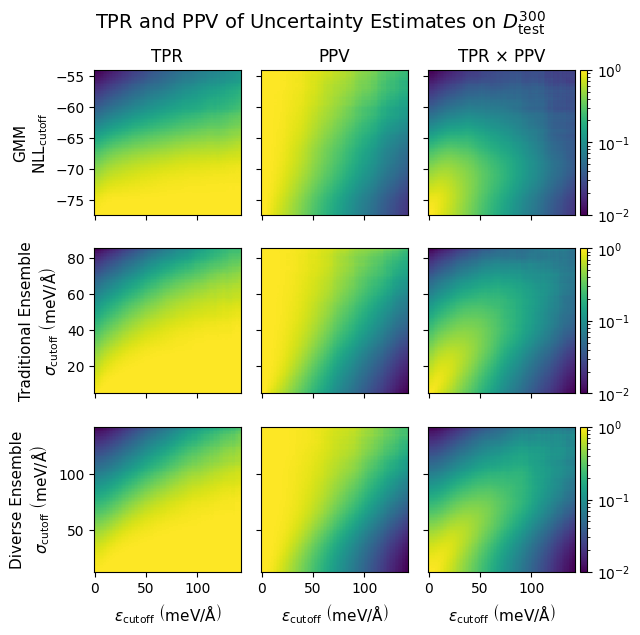}
     \end{subfigure}
     \hfill
     \begin{subfigure}[b]{0.49\textwidth}
         \centering
         \includegraphics[width=\textwidth]{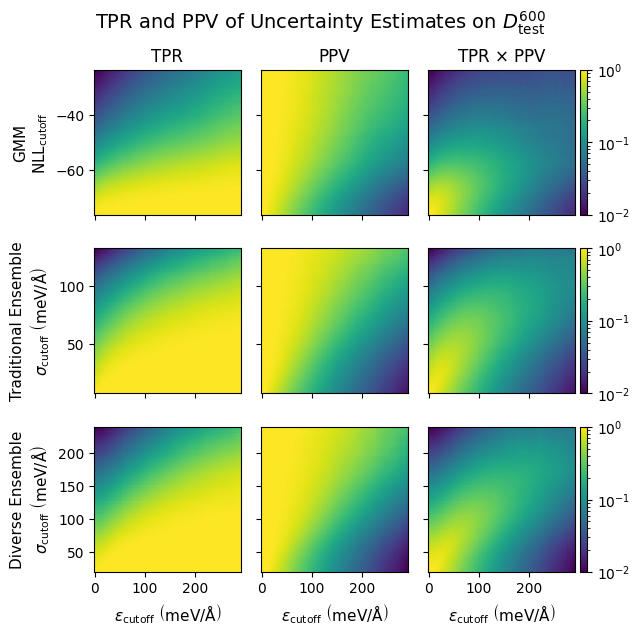}
     \end{subfigure}
     \hfill
     \begin{subfigure}[b]{0.49\textwidth}
         \centering
         \includegraphics[width=\textwidth]{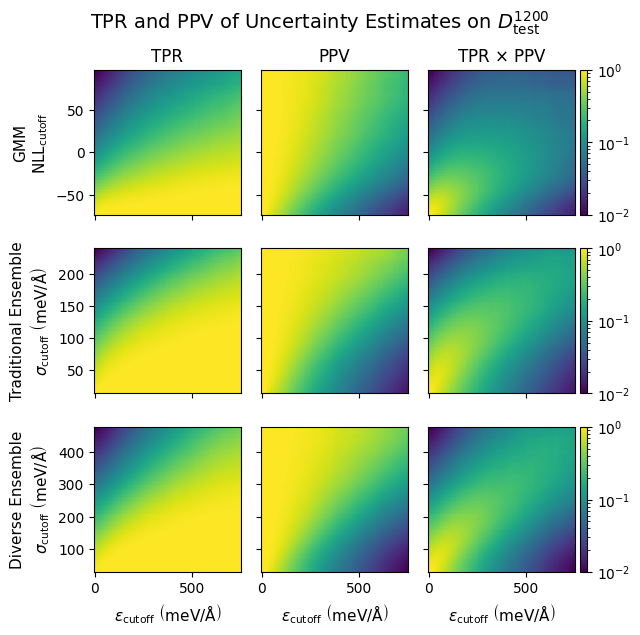}
     \end{subfigure}
     \hfill
     \begin{subfigure}[b]{0.49\textwidth}
         \centering
         \includegraphics[width=\textwidth]{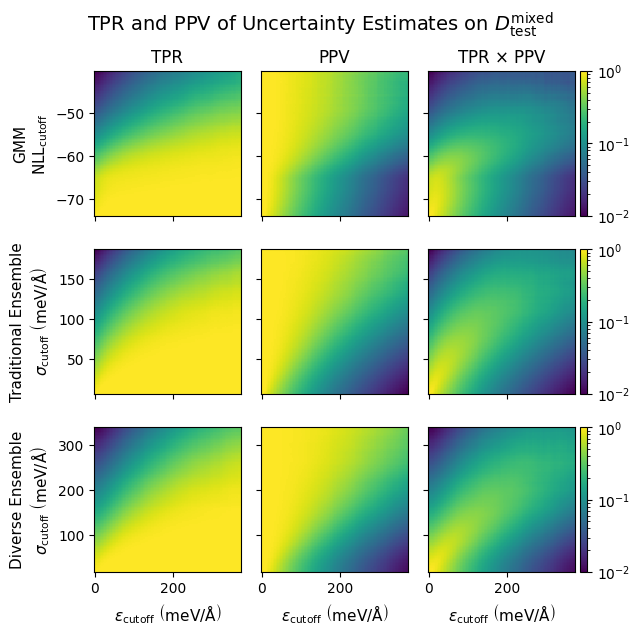}
    \end{subfigure}
    \caption{TPR and PPV profiles of uncertainty estimates on all atoms for models with hidden feature dimension $f=32$ and trained on $D_{\mathrm{train, 50}}^{300}$ for the single-temperature test sets and $D_{\mathrm{train,50}}^{\mathrm{mixed}}$ for the mixed-temperature test set.}
    \label{fig:tpr-ppv-n50-f32-si}
\end{figure*}

\begin{figure*}
\centering
     \begin{subfigure}[b]{0.49\textwidth}
         \centering
         \includegraphics[width=\textwidth]{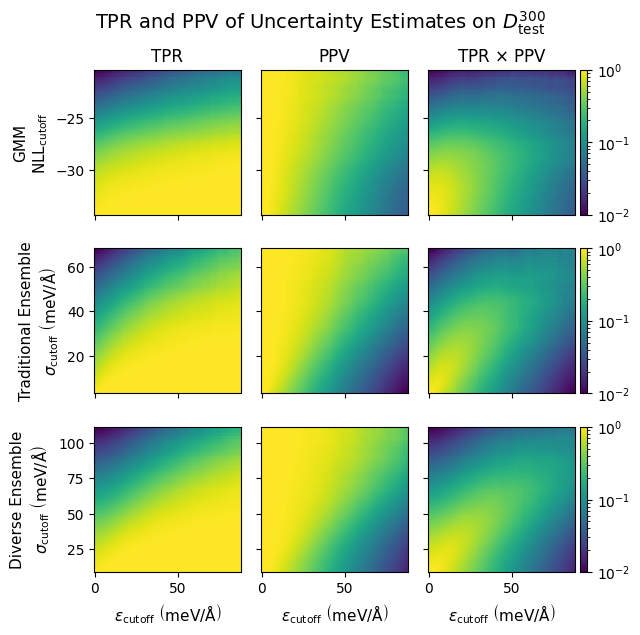}
     \end{subfigure}
     \hfill
     \begin{subfigure}[b]{0.49\textwidth}
         \centering
         \includegraphics[width=\textwidth]{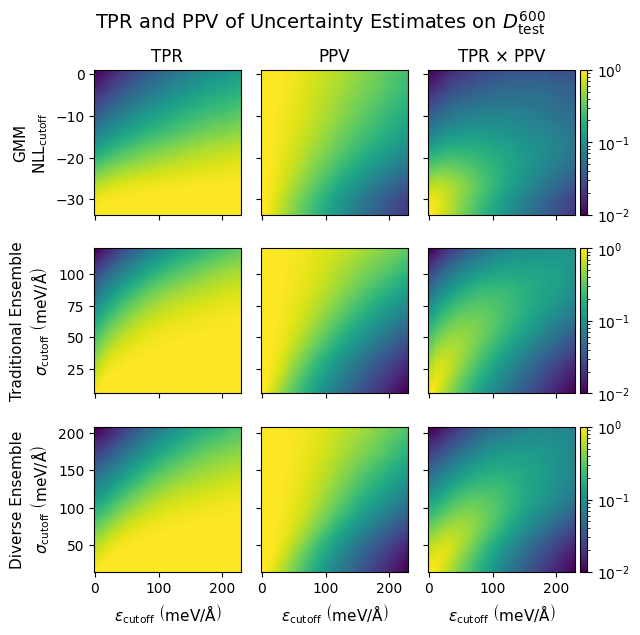}
     \end{subfigure}
     \hfill
     \begin{subfigure}[b]{0.49\textwidth}
         \centering
         \includegraphics[width=\textwidth]{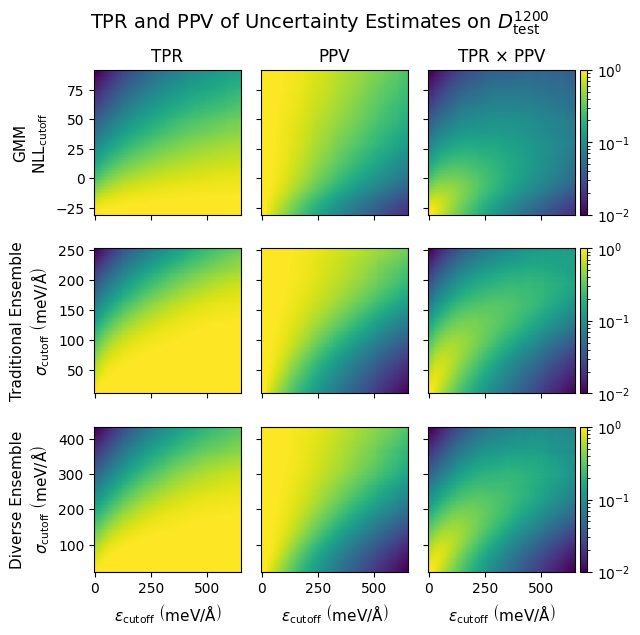}
     \end{subfigure}
     \hfill
     \begin{subfigure}[b]{0.49\textwidth}
         \centering
         \includegraphics[width=\textwidth]{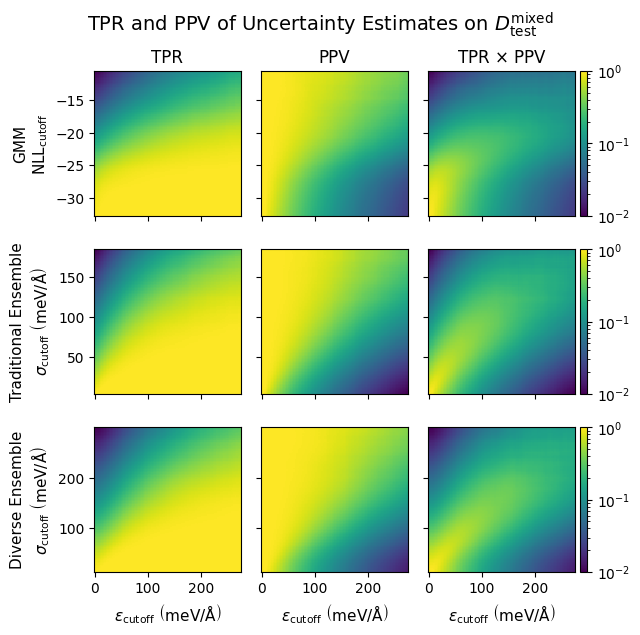}
    \end{subfigure}
    \caption{TPR and PPV profiles of uncertainty estimates on all atoms for models with hidden feature dimension $f=16$ and trained on $D_{\mathrm{train, 100}}^{300}$ for the single-temperature test sets and $D_{\mathrm{train,100}}^{\mathrm{mixed}}$ for the mixed-temperature test set.}
    \label{fig:tpr-ppv-n100-f16-si}
\end{figure*}

\begin{figure*}
\centering
     \begin{subfigure}[b]{0.49\textwidth}
         \centering
         \includegraphics[width=\textwidth]{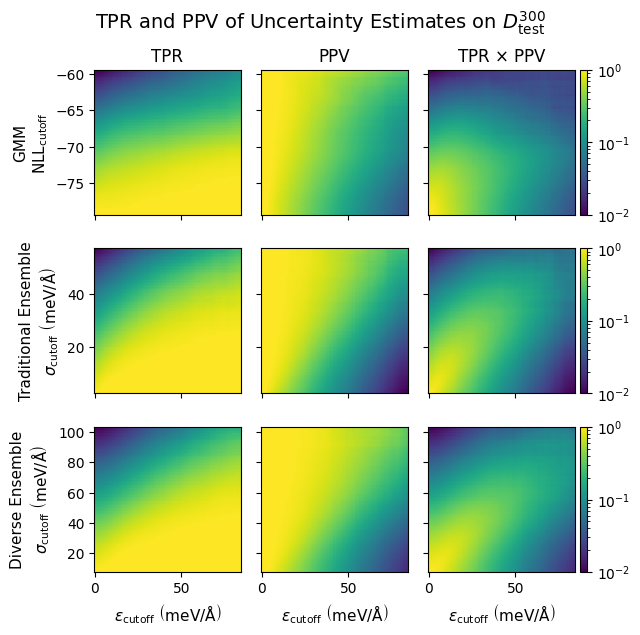}
     \end{subfigure}
     \hfill
     \begin{subfigure}[b]{0.49\textwidth}
         \centering
         \includegraphics[width=\textwidth]{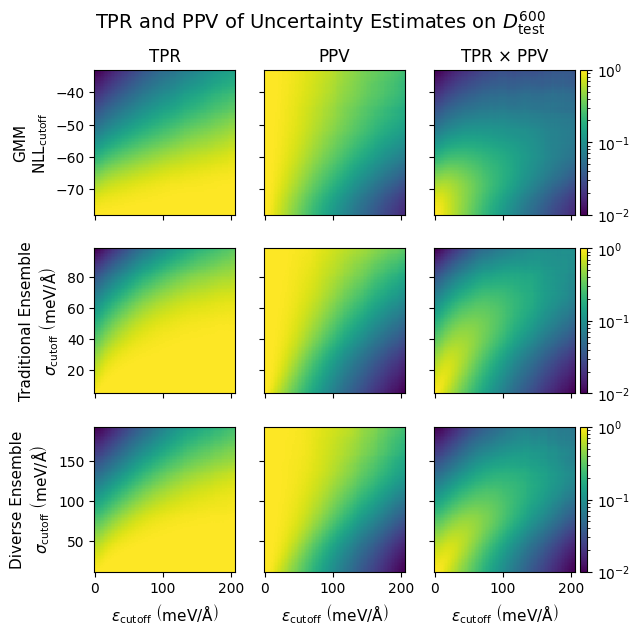}
     \end{subfigure}
     \hfill
     \begin{subfigure}[b]{0.49\textwidth}
         \centering
         \includegraphics[width=\textwidth]{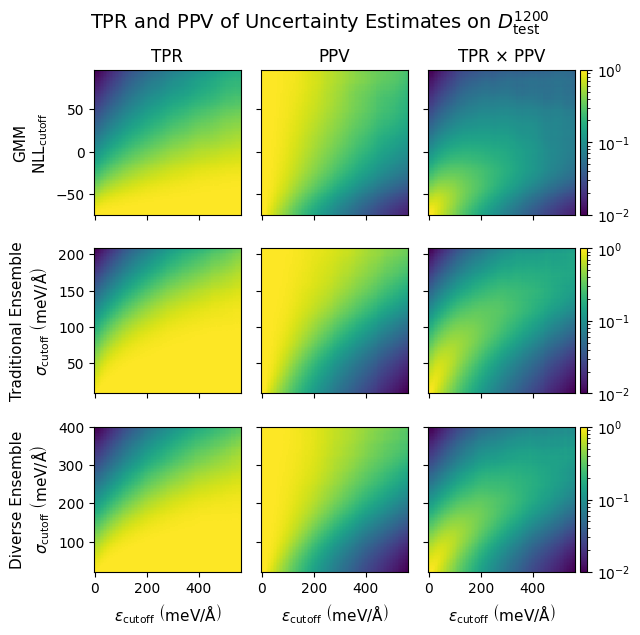}
     \end{subfigure}
     \hfill
     \begin{subfigure}[b]{0.49\textwidth}
         \centering
         \includegraphics[width=\textwidth]{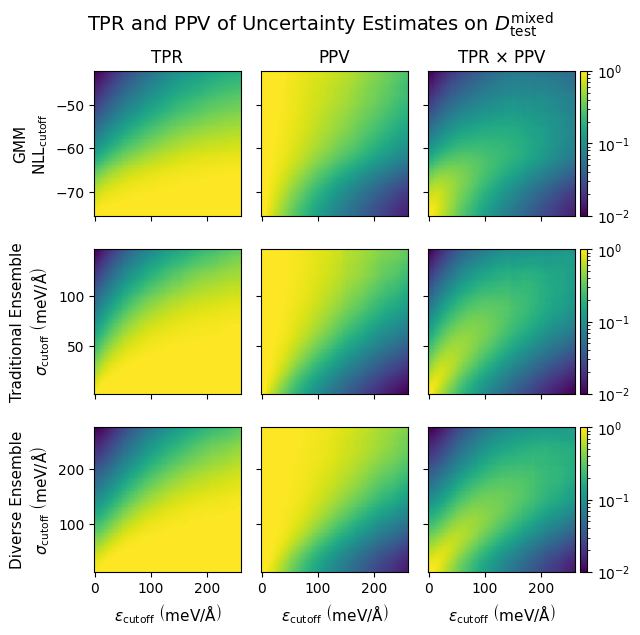}
    \end{subfigure}
    \caption{TPR and PPV profiles of uncertainty estimates on all atoms for models with hidden feature dimension $f=32$ and trained on $D_{\mathrm{train, 100}}^{300}$ for the single-temperature test sets and $D_{\mathrm{train,100}}^{\mathrm{mixed}}$ for the mixed-temperature test set. Plots for $D_{\mathrm{test}}^{1200}$ are presented in the manuscript.}
    \label{fig:tpr-ppv-n100-f32-si}
\end{figure*}

Figure \ref{fig:tpr-ppv-all-atoms} shows the TPR and PPV of the uncertainty metrics for ranges of $U_{\mathrm{cutoff}}$ and $\epsilon_{\mathrm{cutoff}}$ on all atoms in $D_{\mathrm{test}}^{1200}$, along with the product of the TPR and PPV ($\mathrm{TPR \times PPV}$). In each plot, $\epsilon_{\mathrm{cutoff}}$ ranges from 0 to the $99$th percentile of the $\epsilon$'s of the corresponding method on all atoms, and $U_{\mathrm{cutoff}}$ ranges from the $1$st to $99$th percentile of that method's uncertainty estimates. Ranges are chosen to minimize the impacts of outliers on the TPR and PPV. SI figures \ref{fig:tpr-ppv-n50-f16-si}, \ref{fig:tpr-ppv-n50-f32-si}, \ref{fig:tpr-ppv-n100-f16-si}, and \ref{fig:tpr-ppv-n100-f32-si} show plots for the other test sets, training sets, and model hyperparameters.

\begin{figure*}
    \includegraphics[width=\textwidth, draft=false]{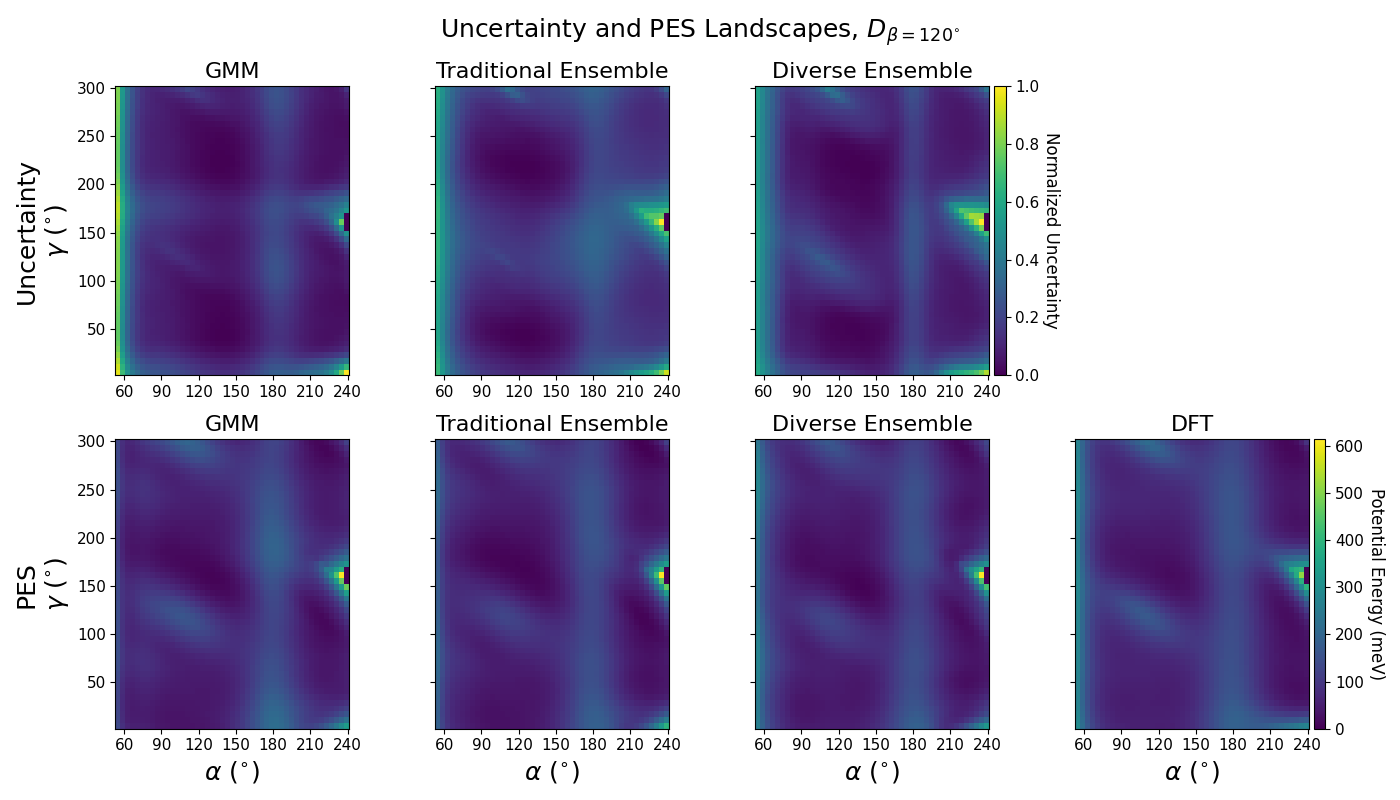}
    \caption{Top row: Uncertainty landscape of the GMM and ensembles for $D_{\beta=120^{\circ}}$. Color scale represents normalized summed force uncertainty over all atoms in a configuration. Bottom row: Potential energy surface (PES) landscape of the GMM, ensembles, and reference DFT calculations for $D_{\beta=120^{\circ}}$. Color scale represents energy relative to the minimum energy over all configurations in units of $\mathrm{meV}$, with yellow indicating higher energy. Purple space at $\alpha \approx 240^{\circ}$, $\gamma \approx 160^{\circ}$ indicates missing data.}
    \label{fig:uncertainty-landscape-beta120}
\end{figure*}

We observe that for a given $\epsilon_{\mathrm{cutoff}}$ in all approaches, lowering $U_{\mathrm{cutoff}}$ captures more high true-error points but incurs more false positives, evidenced by an increase in the TPR but a decrease in the PPV. The GMM achieves $\mathrm{TPR} \approx 1$ for slightly smaller but comparable ranges of $U_{\mathrm{cutoff}}$ and $\epsilon_{\mathrm{cutoff}}$ compared to both ensemble types. All approaches achieve $\mathrm{PPV} \approx 1$ for similar ranges of $U_{\mathrm{cutoff}}$ and $\epsilon_{\mathrm{cutoff}}$, with the GMM's PPV decaying slightly more slowly than that of the ensembles as $\epsilon_{\mathrm{cutoff}}$ increases and $U_{\mathrm{cutoff}}$ decreases. Notably, the similarity in the TPR and PPV profiles between the two ensemble types indicates that diversifying an ensemble increases $\sigma$ for each data point but does not necessarily improve the quality of the uncertainty estimate. Overall, we conclude that for all methods, it is difficult to simultaneously achieve $\mathrm{TPR} \approx 1$ and $\mathrm{PPV} \approx 1$ unless we set a $U_{\mathrm{cutoff}}$ around the 30th percentile of all uncertainty estimates of a given method, marking a majority of configurations for recalculation with DFT. Most importantly, the GMM produces uncertainty estimates comparable in quality to both ensemble types at a much lower computational cost.

\subsection{Uncertainty Landscapes}

To further investigate the similarity between the uncertainty metrics obtained by the ensembles and the GMM, we create ``uncertainty landscapes" in which we plot the uncertainty of each method for configurations of fixed $\beta$ and varying $\alpha$ and $\gamma$, similar to a potential energy landscape. We evaluate the ensembles and a single NequIP model on $D_{\beta=120^{\circ}}$, $D_{\beta=150^{\circ}}$, and $D_{\beta=180^{\circ}}$ and obtain force uncertainties per atom per configuration as usual. For the ensembles, we define a single aggregate uncertainty value for a molecular structure as the square root of the sum of all 27 atomic force variances. For the GMM, we sum all 27 atomic force NLL's for each structure to obtain a single uncertainty value. Lastly, we normalize these aggregate molecular uncertainties for each method to be between 0 and 1 for a clearer comparison. For example, if the uncertainty over all configurations ranges from $U_{\mathrm{low}}$ to $U_{\mathrm{high}}$ for a particular method, a configuration with uncertainty $U$ would have a normalized uncertainty of 
\begin{equation}
    \frac{U - U_{\mathrm{low}}}{U_{\mathrm{high}} - U_{\mathrm{low}}}.
    \label{eqn:colorscale-norm}
\end{equation}

\begin{figure*}
\centering
     \begin{subfigure}[b]{\textwidth}
         \centering
         \includegraphics[width=0.73\textwidth]{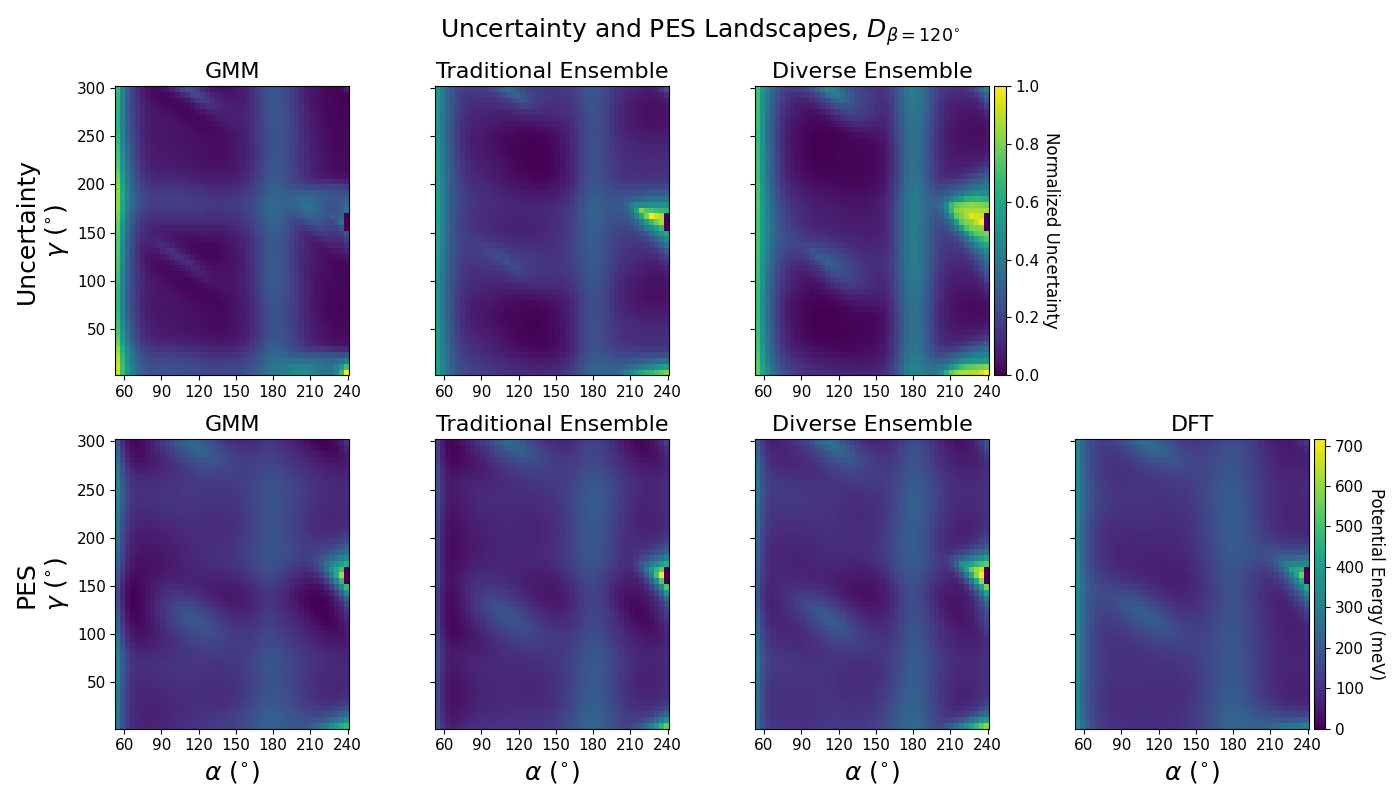}
     \end{subfigure}
     \hfill
     \begin{subfigure}[b]{\textwidth}
         \centering
         \includegraphics[width=0.73\textwidth]{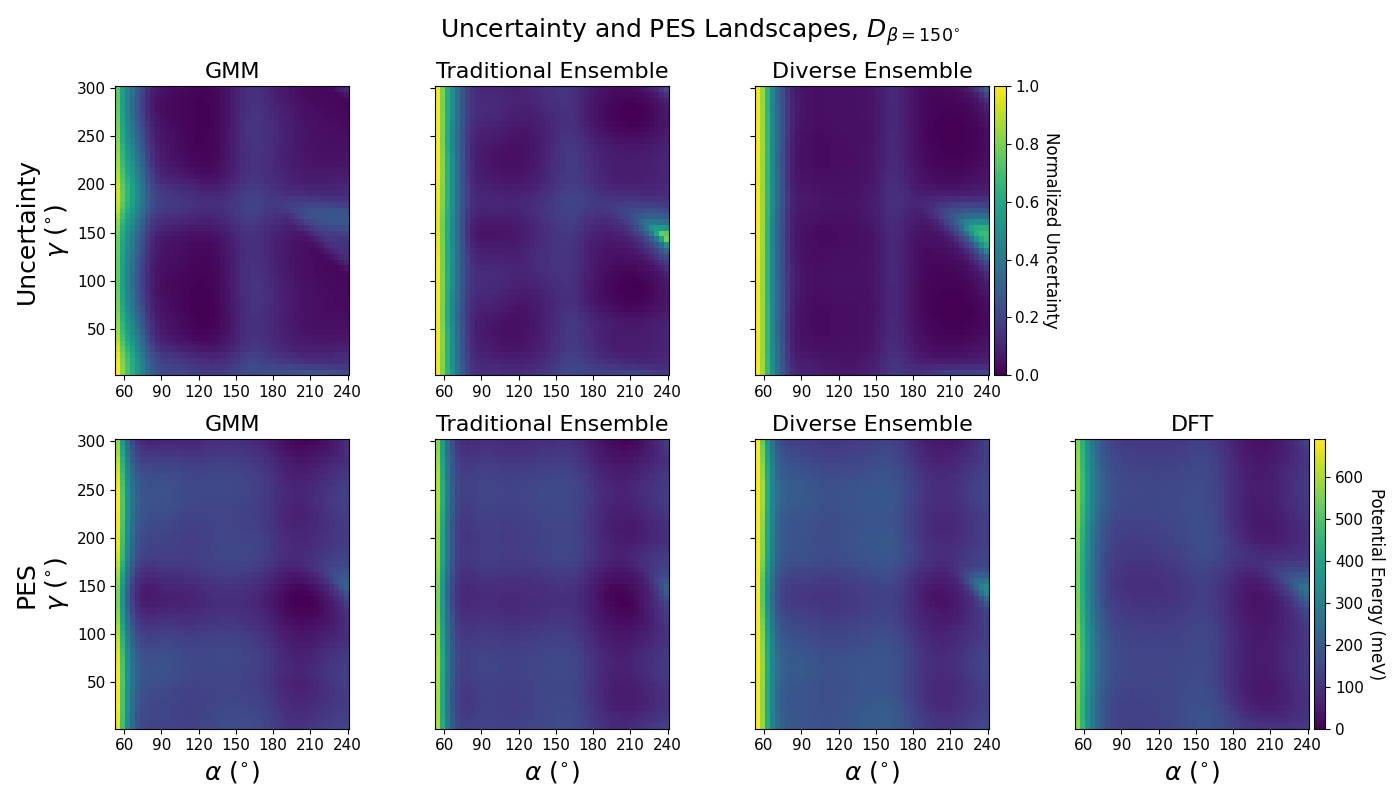}
     \end{subfigure}
     \hfill
    \begin{subfigure}[b]{\textwidth}
         \centering
         \includegraphics[width=0.73\textwidth]{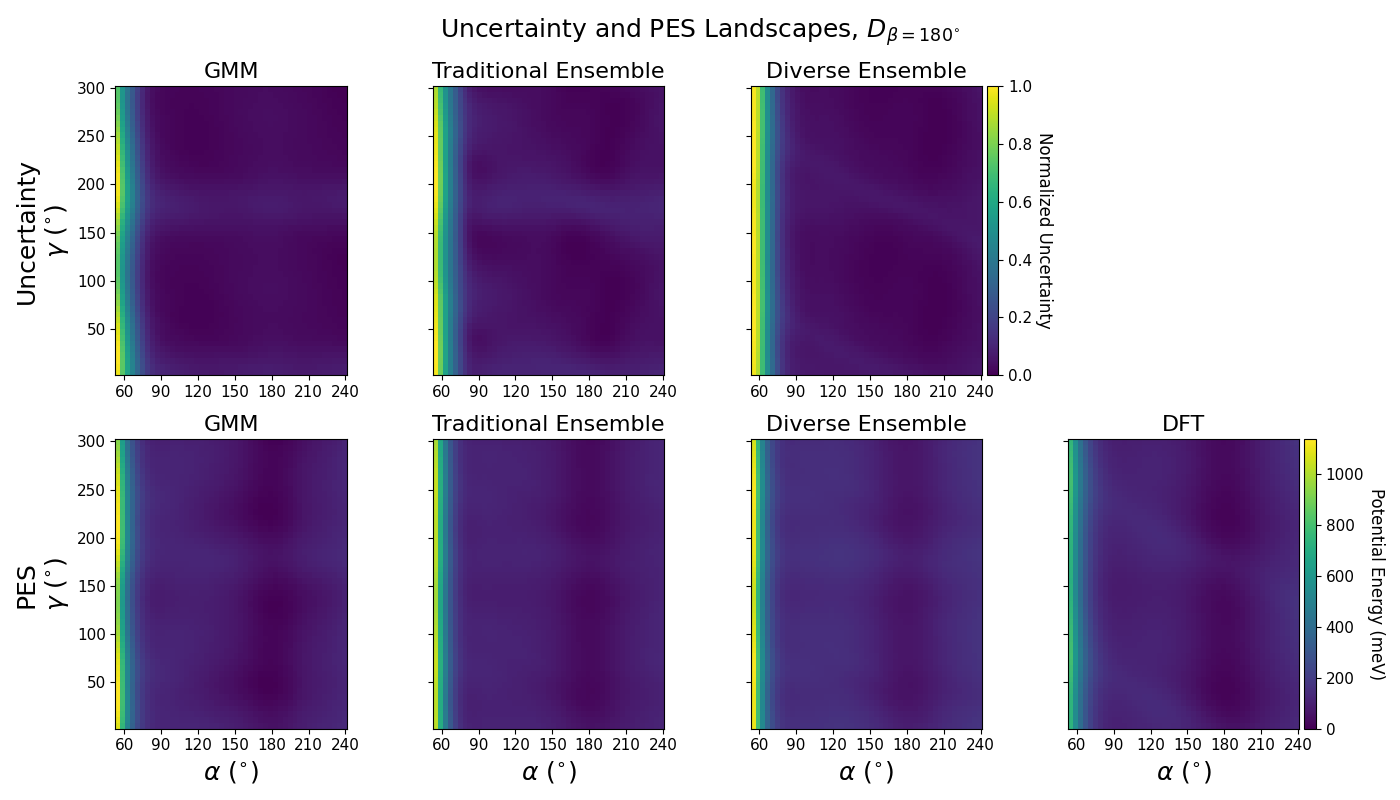}
     \end{subfigure}
    \caption{Uncertainty and PES landscapes for models with hidden feature dimension $f=16$, trained on $D_{\mathrm{train,50}}^{300}$.}
    \label{fig:landscape-n50-f16-si}
\end{figure*}

\begin{figure*}
\centering
     \begin{subfigure}[b]{\textwidth}
         \centering
         \includegraphics[width=0.73\textwidth]{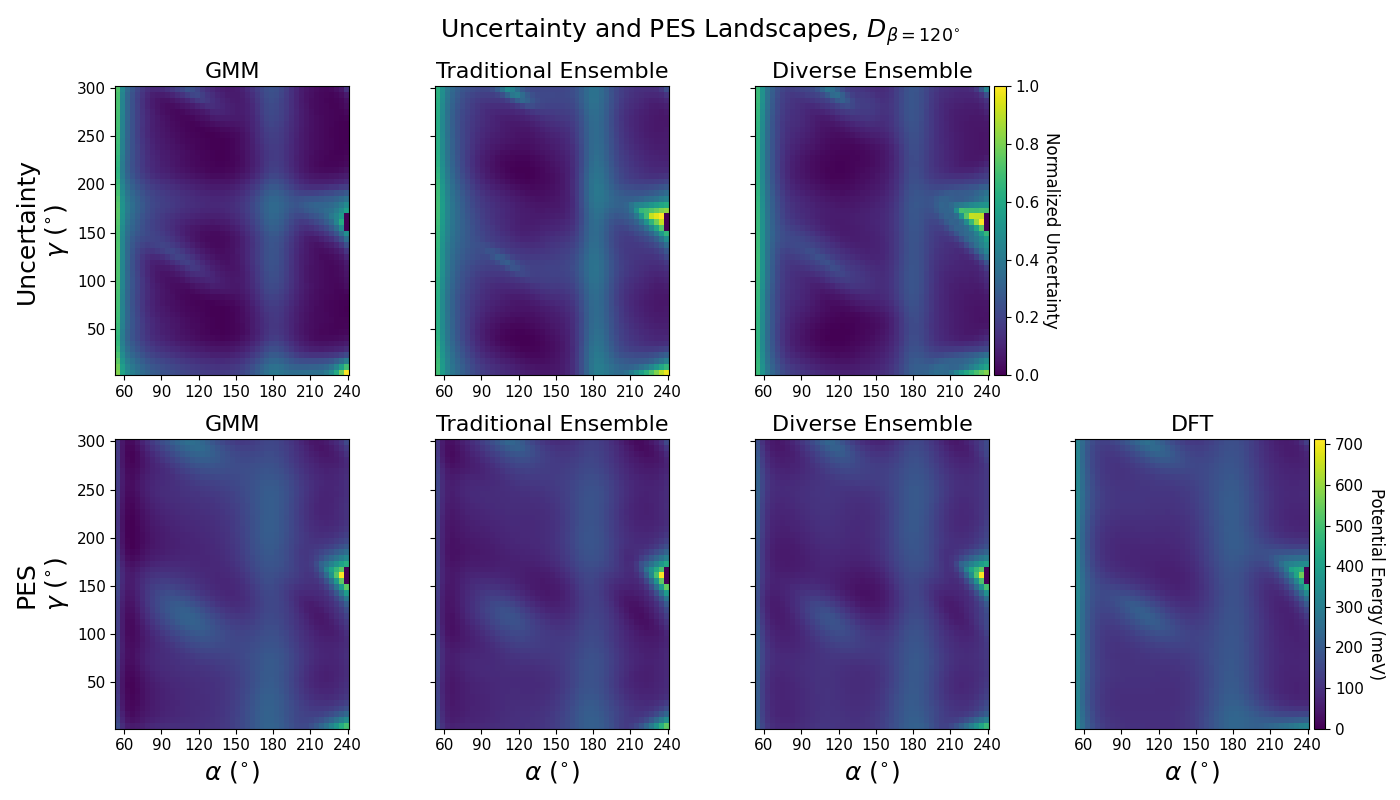}
     \end{subfigure}
     \hfill
     \begin{subfigure}[b]{\textwidth}
         \centering
         \includegraphics[width=0.73\textwidth]{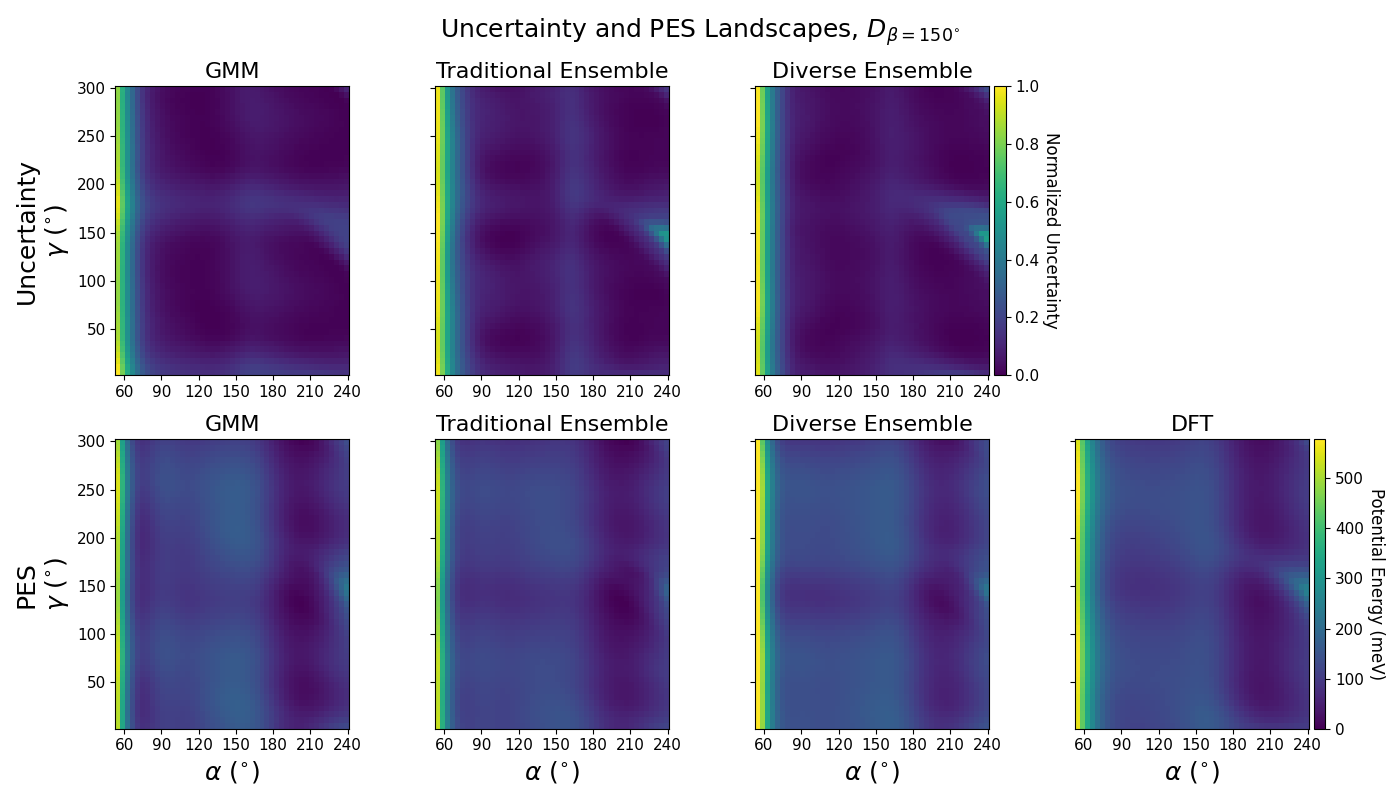}
     \end{subfigure}
     \hfill
    \begin{subfigure}[b]{\textwidth}
         \centering
         \includegraphics[width=0.73\textwidth]{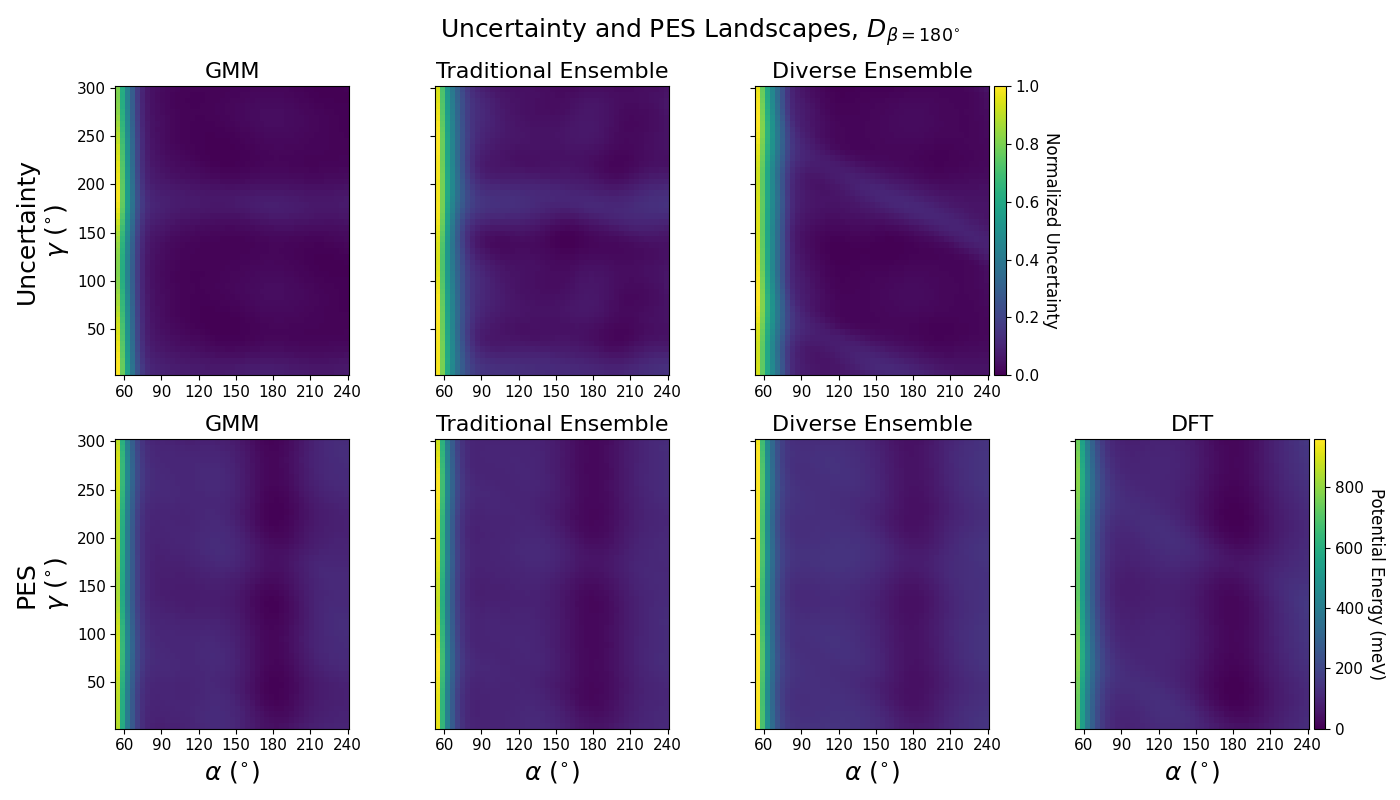}
     \end{subfigure}
    \caption{Uncertainty and PES landscapes for models with hidden feature dimension $f=32$, trained on $D_{\mathrm{train,50}}^{300}$.}
    \label{fig:landscape-n50-f32-si}
\end{figure*}

\begin{figure*}
\centering
     \begin{subfigure}[b]{\textwidth}
         \centering
         \includegraphics[width=0.73\textwidth]{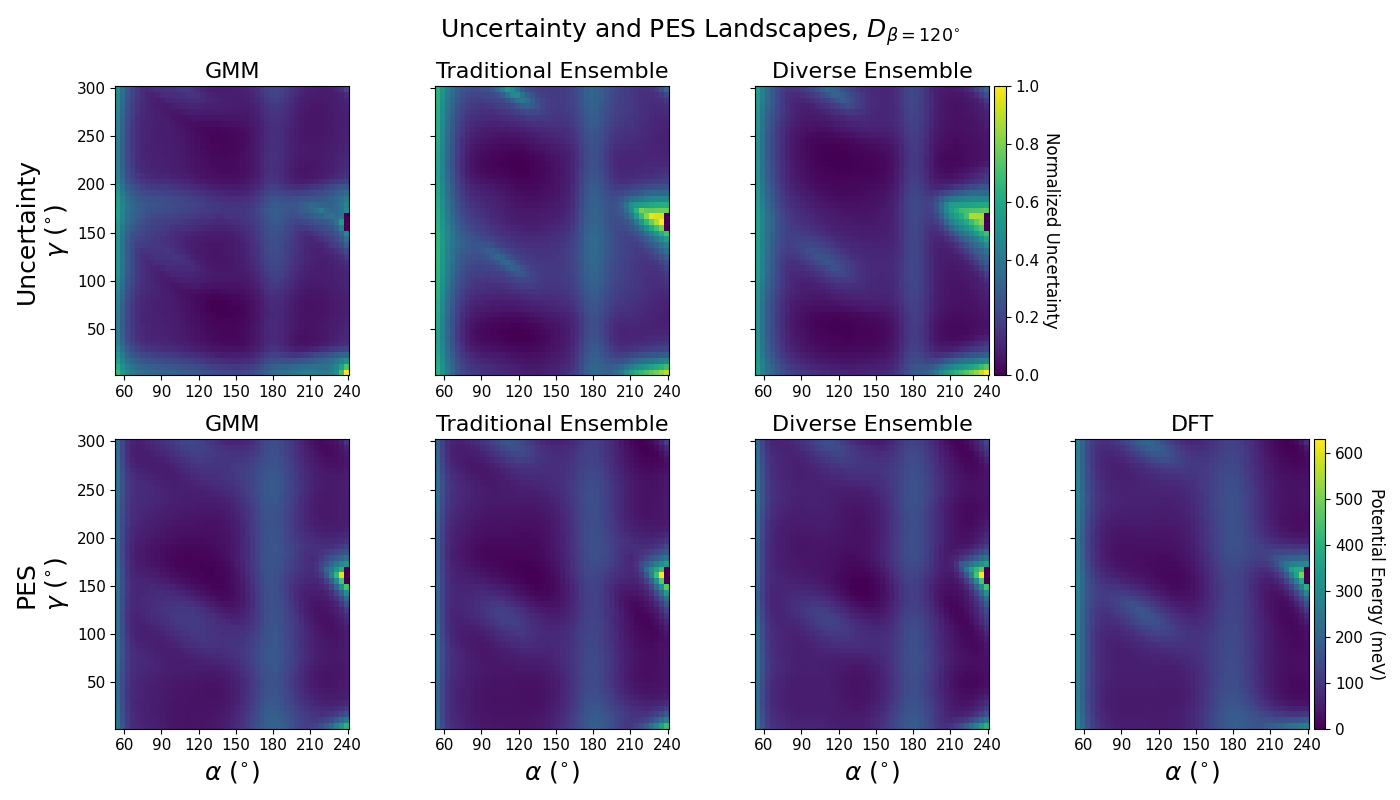}
     \end{subfigure}
     \hfill
     \begin{subfigure}[b]{\textwidth}
         \centering
         \includegraphics[width=0.73\textwidth]{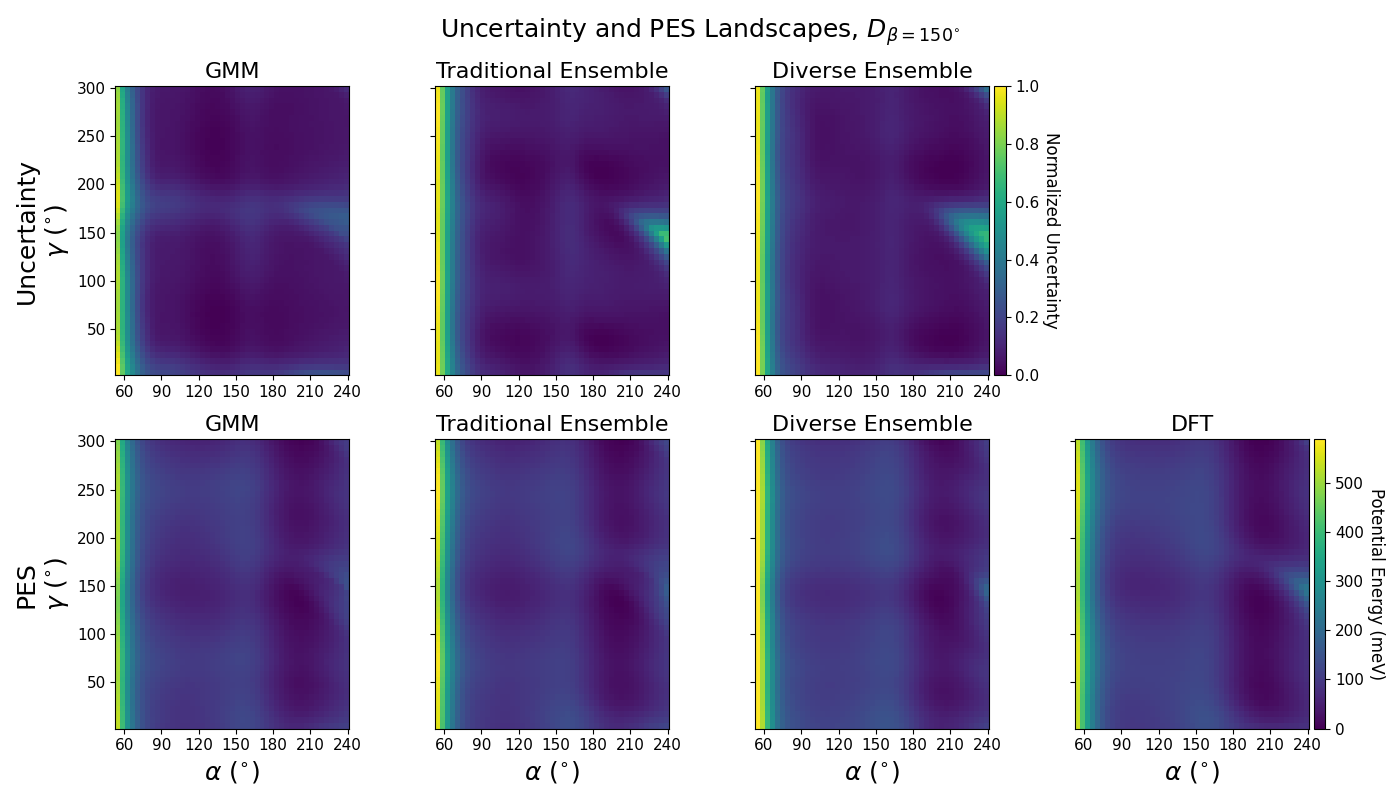}
     \end{subfigure}
     \hfill
    \begin{subfigure}[b]{\textwidth}
         \centering
         \includegraphics[width=0.73\textwidth]{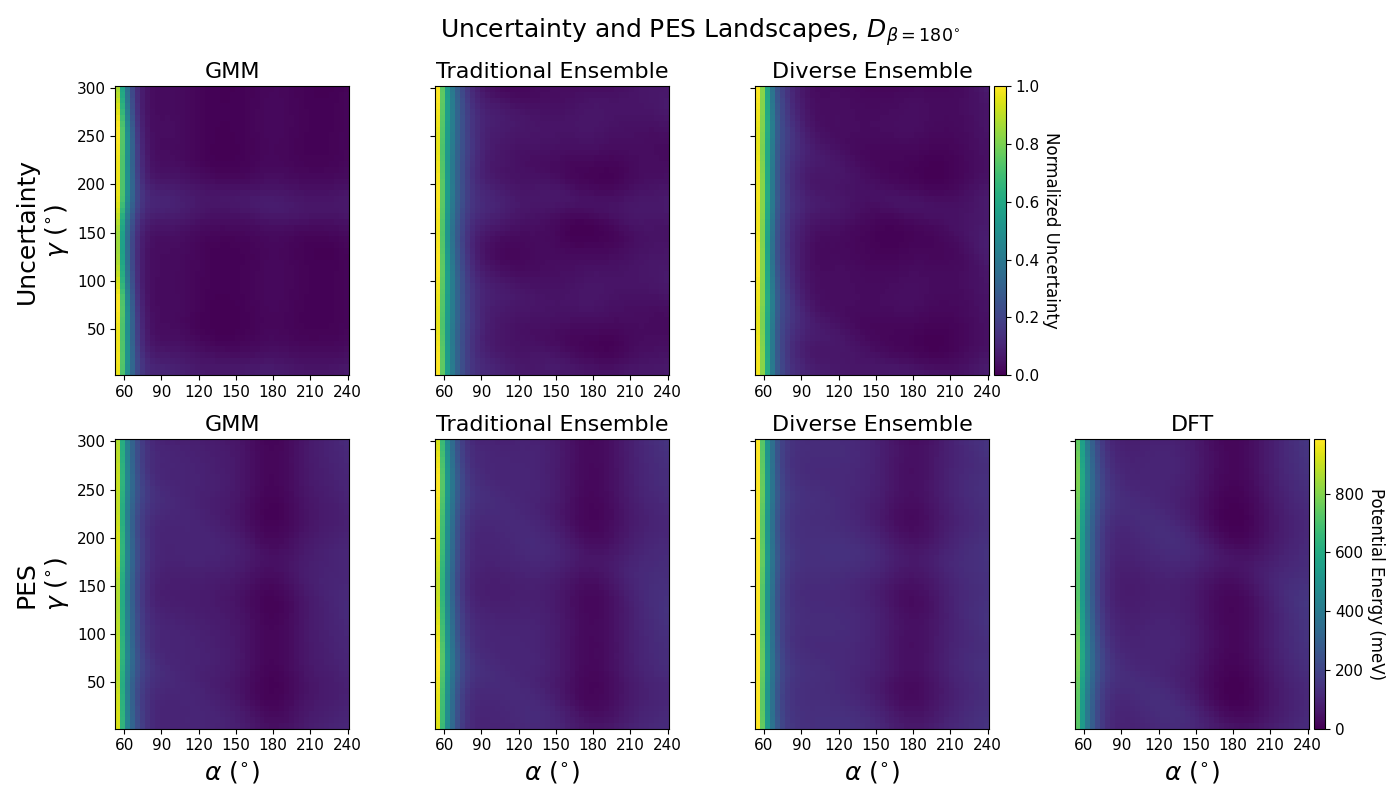}
     \end{subfigure}
    \caption{Uncertainty and PES landscapes for models with hidden feature dimension $f=16$, trained on $D_{\mathrm{train,100}}^{300}$.}
    \label{fig:landscape-n100-f16-si}
\end{figure*}

\begin{figure*}
\centering
     \begin{subfigure}[b]{\textwidth}
         \centering
         \includegraphics[width=0.73\textwidth]{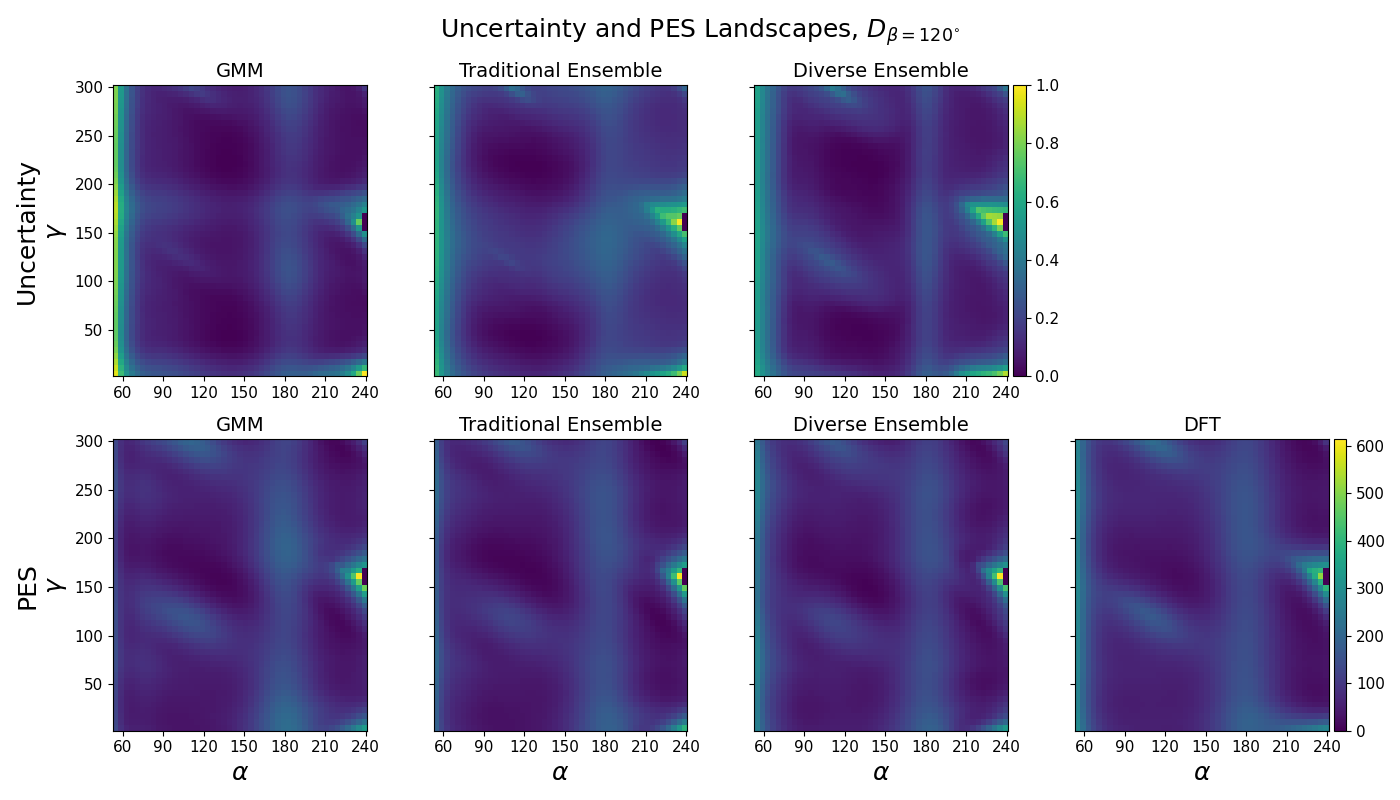}
     \end{subfigure}
     \hfill
     \begin{subfigure}[b]{\textwidth}
         \centering
         \includegraphics[width=0.73\textwidth]{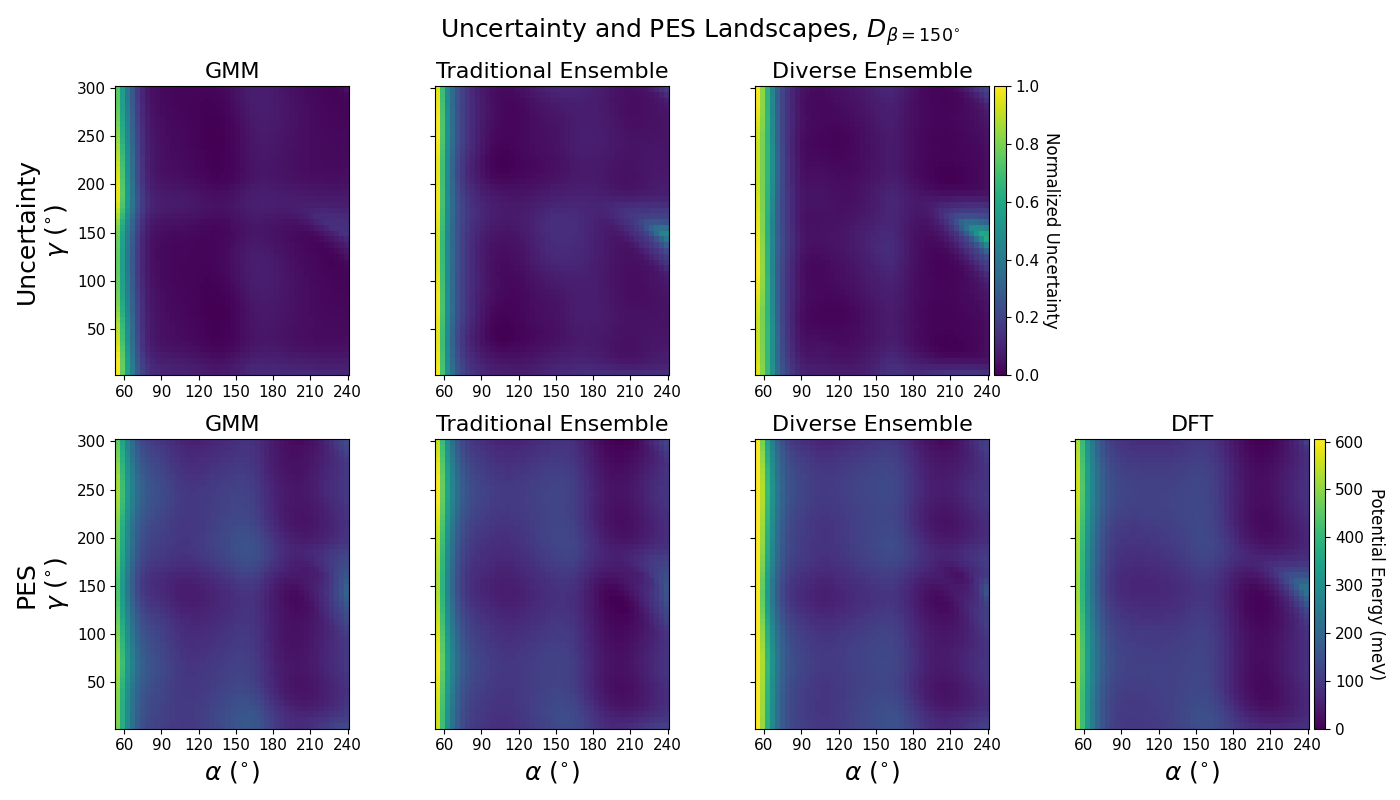}
     \end{subfigure}
     \hfill
    \begin{subfigure}[b]{\textwidth}
         \centering
         \includegraphics[width=0.73\textwidth]{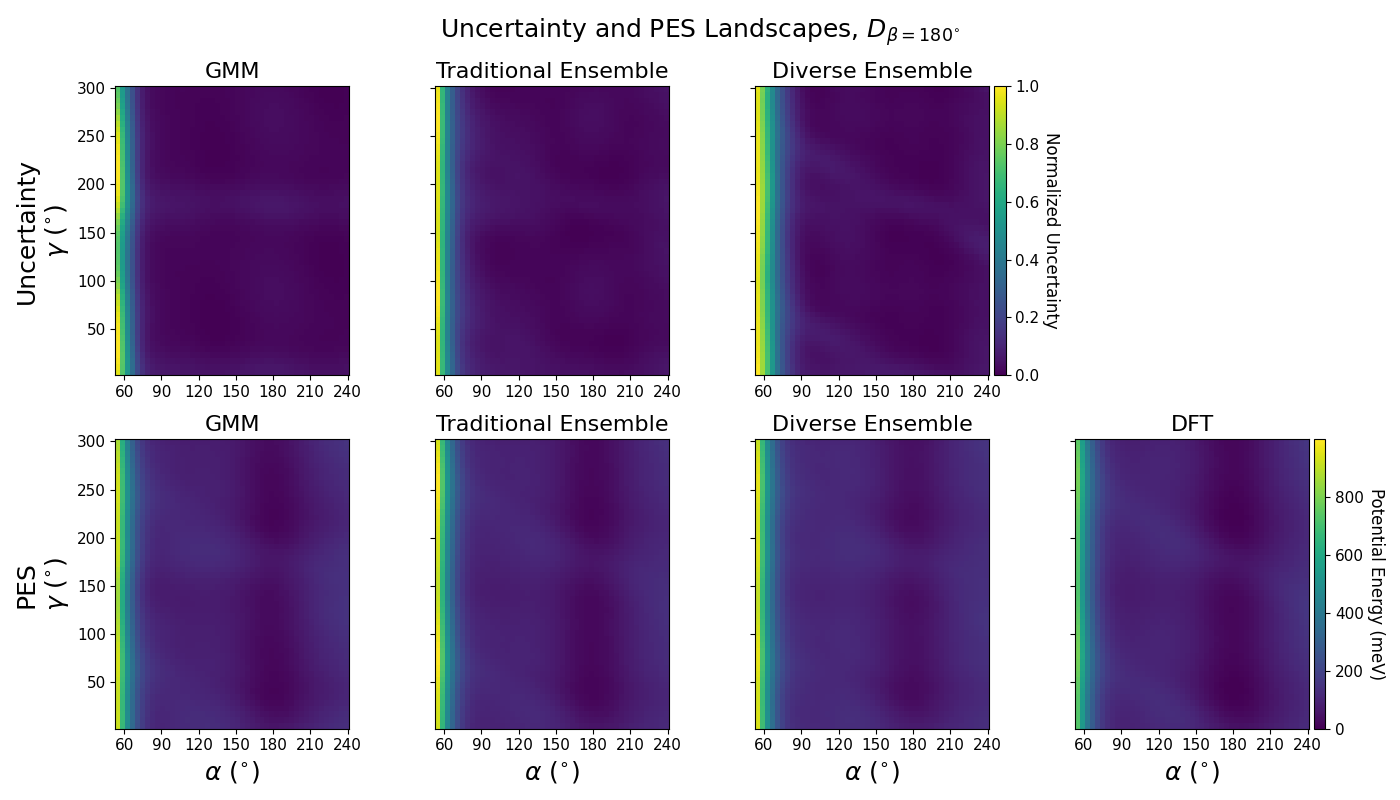}
     \end{subfigure}
    \caption{Uncertainty and PES landscapes for models with hidden feature dimension $f=32$, trained on $D_{\mathrm{train,100}}^{300}$. Plots for $D_{\beta=120^{\circ}}$ are presented in the manuscript.}
    \label{fig:landscape-n100-f32-si}
\end{figure*}

The top row of figure \ref{fig:uncertainty-landscape-beta120} shows the uncertainty landscape of the GMM and both ensemble types with models with $f=32$ trained on $D_{\mathrm{train,100}}^{300}$ and evaluated on $D_{\beta=120^{\circ}}$ (SI figures \ref{fig:landscape-n50-f16-si}, \ref{fig:landscape-n50-f32-si}, \ref{fig:landscape-n100-f16-si}, and \ref{fig:landscape-n100-f32-si} show results for the $D_{\beta=150^{\circ}}$ and $D_{\beta=180^{\circ}}$ test sets, models with $f=16$, and models trained on $D_{\mathrm{train,50}}^{300}$). All three uncertainty landscapes are very similar in that they generally label the same configurations with relatively high uncertainty. For example, all methods label configurations with $\alpha \leq 60^{\circ}$, $\alpha \approx 180^{\circ}$, and ($\alpha \approx 240^{\circ}, \gamma \approx 10^{\circ}$) with high relative uncertainty. Moreover, the uncertainty landscapes of all methods closely resemble their corresponding potential energy landscapes and the reference DFT potential energy landscape (bottom row of figure \ref{fig:uncertainty-landscape-beta120}) \cite{kovacs2021linear}. These similarities further demonstrate that a single NequIP model and GMM evaluation can achieve uncertainty estimates comparable to those of an ensemble at greatly reduced computational cost.

\subsection{Active Learning}

Active learning is a procedure in which a model explores a data distribution and iteratively chooses new data to label and add to its training set to augment it with maximally informative samples. Crucially, in such a setting, one requires a method to decide which data points to label. Using uncertainty estimates from the GMM and the ensembles, we conduct experiments in which we select new data from a set of hold-out examples based on the model's estimated uncertainty on those data points. This uncertainty-based selection should ideally result in a training set that improves the model's generalization error more than by adding an identical number of randomly chosen data points.

\begin{figure*}
    \centering
    \includegraphics[width=\textwidth, draft=false]{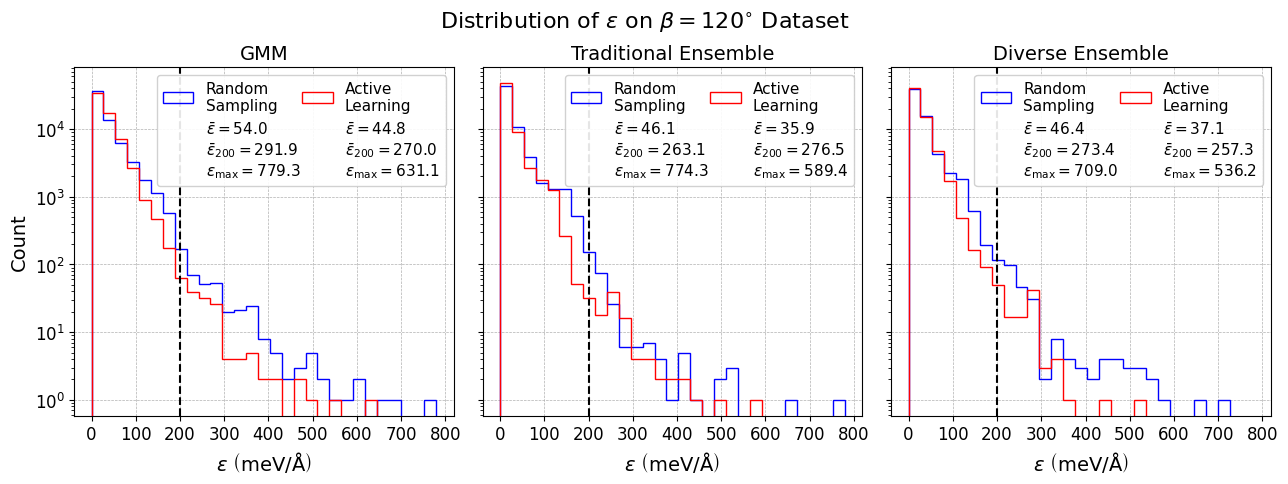}
    \caption{Distribution of $\epsilon$ of the GMM and ensembles on $D_{\beta=120^{\circ}}$ for models with hidden feature dimension $f=16$ and trained on $D_{\mathrm{train,50}}^{300}$. Vertical dashed line in each plot marks the 200 $\mathrm{meV/\AA}$ cutoff for determining the RMSE of outliers.}
     \label{fig:active-learning-n50-f16}
\end{figure*}

To measure the effectiveness of the different uncertainty estimates, we perform one round of this active learning task (which we will refer to as ``active learning") using a single model with GMM uncertainties and compare the results to active learning with uncertainties from both ensemble types. We evaluate a trained ensemble or single NequIP model on a pool of additional training data, obtaining an uncertainty $U_k = \sigma_k$ for each atom $k$ in each molecular structure. Let $k^*$ be the atom within a 3BPA molecule with the highest uncertainty, i.e.:
\begin{equation}
    k^* = \underset{k}{\arg\max}\{U_k\}.
\end{equation}
We select the $N_{\mathrm{train}}$ structures with the highest $U_{k^*}$ to add the original training set of size $N_{\mathrm{train}}$, creating a new training set $D_{\mathrm{train, active}}$. We also randomly select $N_{\mathrm{train}}$ to add to the original training set, creating another training set $D_{\mathrm{train, random}}$. We re-train all models with the same parameters from scratch on each of these two new training sets and evaluate them on various test sets. For further details, see Active Learning section in the SI.

Assuming that atoms with the highest $\epsilon$ (i.e., outliers) are labeled with the highest uncertainties, we ideally expect our active learning procedure to improve generalization error on those atoms the most. Thus, to compare the effectiveness of active learning based on the three uncertainty metrics, we consider not only $\bar{\epsilon}$ (Eq. \ref{eqn:force-rmse}) over all atoms in a test set, but also the distribution of $\epsilon$, the average force RMSE of all $\epsilon$ above 200 $\mathrm{meV/\AA}$ ($\bar{\epsilon}_{200}$), and the maximum force RMSE ($\epsilon_{\mathrm{max}}$) between the three approaches.

\begin{figure*}
\centering
     \begin{subfigure}[b]{\textwidth}
         \centering
         \includegraphics[width=\textwidth]{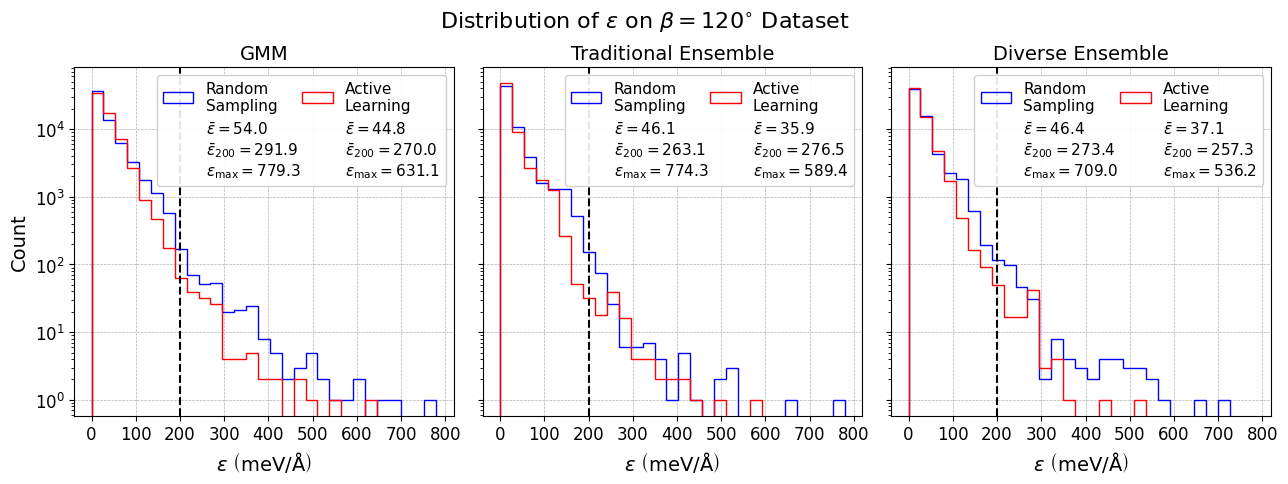}
     \end{subfigure}
     \hfill
     \begin{subfigure}[b]{\textwidth}
         \centering
         \includegraphics[width=\textwidth]{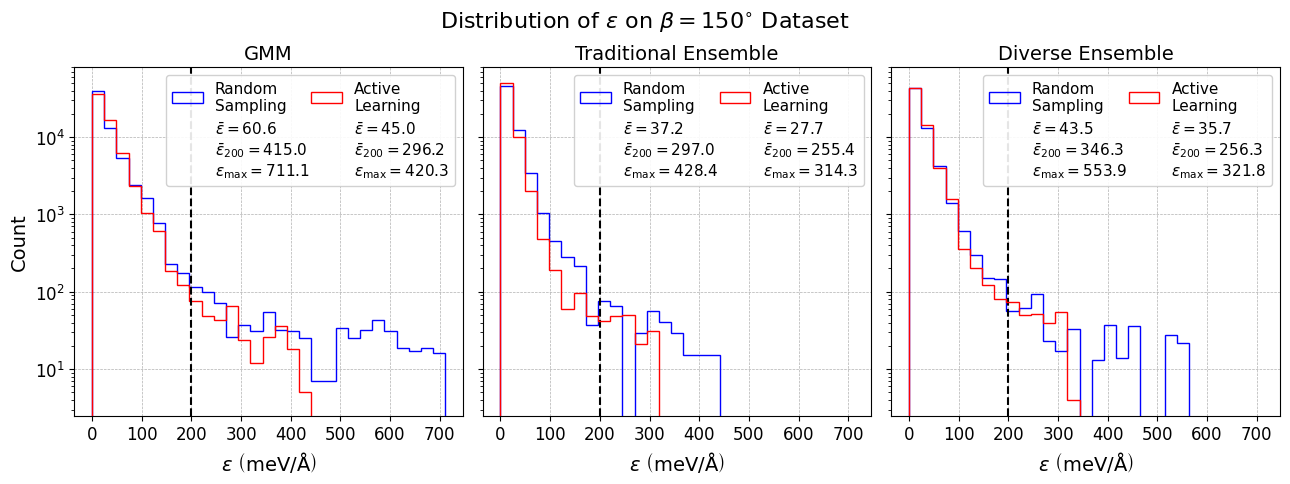}
     \end{subfigure}
     \hfill
    \begin{subfigure}[b]{\textwidth}
         \centering
         \includegraphics[width=\textwidth]{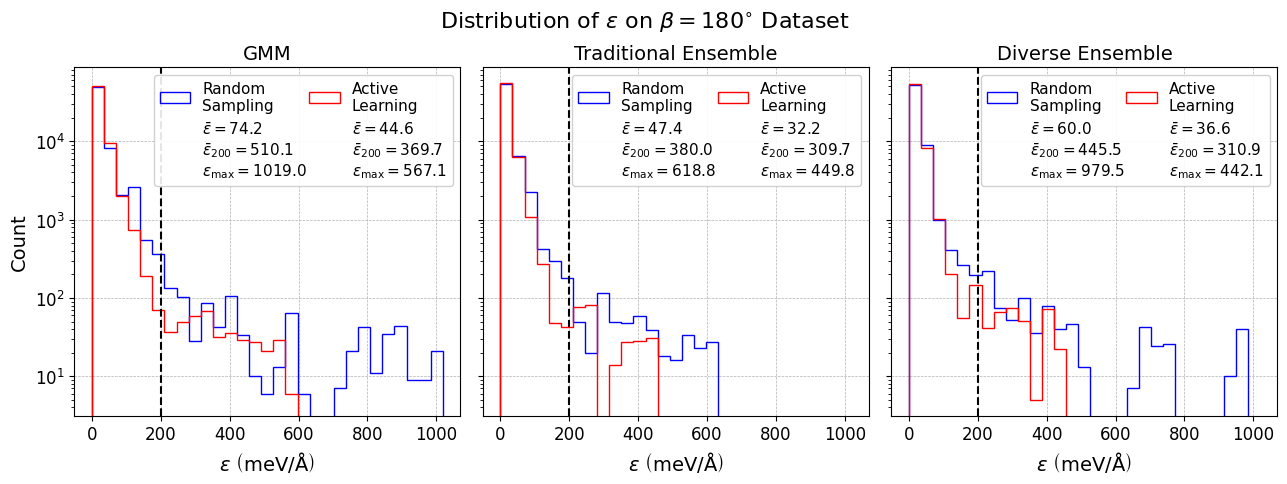}
     \end{subfigure}
    \caption{Distribution of $\epsilon$ of the GMM and ensembles on $D_{\beta=120^{\circ}}$, $D_{\beta=150^{\circ}}$, and $D_{\beta=180^{\circ}}$ for models with $f=16$, trained on $D_{\mathrm{train,50}}^{300}$. Top row of plots is presented in the manuscript.}
    \label{fig:active-learning-betas-n50-f16-si}
\end{figure*}

Figure \ref{fig:active-learning-n50-f16} shows the results of evaluating the ensembles and single NequIP model on $D_{\beta=120^{\circ}}$ after one round of active learning, with all models having hidden feature dimension $f=16$, initially trained on $D_{\mathrm{train,50}}^{300}$, and with new data sampled from $D_{\mathrm{pool}}^{300}$. We observe that active learning improves $\bar{\epsilon}$ on $D_{\beta=120^{\circ}}$ by around 10 $\mathrm{meV/\AA}$ for the three methods, compared to the random selection baseline. The improvement is even greater on $D_{\beta=150^{\circ}}$ and $D_{\beta=180^{\circ}}$, reaching nearly 30 $\mathrm{meV/\AA}$ on $D_{\beta=180^{\circ}}$ using the GMM (SI figure \ref{fig:active-learning-betas-n50-f16-si}).

Additionally, we observe that active learning reduces $\bar{\epsilon}_{200}$ and $\epsilon_{\mathrm{max}}$ by a much larger amount compared to the reduction in $\bar{\epsilon}$, reflected by the presence of fewer high-error data points in the distribution of $\epsilon$ resulting from active learning. On $D_{\beta=150^{\circ}}$ and $D_{\beta=180^{\circ}}$, $\bar{\epsilon}_{200}$ is even reduced by over 100 $\mathrm{meV/\AA}$ with active learning, as seen in SI figure \ref{fig:active-learning-betas-n50-f16-si}. Table \ref{tab:active-learning-improvements} summarizes the absolute and percentage improvements of active learning over random sampling for all three methods on these performance metrics on $D_{\beta=120^{\circ}}$, $D_{\beta=150^{\circ}}$, and $D_{\beta=180^{\circ}}$. From these results, we conclude that the improvement in $\bar{\epsilon}$ is significant for all methods, and in particular, using GMM uncertainties for active learning yields improvements on overall error and outlier generalization error comparable to those achieved with ensemble-based active learning at a significantly lower computational cost.

\begin{table*}[!htbp]
\centering
\resizebox{\textwidth}{!}
{\begin{tabular}{c|c|c|c|c|c|c|c}
\hline \hline
 \multicolumn{2}{c|}{\multirow{2}{*}{}} & \multicolumn{3}{c|}{Absolute Improvement ($\mathrm{meV/\AA}$)} & \multicolumn{3}{c}{Percentage Improvement (\%)} \\
\cline{3-8} 
\multicolumn{2}{c|}{} & GMM & Traditional & Diverse & GMM & Traditional & Diverse \\
\hline \hline
\multirow{3}{*}{$D_{\beta=120^{\circ}}$} & $\bar{\epsilon}$ & 9.2 & 10.2 & 9.3 & 17.0 & 22.1 & 20.0 \\ \cline{2-8}
& $\bar{\epsilon}_{200}$ & 21.9 & -13.4 & 16.1 & 7.5 & -5.1 & 5.9 \\ \cline{2-8}
& $\epsilon_{\mathrm{max}}$ & 148.2 & 184.9 & 172.8 & 19.0 & 23.9 & 24.3 \\ 
\hline \hline
\multirow{3}{*}{$D_{\beta=150^{\circ}}$} & $\bar{\epsilon}$ & 15.6 & 9.5 & 7.8 & 26.0 & 25.5 & 17.9 \\ \cline{2-8}
& $\bar{\epsilon}_{200}$ & 118.8 & 41.6 & 90.0 & 28.6 & 14.0 & 26.0 \\ \cline{2-8}
& $\epsilon_{\mathrm{max}}$ & 290.8 & 114.1 &  232.1 & 40.9 & 41.9 & 41.9 \\ 
\hline \hline
\multirow{3}{*}{$D_{\beta=180^{\circ}}$} & $\bar{\epsilon}$ & 29.6 & 15.2 & 23.4 & 39.9 & 32.1 & 39.0 \\ \cline{2-8}
& $\bar{\epsilon}_{200}$ & 140.4 & 70.3 & 134.6 & 27.5 & 18.5 & 30.2 \\ \cline{2-8}
& $\epsilon_{\mathrm{max}}$ & 451.9 & 169.0 & 537.4 & 44.3 & 27.3 & 54.9 \\ 
\hline \hline
\end{tabular}}
\caption{Improvement of $\bar{\epsilon}$, $\bar{\epsilon}_{200}$, and $\epsilon_{\mathrm{max}}$ with active learning over random sampling for each of the three methods on $D_{\beta=120^{\circ}}$, $D_{\beta=150^{\circ}}$, and $D_{\beta=180^{\circ}}$ (for models with hidden feature dimension $f=16$ and initially trained on $D_{\mathrm{train,50}}^{300}$).}
\label{tab:active-learning-improvements}
\end{table*}

\begin{figure*}
\centering
     \begin{subfigure}[b]{\textwidth}
         \centering
         \includegraphics[width=\textwidth]{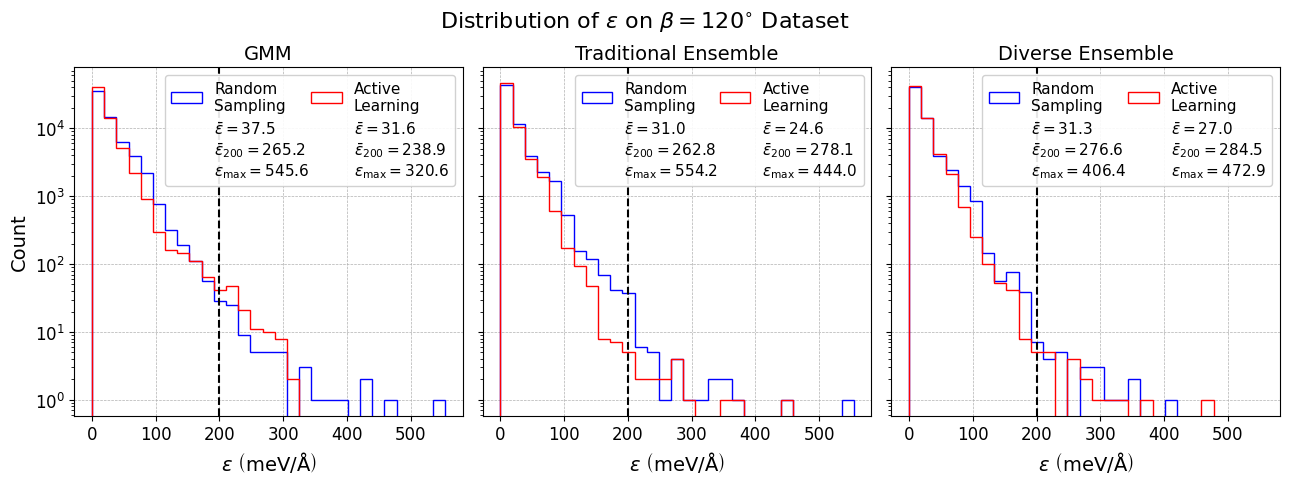}
     \end{subfigure}
     \hfill
     \begin{subfigure}[b]{\textwidth}
         \centering
         \includegraphics[width=\textwidth]{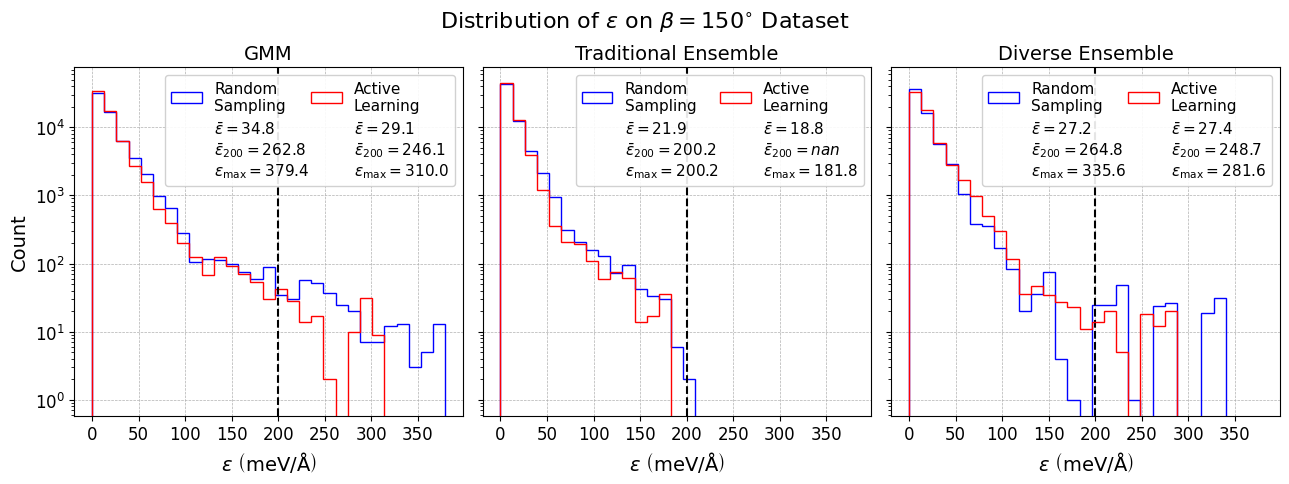}
     \end{subfigure}
     \hfill
    \begin{subfigure}[b]{\textwidth}
         \centering
         \includegraphics[width=\textwidth]{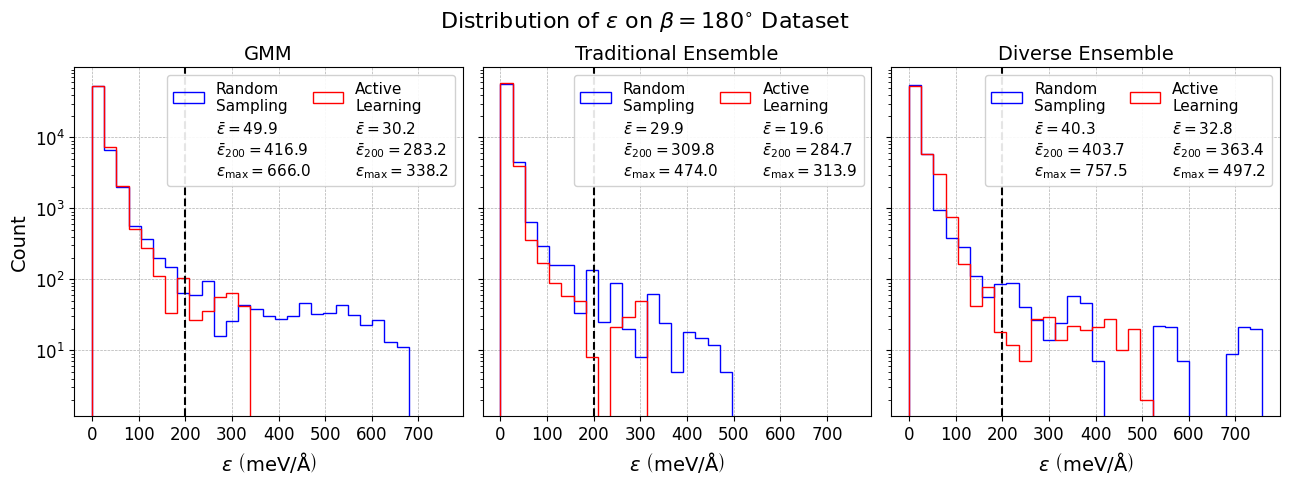}
     \end{subfigure}
    \caption{Distribution of $\epsilon$ of the GMM and ensembles on $D_{\beta=120^{\circ}}$, $D_{\beta=150^{\circ}}$, and $D_{\beta=180^{\circ}}$ for models with $f=16$, trained on $D_{\mathrm{train,100}}^{300}$.}
    \label{fig:active-learning-betas-n100-f16-si}
\end{figure*}

\begin{figure*}
\centering
     \begin{subfigure}[b]{\textwidth}
         \centering
         \includegraphics[width=\textwidth]{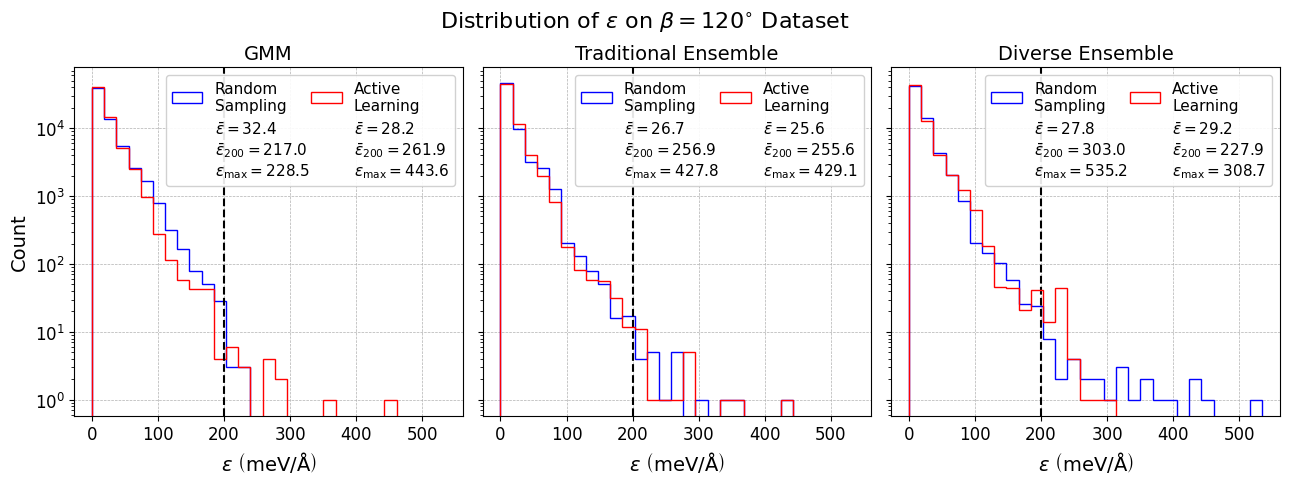}
     \end{subfigure}
     \hfill
     \begin{subfigure}[b]{\textwidth}
         \centering
         \includegraphics[width=\textwidth]{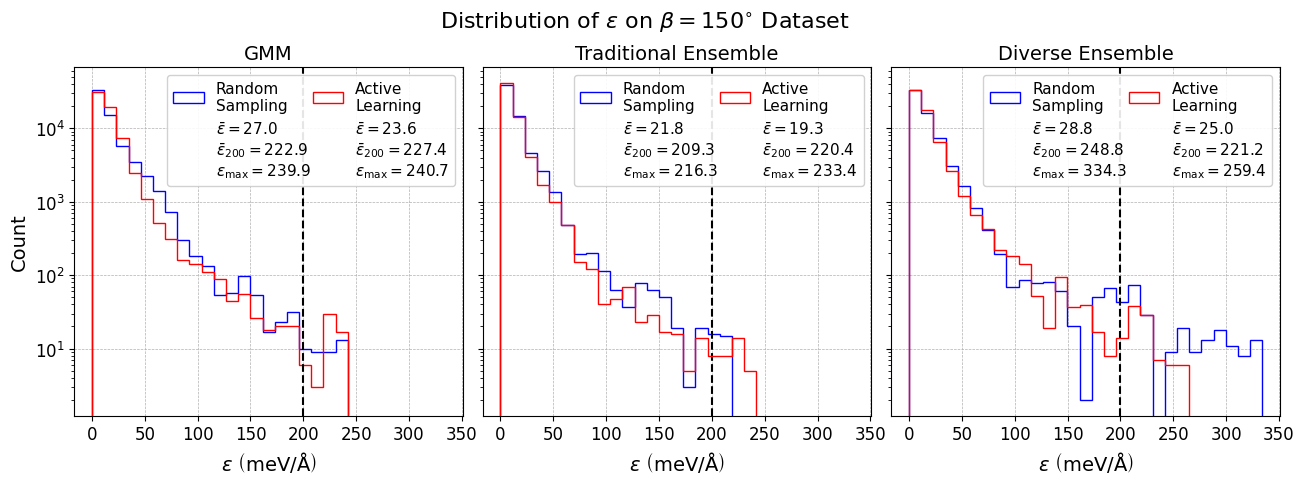}
     \end{subfigure}
     \hfill
    \begin{subfigure}[b]{\textwidth}
         \centering
         \includegraphics[width=\textwidth]{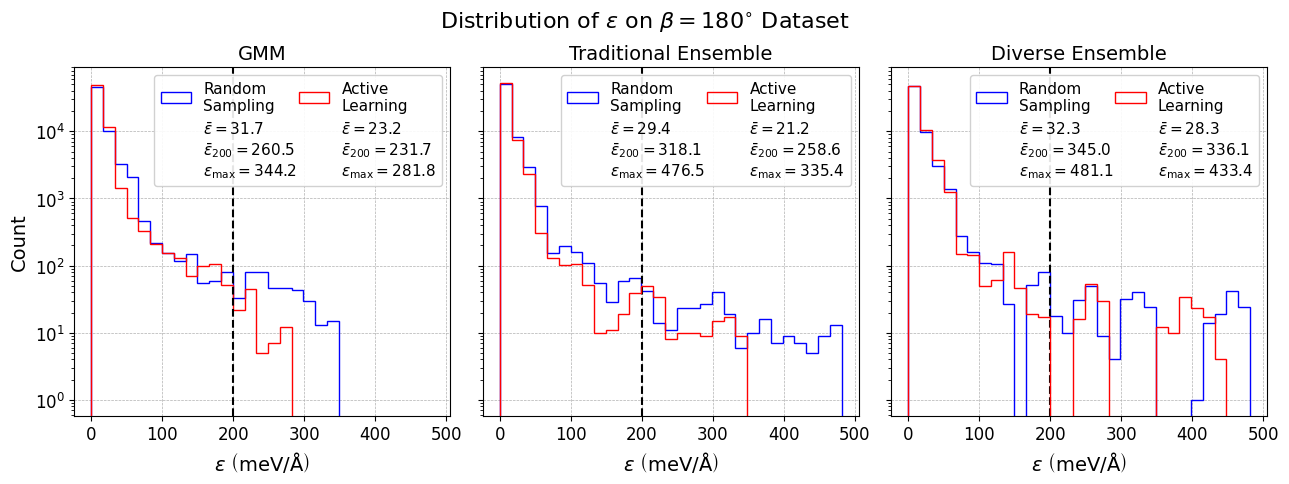}
     \end{subfigure}
    \caption{Distribution of $\epsilon$ of the GMM and ensembles on $D_{\beta=120^{\circ}}$, $D_{\beta=150^{\circ}}$, and $D_{\beta=180^{\circ}}$ for models with $f=32$, trained on $D_{\mathrm{train,100}}^{300}$.}
    \label{fig:active-learning-betas-n100-f32-si}
\end{figure*}

\begin{figure*}
\centering
     \begin{subfigure}[b]{\textwidth}
         \centering
         \includegraphics[width=0.82\textwidth]{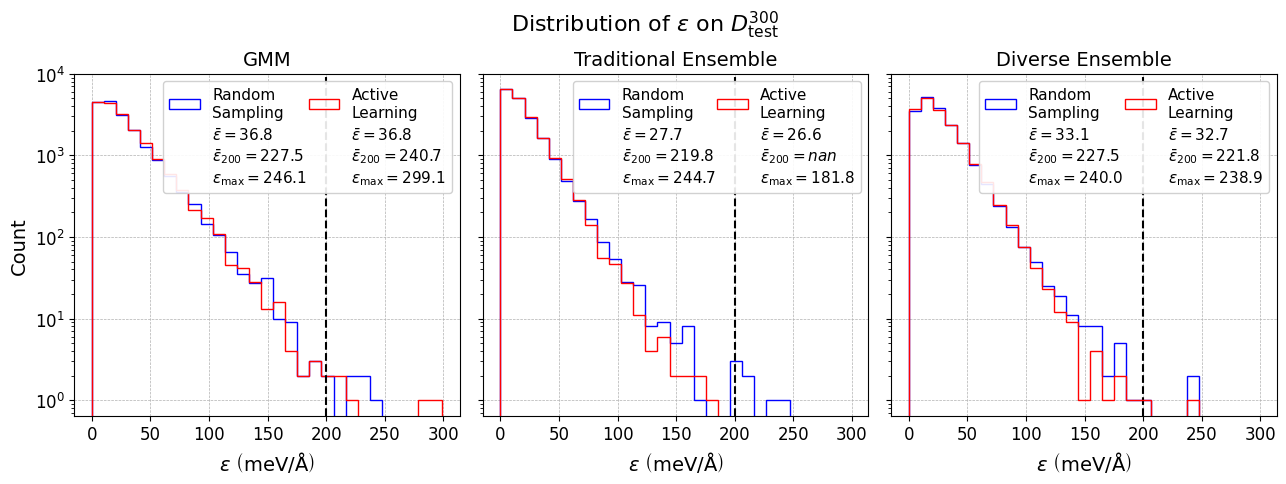}
     \end{subfigure}
     \hfill
     \begin{subfigure}[b]{\textwidth}
         \centering
         \includegraphics[width=0.82\textwidth]{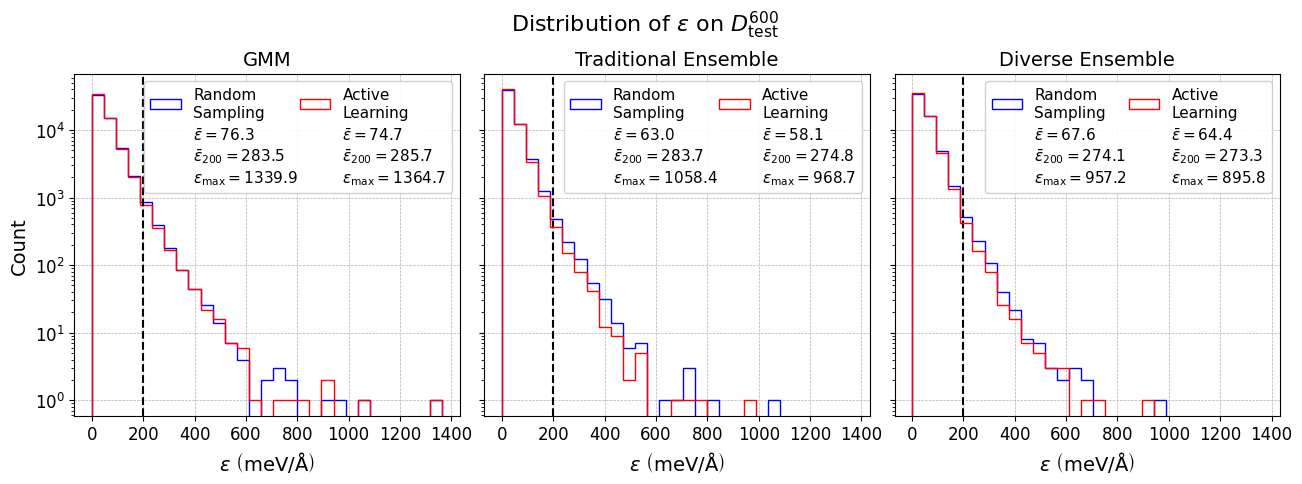}
     \end{subfigure}
     \hfill
     \begin{subfigure}[b]{\textwidth}
         \centering
         \includegraphics[width=0.82\textwidth]{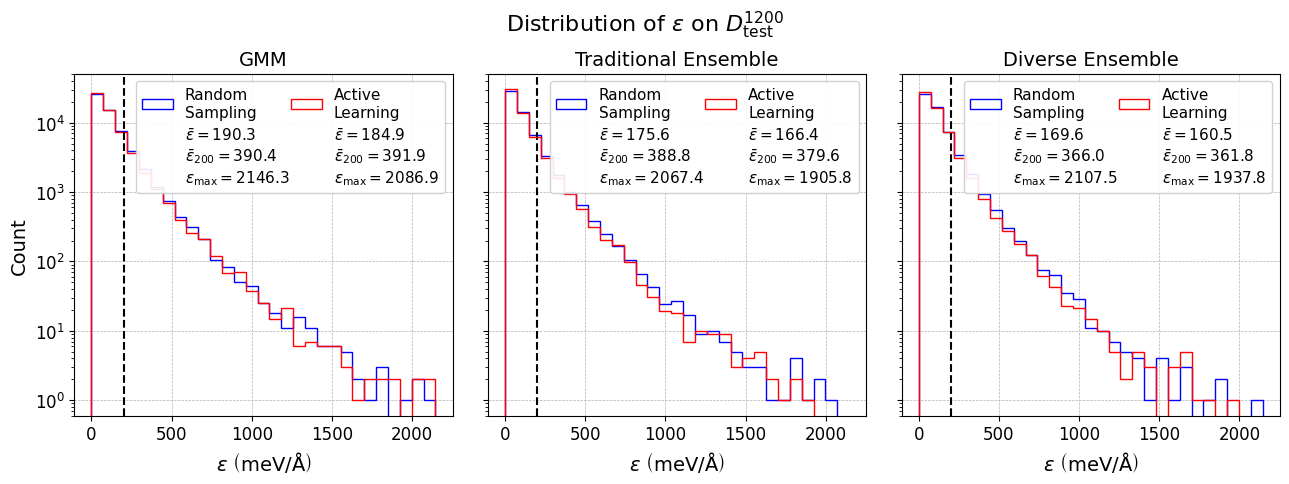}
     \end{subfigure}
     \hfill
    \begin{subfigure}[b]{\textwidth}
         \centering
         \includegraphics[width=0.82\textwidth]{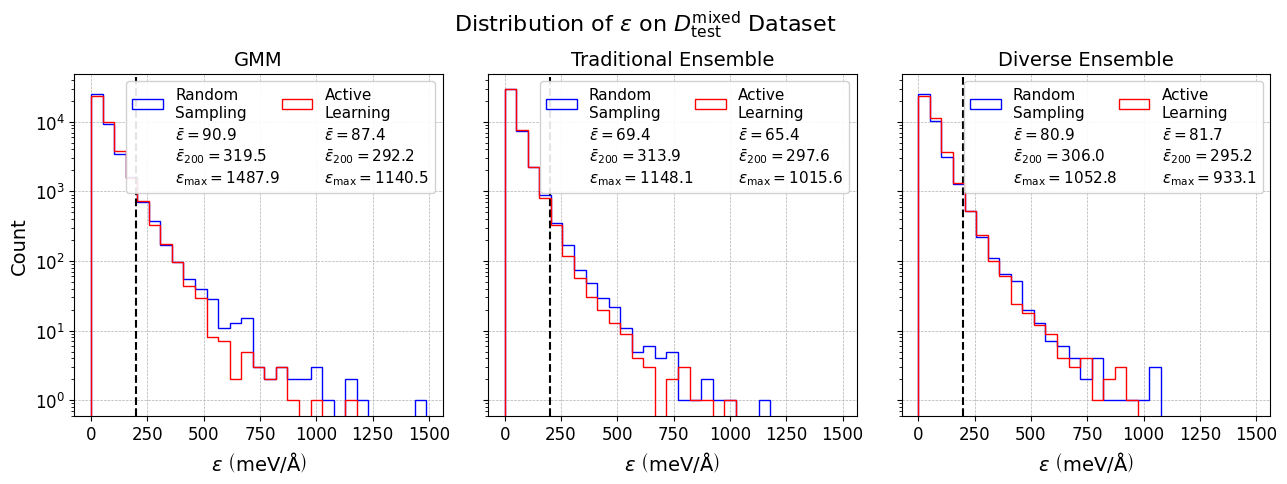}
     \end{subfigure}
    \caption{Distribution of $\epsilon$ of the GMM and ensembles on $D_{\mathrm{test}}^{300}$, $D_{\mathrm{test}}^{600}$, $D_{\mathrm{test}}^{1200}$, and $D_{\mathrm{test}}^{\mathrm{mixed}}$ for models with $f=16$, trained on $D_{\mathrm{train,50}}^{300}$ for the single temperature test sets and $D_{\mathrm{train,50}}^{\mathrm{mixed}}$ for $D_{\mathrm{test}}^{\mathrm{mixed}}$.}
    \label{fig:active-learning-temps-n50-f16-si}
\end{figure*}

\begin{figure*}
\centering
     \begin{subfigure}[b]{\textwidth}
         \centering
         \includegraphics[width=0.82\textwidth]{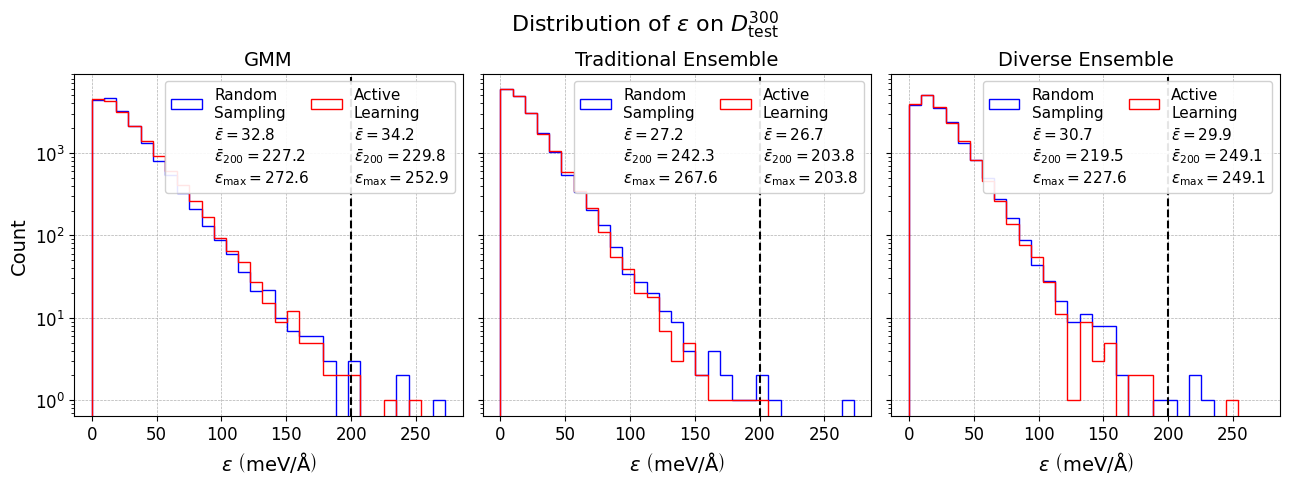}
     \end{subfigure}
     \hfill
     \begin{subfigure}[b]{\textwidth}
         \centering
         \includegraphics[width=0.82\textwidth]{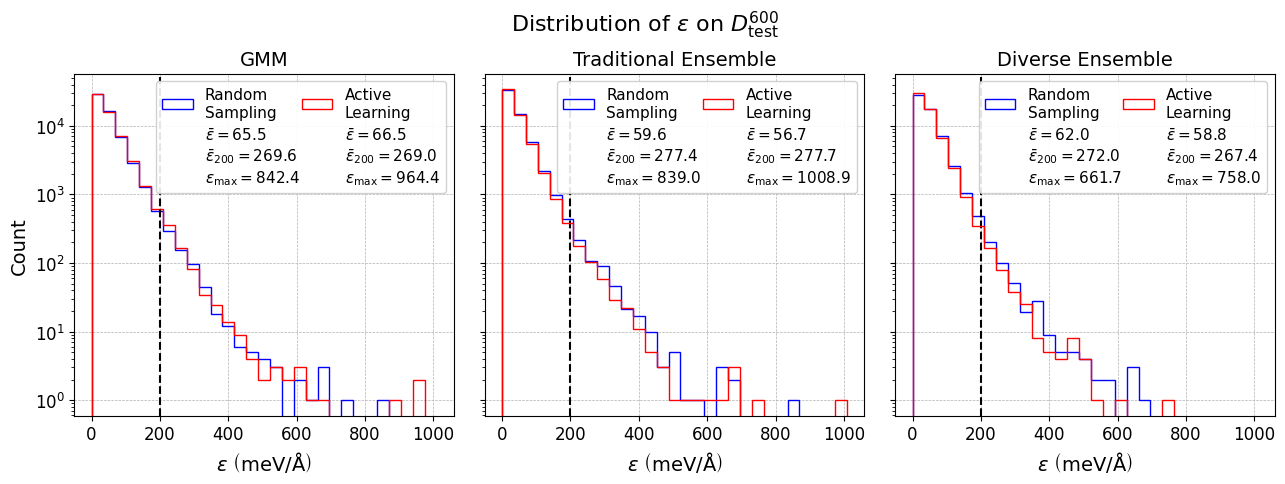}
     \end{subfigure}
     \hfill
     \begin{subfigure}[b]{\textwidth}
         \centering
         \includegraphics[width=0.82\textwidth]{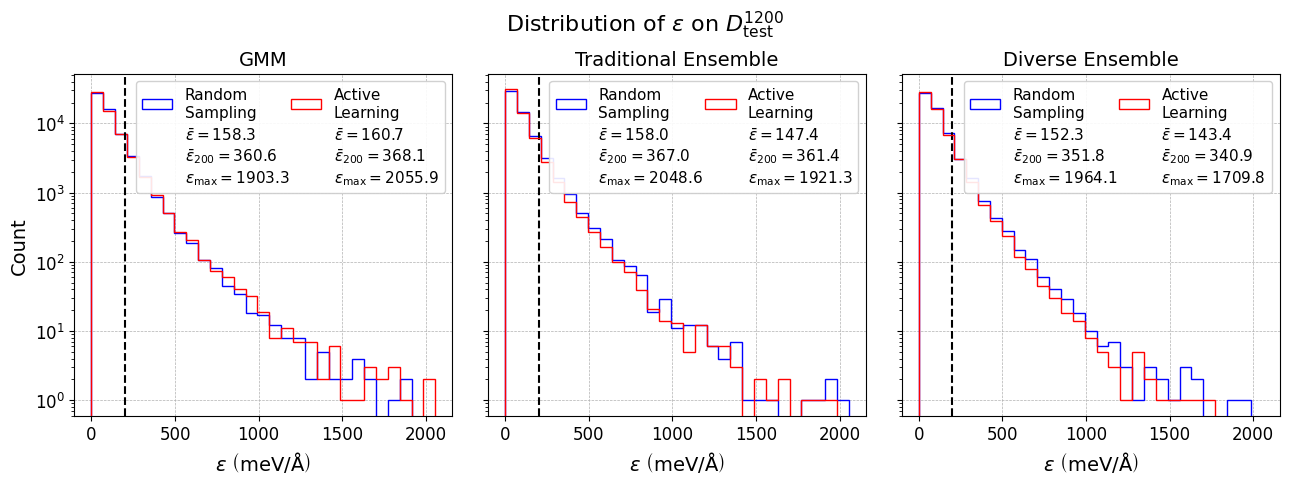}
     \end{subfigure}
     \hfill
    \begin{subfigure}[b]{\textwidth}
         \centering
         \includegraphics[width=0.82\textwidth]{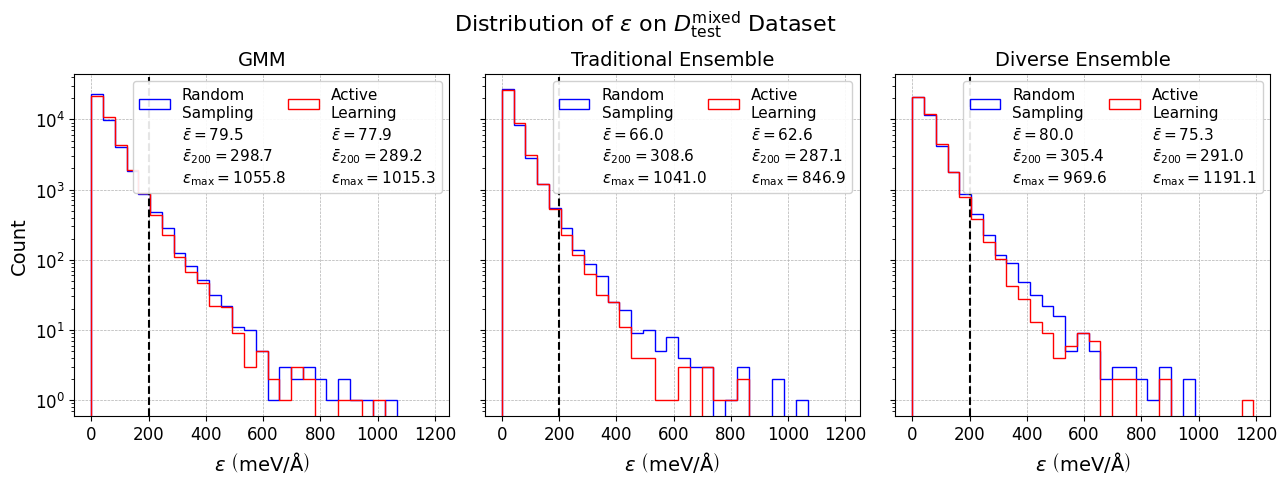}
     \end{subfigure}
    \caption{Distribution of $\epsilon$ of the GMM and ensembles on $D_{\mathrm{test}}^{300}$, $D_{\mathrm{test}}^{600}$, $D_{\mathrm{test}}^{1200}$, and $D_{\mathrm{test}}^{\mathrm{mixed}}$ for models with $f=32$, trained on $D_{\mathrm{train,50}}^{300}$ for the single temperature test sets and $D_{\mathrm{train,50}}^{\mathrm{mixed}}$ for $D_{\mathrm{test}}^{\mathrm{mixed}}$.}
    \label{fig:active-learning-temps-n50-f32-si}
\end{figure*}

\begin{figure*}
\centering
     \begin{subfigure}[b]{\textwidth}
         \centering
         \includegraphics[width=0.82\textwidth]{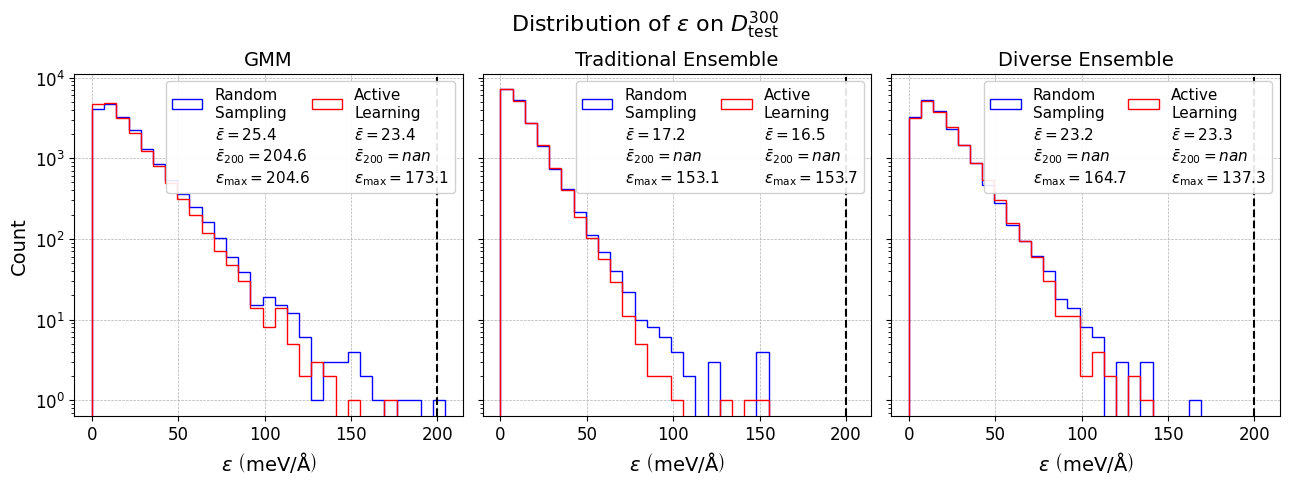}
     \end{subfigure}
     \hfill
     \begin{subfigure}[b]{\textwidth}
         \centering
         \includegraphics[width=0.82\textwidth]{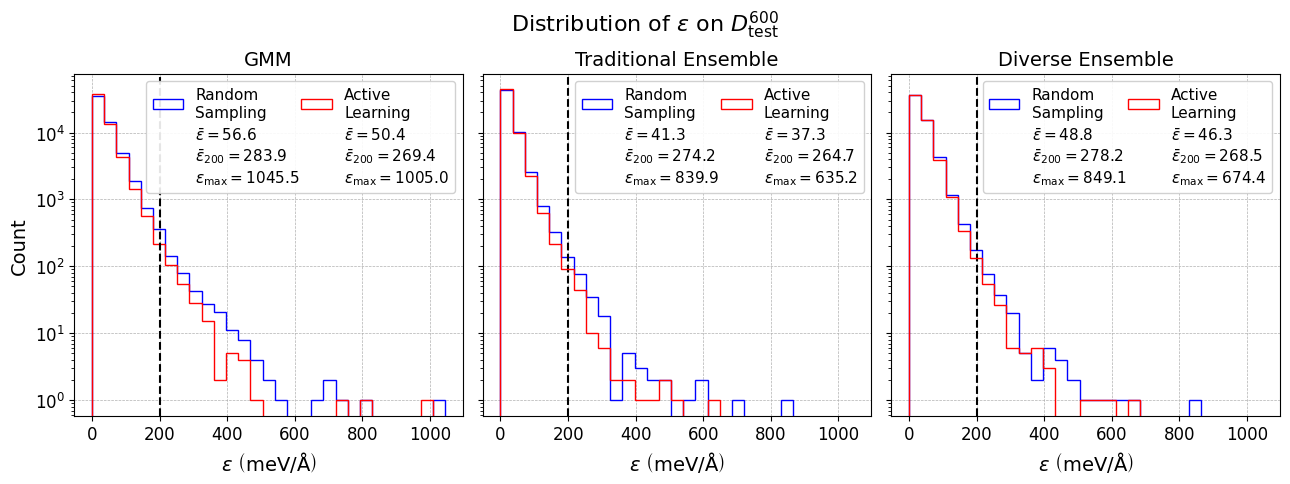}
     \end{subfigure}
     \hfill
     \begin{subfigure}[b]{\textwidth}
         \centering
         \includegraphics[width=0.82\textwidth]{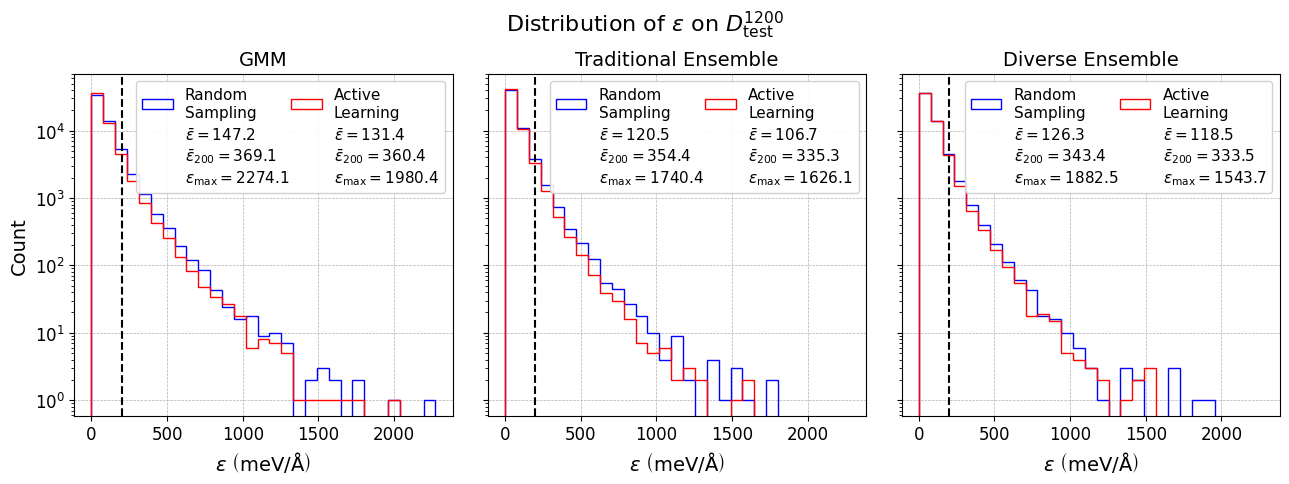}
     \end{subfigure}
     \hfill
    \begin{subfigure}[b]{\textwidth}
         \centering
         \includegraphics[width=0.82\textwidth]{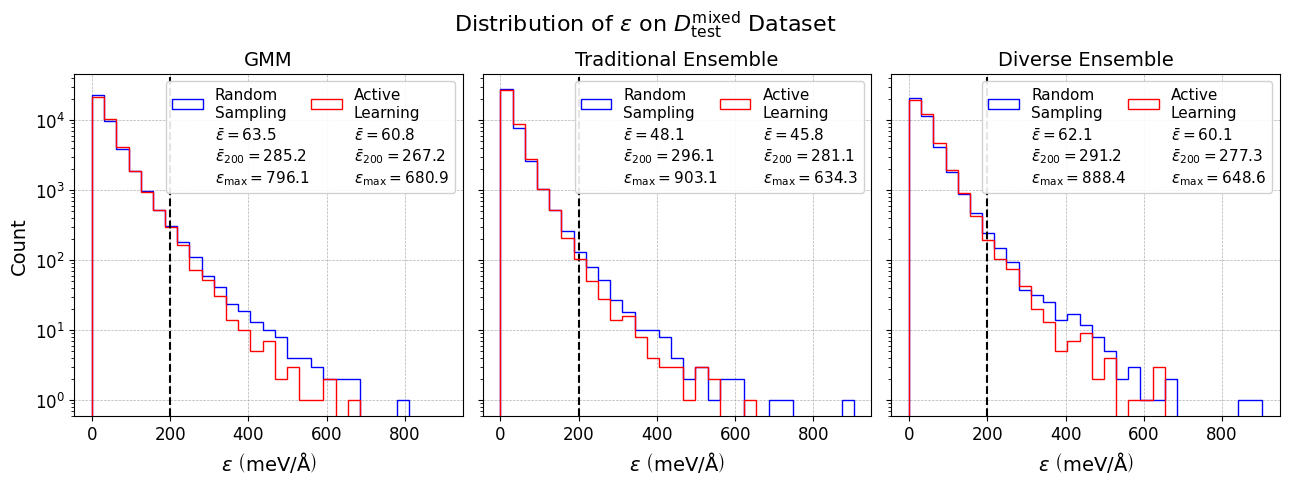}
     \end{subfigure}
    \caption{Distribution of $\epsilon$ of the GMM and ensembles on $D_{\mathrm{test}}^{300}$, $D_{\mathrm{test}}^{600}$, $D_{\mathrm{test}}^{1200}$, and $D_{\mathrm{test}}^{\mathrm{mixed}}$ for models with $f=16$, trained on $D_{\mathrm{train,100}}^{300}$ for the single temperature test sets and $D_{\mathrm{train,100}}^{\mathrm{mixed}}$ for $D_{\mathrm{test}}^{\mathrm{mixed}}$.}
    \label{fig:active-learning-temps-n100-f16-si}
\end{figure*}

\begin{figure*}
\centering
     \begin{subfigure}[b]{\textwidth}
         \centering
         \includegraphics[width=0.82\textwidth]{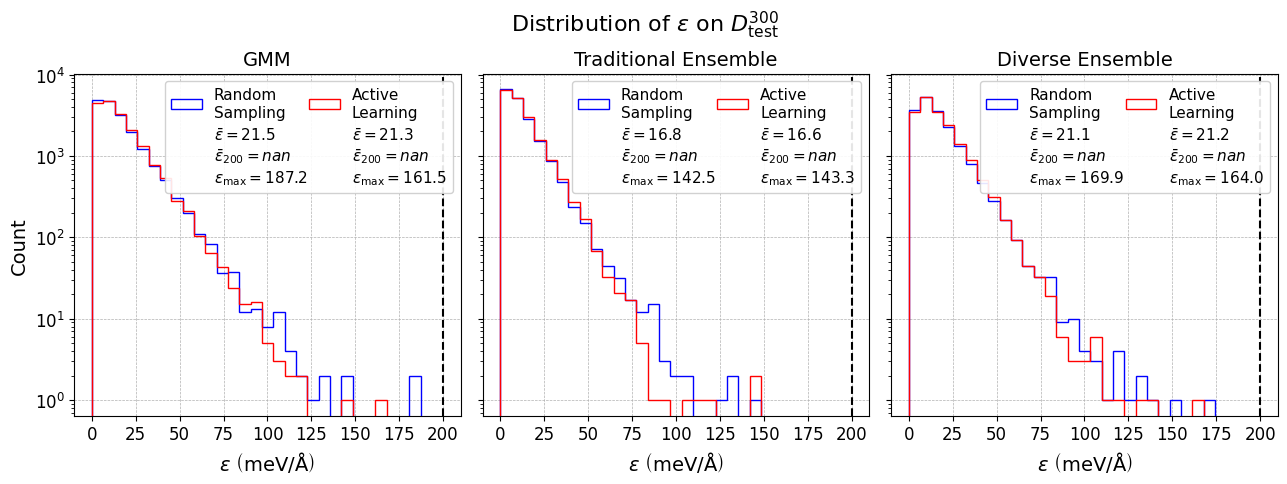}
     \end{subfigure}
     \hfill
     \begin{subfigure}[b]{\textwidth}
         \centering
         \includegraphics[width=0.82\textwidth]{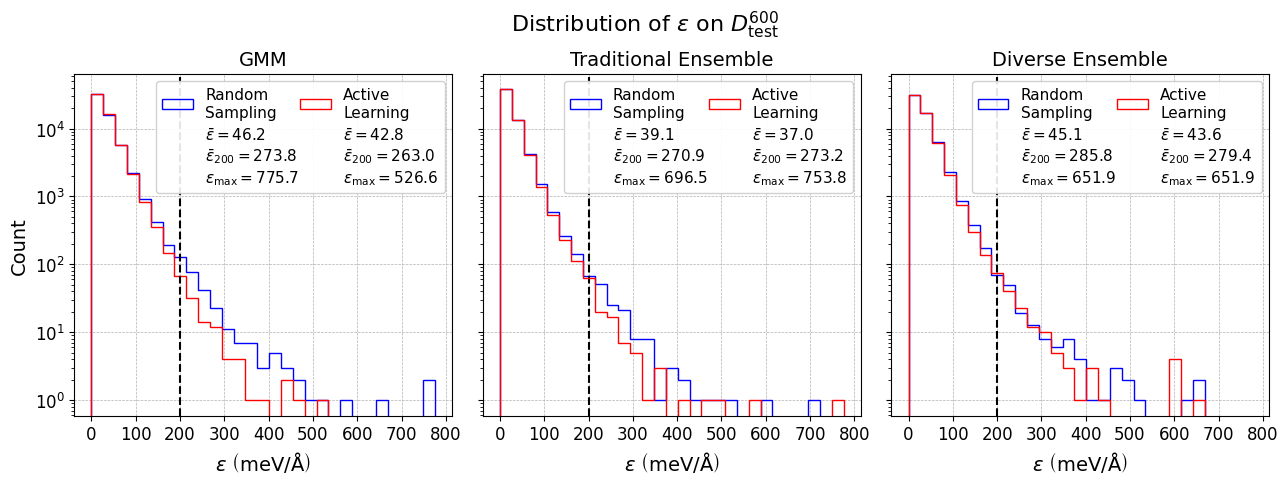}
     \end{subfigure}
     \hfill
     \begin{subfigure}[b]{\textwidth}
         \centering
         \includegraphics[width=0.82\textwidth]{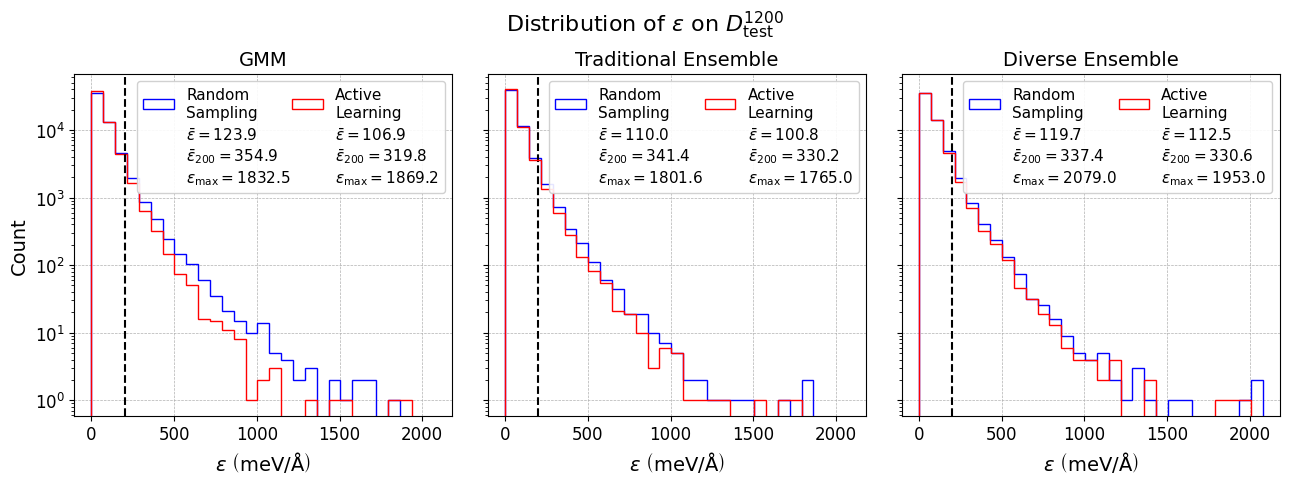}
     \end{subfigure}
     \hfill
    \begin{subfigure}[b]{\textwidth}
         \centering
         \includegraphics[width=0.82\textwidth]{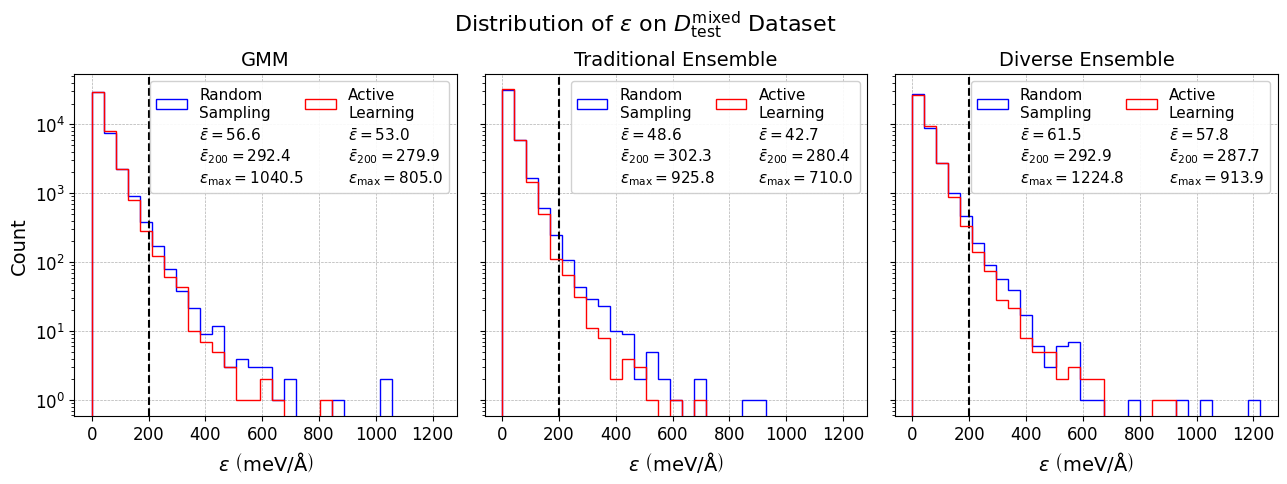}
     \end{subfigure}
    \caption{Distribution of $\epsilon$ of the GMM and ensembles on $D_{\mathrm{test}}^{300}$, $D_{\mathrm{test}}^{600}$, $D_{\mathrm{test}}^{1200}$, and $D_{\mathrm{test}}^{\mathrm{mixed}}$ for models with $f=32$, trained on $D_{\mathrm{train,100}}^{300}$ for the single temperature test sets and $D_{\mathrm{train,100}}^{\mathrm{mixed}}$ for $D_{\mathrm{test}}^{\mathrm{mixed}}$.}
    \label{fig:active-learning-temps-n100-f32-si}
\end{figure*}

We note that assessing the effectiveness of uncertainty quantification and active learning approaches requires configurations of sufficiently high rarity. This ensures that the model is not already fully trained to accurately predict all points in the test set. For instance, active learning achieved minimal improvement over random sampling for all methods when testing on $D_{\mathrm{test}}^{300}$, $D_{\mathrm{test}}^{600}$, $D_{\mathrm{test}}^{1200}$, and $D_{\mathrm{test}}^{\mathrm{mixed}}$ (SI figures \ref{fig:active-learning-temps-n50-f16-si}, \ref{fig:active-learning-temps-n50-f32-si}, \ref{fig:active-learning-temps-n100-f16-si}, \ref{fig:active-learning-temps-n100-f32-si}). $D_{\mathrm{test}}^{300}$ and $D_{\mathrm{test}}^{\mathrm{mixed}}$ contain structures from the same distribution as their respective training sets, rendering them ``easier" test sets that the model marginally improves on with additional training data. Similarly, while $D_{\mathrm{test}}^{600}$ and $D_{\mathrm{test}}^{1200}$ contain more high-energy structures than $D_{\mathrm{train,50}}^{300}$ and $D_{\mathrm{train,100}}^{300}$, their distributions still overlap significantly because structures are Boltzmann-sampled at each temperature. In comparison, $D_{\beta=120^{\circ}}$, $D_{\beta=150^{\circ}}$, and $D_{\beta=180^{\circ}}$ contain structures spanning a much wider range of $\alpha$ and $\gamma$ angles for a fixed $\beta$ \cite{kovacs2021linear}. Furthermore, compared to active learning with models initially trained on $D_{\mathrm{train,50}}^{300}$, active learning with models initially trained on $D_{\mathrm{train,100}}^{300}$ generally achieves smaller improvement over random sampling (SI figures \ref{fig:active-learning-betas-n100-f16-si}, \ref{fig:active-learning-betas-n100-f32-si}). In summary, we emphasize that for developing and benchmarking uncertainty quantification and active learning approaches, one must establish that active learning methods are justified and statistically distinguishable in performance from random sampling.

\section{Conclusion}
Algorithmic advances in fast and accurate uncertainty quantification for deep neural network interatomic potentials are needed to enable robust large-scale uncertainty-aware simulations. While an efficient method that can give access to predictive uncertainties in deep learning interatomic potentials has been a long-standing goal, so far it has not been achieved and current methods still rely on leveraging an ensemble of networks, thereby incurring a massive computational overhead. Here---by training a probabilistic model on the feature space of the neural network---we show that it is possible to retain the accuracy of ensemble uncertainty estimates with a single neural network evaluation, resulting in large computational savings in training and inference. In particular, we show that a Gaussian Mixture Model trained on NequIP features produces uncertainty estimates of similar quality to deep ensembles while requiring only a single model evaluation, resulting in a significant reduction of computational cost in both training and inference. While we find that the GMM models are competitive with ensembles, which are currently the state of the art methodology, significant improvement is desired for reliable predictions of quantitative uncertainties in both types of approaches. One future direction is to compare the methods explored here with rigorous Bayesian inference techniques, both in neural network and kernel-based learning models.

\section{Data Availability}

The data set of structures for the 3BPA molecule is publicly available at \url{https://pubs.acs.org/doi/full/10.1021/acs.jctc.1c00647}

\section{Code Availability}

An open-source software implementation of NequIP is available at \url{https://github.com/mir-group/nequip}. 

\section{References}
\bibliographystyle{naturemag}
\bibliography{bib.bib} 

\section{Acknowledgements}

We thank Yu Xie and Julia Yang for helpful discussions.\\

Work at Harvard University was supported by Bosch Research, the US Department of Energy, Office of Basic Energy Sciences Award No. DE-SC0022199 and by the NSF through the Harvard University Materials Research Science and Engineering Center Grant No. DMR-2011754. A.M is supported by U.S. Department of Energy, Office of Science, Office of Advanced Scientific Computing Research, Computational Science Graduate Fellowship under Award Number(s) DE-SC0021110. A.Z was supported by the Harvard College Research Program. The authors acknowledge computing resources provided by the Harvard University FAS Division of Science Research Computing Group. \\

\section{Author contributions}

A.Z. trained the NequIP networks, conducted all uncertainty experiments, active learning experiments, analysis of models, and wrote the first version of the manuscript. S.B. also trained some of the NequIP networks, proposed to use a GMM on final layer features, and contributed to the first version of the manuscript. A.M. contributed to the software development for ensembles and GMM-based evaluations. All authors discussed results and designed the experiments. B.K. supervised and guided the project from conception to design of experiments, implementation, theory, as well as analysis of data. All authors contributed to the manuscript. 
\section{Competing interests}
The authors declare no competing interests.

\section{Supplemental Information}
\subsection{Training details}

All experiments were run with version \texttt{0.5.4}, git commit \texttt{9bd9e3027b8756ea2ab2d50d0b8396694962d2a5}. Additionally, we used the \texttt{e3nn} code under version \texttt{0.4.4} \cite{mario_geiger_2021_5785497}, as well as PyTorch \cite{paszke2019pytorch} using version 1.10.0. NequIP models were trained in single-GPU training on a NVIDIA V100 GPU. We train one set of models on $D_{\mathrm{train,100}}^{300}$ and further split into 90 for training and 10 for validation. We train another set of models on $D_{\mathrm{train,50}}^{300}$, a subset of $D_{\mathrm{train,100}}^{300}$, split into 40 for training and 10 for validation. We repeat these trainings using training data sampled from a mixed-temperature data set (consisting of structures at 300K, 600K, and 1200K). We did not sample different training sets or different train/val splits for different ensembles members. The ensemble members have different weight initializations based on a random seed as well as a different random sampling of batches, i.e. a different order in which training points are seen over the course of training. Models were trained with float64 precision.

In the active learning experiments, models were trained with the same parameters as above, including the same validation set. The new training set consisted of the 90 structures from the original training set and 100 additional structures sampled from training pool sets for settings in which the new training set has 200 structures; in settings in which the new training set has 100 structures, the new training set consisted of 40 structures from the original training set and 50 additional structures.

\subsubsection{Traditional Ensembles}
The traditional ensemble consists of networks with 6 layers, $\ell_{max}=2$, 32 hidden features of both even and odd parity ($f=32$), a radial network with 3 layers of 64 neurons each and SiLU nonlinearities, a radial cutoff of 5 \AA\ , a trainable Bessel basis using 8 basis functions, and the previously discussed polynomial envelope function using $p=2$ \cite{nequip, klicpera2020directional}. We train a second traditional ensemble consisting of networks with all of the previous parameters except with 16 hidden features of both even and odd parity ($f=16$). Networks in the traditional ensembles differ only by weight initialization and batch sampling sequence. We use gated equivariant nonlinearities as outlined in \cite{nequip}. We use the Adam optimizers with the AMSGrad variant \cite{kingma2014adam, loshchilov2017decoupled, reddi2019convergence} as implemented in PyTorch \cite{paszke2019pytorch}, with default parameters of $\beta_1=0.9$, $\beta_2=0.999$, and $\epsilon=10^{-8}$ without weight decay. We use a batch size of 5, a learning rate of 0.01, as well as a joint loss function using both energies and forces. In particular, we use a per-atom MSE term, in which both the energy and force weights are set to 1:

\begin{equation}
    \mathcal{L} = \frac{\lambda_E}{B} \sum_{b}^{B}{\left( \frac{\hat{E}_b - E_b}{N} \right)^2} +  \frac{\lambda_F}{3BN} \sum_{i=1}^{BN} \sum_{\alpha=1}^3 \lVert -\frac{\partial \hat{E}}{\partial r_{i, \alpha}}  - F_{i, \alpha} \rVert^2
\end{equation}

We multiply the learning rate by 0.5 whenever no improvement in the validation loss was measured for 50 epochs. Further, we make use of an exponential moving average of the weights used for validation and testing with a weight of 0.99. We stop the training when the first of the following conditions is reached: a) a maximum training time of 5 days, b)  maximum number of epochs of 100,000, c) no improvement in the validation loss for 1,000 epochs, d) the learning rate dropped lower than 1e-6. We make use of the previously reported \cite{nequip, allegro} per-atom shift $\mu_{Z_i}$ which we set to the average per-atom potential energy computed over all training frames and a per-atom scale $\sigma_{Z_i}$ which we set to the root-mean-square of the components of the forces over the training set.\\

\subsubsection{Diverse Ensembles}

The diverse ensemble consist of 10 networks with identical parameters and training details as the traditional ensemble members described above, differing only in the parameters outlined in table \ref{tab:ensemble-config} (and similarly, we train one diverse ensemble with $f=32$ and a second with $f=16$). We vary a combination of $\ell_{max}$, the number of message passing layers, and seed (affecting both weight initialization and order in which batches are seen during training).

\subsection{Active Learning}

To perform active learning with ensembles of models trained on $D_{\mathrm{train,100}}^{300}$, we evaluate each ensemble on $D_{\mathrm{pool}}^{300}$, obtaining a force uncertainty $U_k = \sigma_k$ for each atom $k$ in each configuration. To perform active learning with a single NequIP model and the GMM uncertainties, we evaluate a NequIP model trained on $D_{\mathrm{train,100}}^{300}$ on $D_{\mathrm{pool}}^{300}$ and subsequently evaluate the fitted GMM on the predicted feature vectors of each atom $k$ in $D_{\mathrm{pool}}^{300}$, obtaining uncertainties $U_{k} = \mathrm{NLL}_{k}$. As stated in the main text, let $k^*$ be the atom within a 3BPA structure with the highest uncertainty, i.e.:
\begin{equation}
    k^* = \underset{k \in [1, K]}{\arg\max}\{U_k\}.
\end{equation}
Of all structures in $D_{\mathrm{pool}}^{300}$, we select the 100 structures with the highest $U_{k^*}$ to add to $D_{\mathrm{train,100}}^{300}$, creating a new training set $D_{\mathrm{a,200}}^{300}$ of size 200. For comparison, we also select 100 structures uniformly at random, without replacement, from $D_{\mathrm{pool}}^{300}$ to add to $D_{\mathrm{train,100}}^{300}$, creating another new training set $D_{\mathrm{r,200}}^{300}$ of size 200 (we refer to this process as ``random sampling"). We re-train all models with the same parameters from scratch on each of these two new training sets and evaluate them on the following test sets: $D_{\mathrm{test}}^{300}$, $D_{\mathrm{test}}^{600}$, $D_{\mathrm{test}}^{1200}$, $D_{\beta=120^{\circ}}$, $D_{\beta=150^{\circ}}$, and $D_{\beta=180^{\circ}}$.

We repeat this active learning procedure for models trained on $D_{\mathrm{train,100}}^{\mathrm{mixed}}$ using the $D_{\mathrm{pool}}^{\mathrm{mixed}}$ for structure exploration and $D_{\mathrm{test}}^{\mathrm{mixed}}$ for testing. We also repeat this entire procedure for models with $f=16$ trained on $D_{\mathrm{train,100}}^{300}$ and $D_{\mathrm{train,100}}^{\mathrm{mixed}}$. Finally, we repeat the active learning procedure with the models trained on $D_{\mathrm{train,50}}^{300}$ and $D_{\mathrm{train,50}}^{\mathrm{mixed}}$, with the adjustment that we select only 50 structures to add to the training sets (i.e., we create $D_{\mathrm{a,100}}^{300}$ of size 100 by adding the 50 structures from $D_{\mathrm{pool}}^{300}$ with the highest $U_{k^*}$ to $D_{\mathrm{train,50}}^{300}$, and we create $D_{\mathrm{r,100}}^{300}$ of size 100 by adding 50 randomly sampled structures from $D_{\mathrm{pool}}^{300}$ to $D_{\mathrm{train,50}}^{300}$).

\begin{table*}[!htbp]
\centering
\resizebox{\textwidth}{!}{\begin{tabular}{cccccccccc}
\hline \hline
 Seed & $\mathrm{T_{train}}$ & $\ell_{max}$ & Number of layers & 300K, E & 300K, F & 600K, E & 600K, F & 1200K, E & 1200K, F\\
\hline 
\rule{0pt}{3ex}     \textbf{Traditional Ensemble}  \vspace{4.0pt} \\ 
23456 & 300K & 2 & 6 & 24.5 & 66.4 & 52.4 & 125.3 & 206.3 & 308.8 \\
27617 & 300K & 2 & 6 & 22.3 & 69.7 & 53.5 & 130.1 & 206.0 & 311.8 \\
34567 & 300K & 2 & 6 & 21.1 & 65.3 & 54.0 & 129.8 & 207.1 & 323.1 \\
45678 & 300K & 2 & 6 & 25.0 & 70.0 & 55.5 & 122.9 & 184.3 & 286.0 \\
56789 & 300K & 2 & 6 & 23.6 & 68.3 & 61.6 & 137.2 & 275.4 & 360.0 \\
78901 & 300K & 2 & 6 & 22.2 & 66.4 & 52.5 & 124.4 & 190.1 & 293.2 \\
87654 & 300K & 2 & 6 & 20.6 & 58.1 & 44.5 & 108.1 & 153.8 & 251.5 \\
89012 & 300K & 2 & 6 & 21.6 & 63.7 & 57.9 & 127.3 & 228.1 & 328.2 \\
90123 & 300K & 2 & 6 & 23.1 & 67.2 & 53.2 & 126.2 & 203.8 & 309.5 \\
98765 & 300K & 2 & 6 & 21.1 & 64.9 & 53.7 & 129.5 & 237.7 & 327.3 \\
\hline  
\rule{0pt}{3ex}     \textbf{Diverse Ensemble}  \vspace{4.0pt} \\ 
27617 & 300K & 0 & 6 & 77.2 & 143.1 & 179.1 & 233.4 & 406.5 & 444.5 \\
27617 & 300K & 1 & 2 & 110.8 & 98.5 & 135.0 & 169.3 & 231.3 & 326.3 \\
27617 & 300K & 1 & 4 & 34.5 & 82.4 & 71.2 & 148.6 & 209.7 & 319.8 \\
27617 & 300K & 1 & 6 & 52.2 & 98.8 & 94.2 & 171.4 & 292.5 & 373.7 \\
27617 & 300K & 2 & 2 & 37.9 & 77.9 & 67.2 & 135.0 & 165.5 & 277.2 \\
27617 & 300K & 2 & 4 & 20.0 & 58.1 & 45.2 & 104.3 & 123.2 & 241.3 \\
23456 & 300K & 2 & 6 & 24.5 & 66.4 & 52.4 & 125.3 & 206.3 & 308.8 \\
27617 & 300K & 2 & 6 & 22.3 & 69.7 & 53.5 & 130.1 & 206.0 & 311.8 \\
34567 & 300K & 2 & 6 & 21.1 & 65.3 & 54.0 & 129.8 & 207.1 & 323.1 \\
27617 & 300K & 3 & 6 & 19.0 & 58.1 & 45.0 & 111.2 & 174.5 & 283.9 \\
\hline \hline
 Seed & $\mathrm{T_{train}}$ & $\ell_{max}$ & Number of layers & Mixed-T, E & Mixed-T, F & & & & \\
\hline 
\rule{0pt}{3ex}     \textbf{Traditional Ensemble}  \vspace{4.0pt} \\ 
23456 & Mixed-T & 2 & 6 & 67.6 & 130.2 & & & & \\
27617 & Mixed-T & 2 & 6 & 67.4 & 136.0 & & & & \\
34567 & Mixed-T & 2 & 6 & 65.9 & 130.5 & & & & \\
45678 & Mixed-T & 2 & 6 & 79.5 & 143.1 & & & & \\
56789 & Mixed-T & 2 & 6 & 65.3 & 133.2 & & & & \\
78901 & Mixed-T & 2 & 6 & 66.9 & 137.7 & & & & \\
87654 & Mixed-T & 2 & 6 & 66.8 & 134.1 & & & & \\
89012 & Mixed-T & 2 & 6 & 69.9 & 132.1 & & & & \\
90123 & Mixed-T & 2 & 6 & 71.2 & 138.9 & & & & \\
98765 & Mixed-T & 2 & 6 & 68.6 & 135.2 & & & & \\
\hline  
\rule{0pt}{3ex}     \textbf{Diverse Ensemble}  \vspace{4.0pt} \\ 
27617 & Mixed-T & 0 & 6 & 269.2 & 290.6 & & & & \\
27617 & Mixed-T & 1 & 2 & 162.8 & 220.1 & & & & \\ 
27617 & Mixed-T & 1 & 4 & 111.7 & 189.7 & & & & \\
27617 & Mixed-T & 1 & 6 & 104.2 & 190.8 & & & & \\
27617 & Mixed-T & 2 & 2 & 122.3 & 169.7 & & & & \\
27617 & Mixed-T & 2 & 4 & 70.2 & 137.0 & & & & \\
23456 & Mixed-T & 2 & 6 & 67.6 & 130.2 & & & & \\
27617 & Mixed-T & 2 & 6 & 67.4 & 136.0 & & & & \\
34567 & Mixed-T & 2 & 6 & 65.9 & 130.5 & & & & \\
27617 & Mixed-T & 3 & 6 & 72.0 & 130.2 & & & & \\
\hline \hline
\end{tabular}}
\caption{Configuration of the ensembles together with test errors for models with hidden feature dimension $f=16$ trained on $N=50$ structures at T=300K and Mixed-T, measured via the RMSE of energies and forces in units of [meV] and [meV/\AA], respectively.}
\label{tab:ensemble-config}
\end{table*}

\begin{table*}[!htbp]
\centering
\resizebox{\textwidth}{!}{\begin{tabular}{cccccccccc}
\hline \hline
 Seed & $\mathrm{T_{train}}$ & $\ell_{max}$ & Number of layers & 300K, E & 300K, F & 600K, E & 600K, F & 1200K, E & 1200K, F\\
\hline 
\rule{0pt}{3ex}     \textbf{Traditional Ensemble}  \vspace{4.0pt} \\ 
23456 & 300K & 2 & 6 & 21.6 & 58.4 & 50.0 & 111.1 & 186.0 & 269.9 \\
27617 & 300K & 2 & 6 & 18.0 & 53.8 & 43.5 & 101.4 & 155.8 & 239.6 \\
34567 & 300K & 2 & 6 & 20.1 & 56.9 & 45.8 & 102.1 & 147.1 & 226.0 \\
45678 & 300K & 2 & 6 & 19.8 & 56.5 & 47.6 & 104.4 & 157.1 & 238.2 \\
56789 & 300K & 2 & 6 & 17.6 & 56.6 & 43.5 & 103.7 & 156.5 & 240.8 \\
78901 & 300K & 2 & 6 & 17.4 & 55.3 & 43.8 & 102.4 & 139.3 & 238.8 \\
87654 & 300K & 2 & 6 & 18.3 & 57.3 & 49.1 & 112.5 & 197.8 & 273.6 \\
89012 & 300K & 2 & 6 & 18.8 & 54.9 & 47.9 & 102.4 & 153.7 & 239.1 \\
90123 & 300K & 2 & 6 & 17.8 & 57.5 & 49.2 & 109.4 & 159.5 & 255.1 \\
98765 & 300K & 2 & 6 & 22.2 & 63.6 & 50.4 & 117.2 & 165.0 & 268.7 \\
\hline  
\rule{0pt}{3ex}     \textbf{Diverse Ensemble}  \vspace{4.0pt} \\ 
27617 & 300K & 0 & 6 & 70.9 & 140.0 & 161.7 & 224.6 & 427.0 & 420.8 \\
27617 & 300K & 1 & 2 & 72.4 & 107.5 & 157.7 & 169.3 & 238.0 & 307.4 \\
27617 & 300K & 1 & 4 & 42.4 & 76.0 & 59.9 & 134.3 & 189.8 & 305.2 \\
27617 & 300K & 1 & 6 & 41.5 & 88.6 & 81.5 & 154.6 & 274.1 & 334.1 \\
27617 & 300K & 2 & 2 & 33.9 & 71.2 & 67.8 & 122.9 & 154.5 & 247.6 \\
27617 & 300K & 2 & 4 & 18.1 & 49.6 & 36.4 & 87.7 & 90.5 & 177.7 \\
23456 & 300K & 2 & 6 & 21.6 & 58.4 & 50.0 & 111.1 & 186.0 & 269.9 \\
27617 & 300K & 2 & 6 & 18.0 & 53.8 & 43.5 & 101.4 & 155.8 & 239.6 \\
34567 & 300K & 2 & 6 & 20.1 & 56.9 & 45.8 & 102.1 & 147.1 & 226.0 \\
27617 & 300K & 3 & 6 & 18.2 & 42.7 & 45.0 & 94.0 & 155.2 & 213.1 \\
\hline \hline
 Seed & $\mathrm{T_{train}}$ & $\ell_{max}$ & Number of layers & Mixed-T, E & Mixed-T, F & & & & \\
\hline 
\rule{0pt}{3ex}     \textbf{Traditional Ensemble}  \vspace{4.0pt} \\ 
23456 & Mixed-T & 2 & 6 & 65.5 & 126.4 & & & & \\
27617 & Mixed-T & 2 & 6 & 60.4 & 115.4 & & & & \\
34567 & Mixed-T & 2 & 6 & 66.2 & 121.9 & & & & \\
45678 & Mixed-T & 2 & 6 & 56.1 & 114.1 & & & & \\
56789 & Mixed-T & 2 & 6 & 59.8 & 120.2 & & & & \\
78901 & Mixed-T & 2 & 6 & 64.8 & 119.9 & & & & \\
87654 & Mixed-T & 2 & 6 & 68.7 & 130.4 & & & & \\
89012 & Mixed-T & 2 & 6 & 59.0 & 118.4 & & & & \\
90123 & Mixed-T & 2 & 6 & 61.5 & 122.9 & & & & \\
98765 & Mixed-T & 2 & 6 & 68.2 & 129.7 & & & & \\
\hline  
\rule{0pt}{3ex}     \textbf{Diverse Ensemble}  \vspace{4.0pt} \\ 
27617 & Mixed-T & 0 & 6 & 269.2 & 273.5 & & & & \\
27617 & Mixed-T & 1 & 2 & 167.9 & 204.5 & & & & \\ 
27617 & Mixed-T & 1 & 4 & 110.6 & 170.9 & & & & \\
27617 & Mixed-T & 1 & 6 & 107.5 & 179.5 & & & & \\
27617 & Mixed-T & 2 & 2 & 104.7 & 161.5 & & & & \\
27617 & Mixed-T & 2 & 4 & 58.0 & 110.9 & & & & \\
23456 & Mixed-T & 2 & 6 & 65.5 & 126.4 & & & & \\
27617 & Mixed-T & 2 & 6 & 60.4 & 115.4 & & & & \\
34567 & Mixed-T & 2 & 6 & 66.2 & 121.9 & & & & \\
27617 & Mixed-T & 3 & 6 & 59.6 & 104.6 & & & & \\
\hline \hline
\end{tabular}}
\caption{Configuration of the ensembles together with test errors for models with hidden feature dimension $f=32$ trained on $N=50$ structures at T=300K and Mixed-T, measured via the RMSE of energies and forces in units of [meV] and [meV/\AA], respectively.}
\end{table*}

\begin{table*}[!htbp]
\centering
\resizebox{\textwidth}{!}{\begin{tabular}{cccccccccc}
\hline \hline
 Seed & $\mathrm{T_{train}}$ & $\ell_{max}$ & Number of layers & 300K, E & 300K, F & 600K, E & 600K, F & 1200K, E & 1200K, F\\
\hline 
\rule{0pt}{3ex}     \textbf{Traditional Ensemble}  \vspace{4.0pt} \\ 
23456 & 300K & 2 & 6 & 12.6 & 36.2 & 28.5 & 71.3 & 87.6 & 171.8 \\
27617 & 300K & 2 & 6 & 14.0 & 39.2 & 32.3 & 83.0 & 115.2 & 218.6 \\
34567 & 300K & 2 & 6 & 11.9 & 37.8 & 35.4 & 81.7 & 118.2 & 213.7 \\
45678 & 300K & 2 & 6 & 13.3 & 38.5 & 30.4 & 76.8 & 93.5 & 190.3 \\
56789 & 300K & 2 & 6 & 10.5 & 35.0 & 30.8 & 76.5 & 126.5 & 205.0 \\
78901 & 300K & 2 & 6 & 15.3 & 42.9 & 33.8 & 86.6 & 190.1 & 219.8 \\
87654 & 300K & 2 & 6 & 13.1 & 37.4 & 31.3 & 74.5 & 96.4 & 182.3 \\
89012 & 300K & 2 & 6 & 14.7 & 38.7 & 34.0 & 79.5 & 110.7 & 200.6 \\
90123 & 300K & 2 & 6 & 11.7 & 40.4 & 32.9 & 83.5 & 129.1 & 220.2 \\
98765 & 300K & 2 & 6 & 11.1 & 37.1 & 33.9 & 79.6 & 129.1 & 213.1 \\
\hline  
\rule{0pt}{3ex}     \textbf{Diverse Ensemble}  \vspace{4.0pt} \\ 
27617 & 300K & 0 & 6 & 71.9 & 104.0 & 127.8 & 181.5 & 258.7 & 354.9 \\
27617 & 300K & 1 & 2 & 59.2 & 71.5 & 81.2 & 125.7 & 171.9 & 255.8 \\
27617 & 300K & 1 & 4 & 28.4 & 56.2 & 45.7 & 105.5 & 114.2 & 235.2 \\
27617 & 300K & 1 & 6 & 30.8 & 58.6 & 55.3 & 112.7 & 179.8 & 268.1 \\
27617 & 300K & 2 & 2 & 32.7 & 59.5 & 58.2 & 106.4 & 135.6 & 223.3 \\
27617 & 300K & 2 & 4 & 18.9 & 42.6 & 35.8 & 83.8 & 111.0 & 209.8 \\
23456 & 300K & 2 & 6 & 12.6 & 36.2 & 28.5 & 71.3 & 87.6 & 171.8 \\
27617 & 300K & 2 & 6 & 14.0 & 39.2 & 32.3 & 83.0 & 115.2 & 218.6 \\
34567 & 300K & 2 & 6 & 11.9 & 37.8 & 35.4 & 81.7 & 118.2 & 213.7 \\
27617 & 300K & 3 & 6 & 13.6 & 37.3 & 32.7 & 80.8 & 90.5 & 157.1 \\
\hline \hline
 Seed & $\mathrm{T_{train}}$ & $\ell_{max}$ & Number of layers & Mixed-T, E & Mixed-T, F & & & & \\
\hline 
\rule{0pt}{3ex}     \textbf{Traditional Ensemble}  \vspace{4.0pt} \\ 
23456 & Mixed-T & 2 & 6 & 42.9 & 86.8 & & & & \\
27617 & Mixed-T & 2 & 6 & 41.0 & 97.0 & & & & \\
34567 & Mixed-T & 2 & 6 & 39.0 & 94.4 & & & & \\
45678 & Mixed-T & 2 & 6 & 48.5 & 95.9 & & & & \\
56789 & Mixed-T & 2 & 6 & 40.6 & 93.0 & & & & \\
78901 & Mixed-T & 2 & 6 & 47.9 & 105.0 & & & & \\
87654 & Mixed-T & 2 & 6 & 39.0 & 88.9 & & & & \\
89012 & Mixed-T & 2 & 6 & 43.0 & 91.3 & & & & \\
90123 & Mixed-T & 2 & 6 & 47.8 & 97.1 & & & & \\
98765 & Mixed-T & 2 & 6 & 37.7 & 90.9 & & & & \\
\hline  
\rule{0pt}{3ex}     \textbf{Diverse Ensemble}  \vspace{4.0pt} \\ 
27617 & Mixed-T & 0 & 6 & 224.5 & 228.7 & & & & \\
27617 & Mixed-T & 1 & 2 & 137.8 & 172.0 & & & & \\ 
27617 & Mixed-T & 1 & 4 & 94.5 & 146.6 & & & & \\
27617 & Mixed-T & 1 & 6 & 73.7 & 142.4 & & & & \\
27617 & Mixed-T & 2 & 2 & 87.7 & 130.9 & & & & \\
27617 & Mixed-T & 2 & 4 & 45.7 & 94.5 & & & & \\
23456 & Mixed-T & 2 & 6 & 42.9 & 86.8 & & & & \\
27617 & Mixed-T & 2 & 6 & 41.0 & 97.0 & & & & \\
34567 & Mixed-T & 2 & 6 & 39.0 & 94.4 & & &  & \\
27617 & Mixed-T & 3 & 6 & 42.9 & 89.2 & & & & \\
\hline \hline
\end{tabular}}
\caption{Configuration of the ensembles together with test errors for models with hidden feature dimension $f=16$ trained on $N=100$ structures at T=300K and Mixed-T, measured via the RMSE of energies and forces in units of [meV] and [meV/\AA], respectively.}
\end{table*}

\begin{table*}[!htbp]
\centering
\resizebox{\textwidth}{!}{\begin{tabular}{cccccccccc}
\hline \hline
 Seed & $\mathrm{T_{train}}$ & $\ell_{max}$ & Number of layers & 300K, E & 300K, F & 600K, E & 600K, F & 1200K, E & 1200K, F\\
\hline 
\rule{0pt}{3ex}     \textbf{Traditional Ensemble}  \vspace{4.0pt} \\ 
23456 & 300K & 2 & 6 & 12.9 & 37.2 & 31.3 & 74.8 & 106.0 & 189.3 \\
27617 & 300K & 2 & 6 & 10.9 & 35.4 & 27.4 & 71.6 & 90.2 & 176.0 \\
34567 & 300K & 2 & 6 & 10.5 & 34.8 & 21.1 & 69.0 & 77.9 & 157.0 \\
45678 & 300K & 2 & 6 & 9.6 & 33.9 & 25.4 & 67.3 & 77.2 & 158.8 \\
56789 & 300K & 2 & 6 & 9.6 & 33.6 & 28.5 & 68.1 & 87.8 & 167.4 \\
78901 & 300K & 2 & 6 & 10.9 & 35.0 & 27.7 & 68.4 & 70.0 & 155.7 \\
87654 & 300K & 2 & 6 & 10.6 & 34.9 & 27.8 & 72.2 & 102.5 & 179.8 \\
89012 & 300K & 2 & 6 & 12.2 & 34.5 & 31.6 & 68.4 & 88.3 & 165.5 \\
90123 & 300K & 2 & 6 & 9.6 & 34.3 & 29.4 & 71.3 & 97.7 & 177.6 \\
98765 & 300K & 2 & 6 & 10.4 & 36.0 & 28.2 & 75.2 & 98.0 & 189.9 \\
\hline  
\rule{0pt}{3ex}     \textbf{Diverse Ensemble}  \vspace{4.0pt} \\ 
27617 & 300K & 0 & 6 & 49.6 & 99.0 & 105.9 & 175.1 & 216.1 & 343.4 \\
27617 & 300K & 1 & 2 & 38.3 & 62.8 & 68.6 & 115.2 & 146.1 & 244.3 \\
27617 & 300K & 1 & 4 & 27.6 & 51.3 & 44.4 & 96.1 & 137.6 & 225.7 \\
27617 & 300K & 1 & 6 & 25.2 & 55.1 & 48.2 & 104.9 & 161.3 & 247.9 \\
27617 & 300K & 2 & 2 & 29.6 & 53.2 & 55.1 & 97.0 & 126.0 & 206.2 \\
27617 & 300K & 2 & 4 & 12.7 & 33.9 & 26.5 & 63.4 & 65.1 & 138.1 \\
23456 & 300K & 2 & 6 & 12.9 & 37.2 & 31.3 & 74.8 & 106.0 & 189.3 \\
27617 & 300K & 2 & 6 & 10.9 & 35.4 & 27.4 & 71.6 & 90.2 & 176.0 \\
34567 & 300K & 2 & 6 & 10.5 & 34.8 & 21.1 & 69.0 & 77.9 & 157.0 \\
27617 & 300K & 3 & 6 & 11.1 & 31.6 & 28.9 & 64.2 & 90.5 & 157.1 \\
\hline \hline
 Seed & $\mathrm{T_{train}}$ & $\ell_{max}$ & Number of layers & Mixed-T, E & Mixed-T, F & & & & \\
\hline 
\rule{0pt}{3ex}     \textbf{Traditional Ensemble}  \vspace{4.0pt} \\ 
23456 & Mixed-T & 2 & 6 & 36.9 & 86.5 & & & & \\
27617 & Mixed-T & 2 & 6 & 36.8 & 80.1 & & & & \\
34567 & Mixed-T & 2 & 6 & 36.7 & 82.6 & & & & \\
45678 & Mixed-T & 2 & 6 & 37.6 & 82.3 & & & & \\
56789 & Mixed-T & 2 & 6 & 36.4 & 82.2 & & & & \\
78901 & Mixed-T & 2 & 6 & 36.3 & 83.4 & & & & \\
87654 & Mixed-T & 2 & 6 & 39.6 & 85.3 & & & & \\
89012 & Mixed-T & 2 & 6 & 34.6 & 79.1 & & & & \\
90123 & Mixed-T & 2 & 6 & 36.7 & 83.7 & & & & \\
98765 & Mixed-T & 2 & 6 & 42.4 & 90.0 & & & & \\
\hline  
\rule{0pt}{3ex}     \textbf{Diverse Ensemble}  \vspace{4.0pt} \\ 
27617 & Mixed-T & 0 & 6 & 235.4 & 223.2 & & & & \\
27617 & Mixed-T & 1 & 2 & 124.5 & 161.1 & & & & \\ 
27617 & Mixed-T & 1 & 4 & 69.7 & 126.0 & & & & \\
27617 & Mixed-T & 1 & 6 & 78.4 & 132.2 & & & & \\
27617 & Mixed-T & 2 & 2 & 80.7 & 122.5 & & & & \\
27617 & Mixed-T & 2 & 4 & 40.2 & 83.3 & & & & \\
23456 & Mixed-T & 2 & 6 & 36.9 & 86.5 & & & & \\
27617 & Mixed-T & 2 & 6 & 36.8 & 80.1 & & & & \\
34567 & Mixed-T & 2 & 6 & 36.7 & 82.6 & & & & \\
27617 & Mixed-T & 3 & 6 & 38.9 & 77.0 & & & & \\
\hline \hline
\end{tabular}}
\caption{Configuration of the ensembles together with test errors for models with hidden feature dimension $f=32$ trained on $N=100$ structures at T=300K and Mixed-T, measured via the RMSE of energies and forces in units of [meV] and [meV/\AA], respectively.}
\end{table*}

\clearpage

\begin{figure*}
\centering
     \begin{subfigure}[b]{\textwidth}
         \centering
         \includegraphics[width=\textwidth]{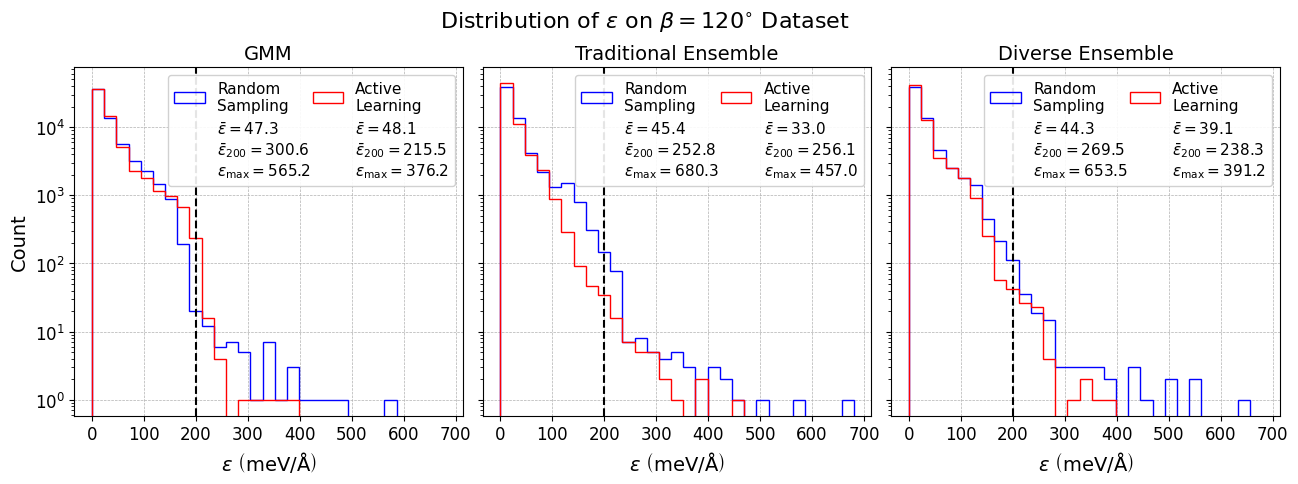}
     \end{subfigure}
     \hfill
     \begin{subfigure}[b]{\textwidth}
         \centering
         \includegraphics[width=\textwidth]{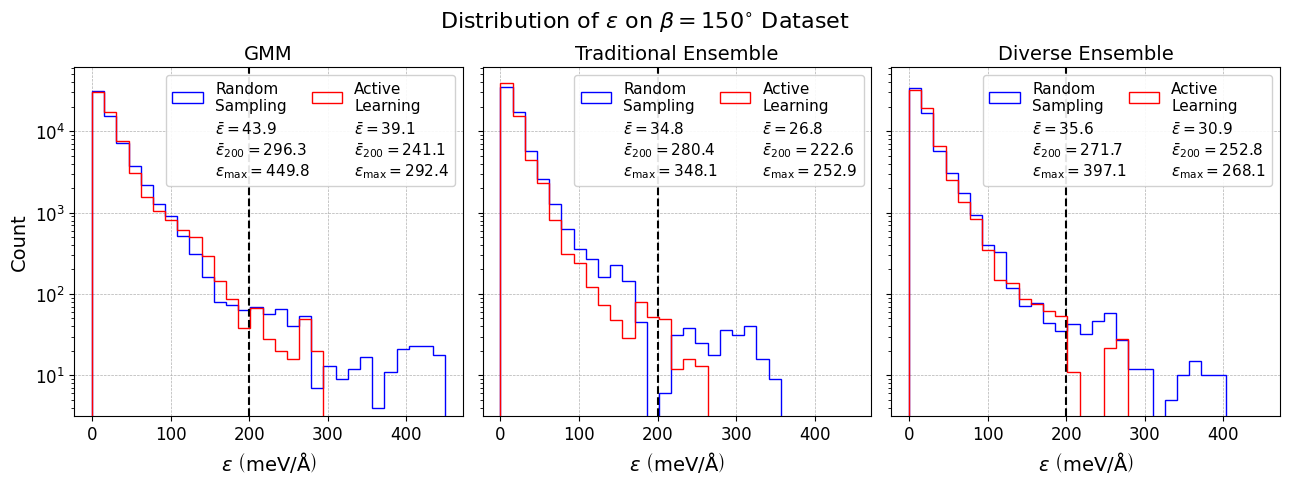}
     \end{subfigure}
     \hfill
    \begin{subfigure}[b]{\textwidth}
         \centering
         \includegraphics[width=\textwidth]{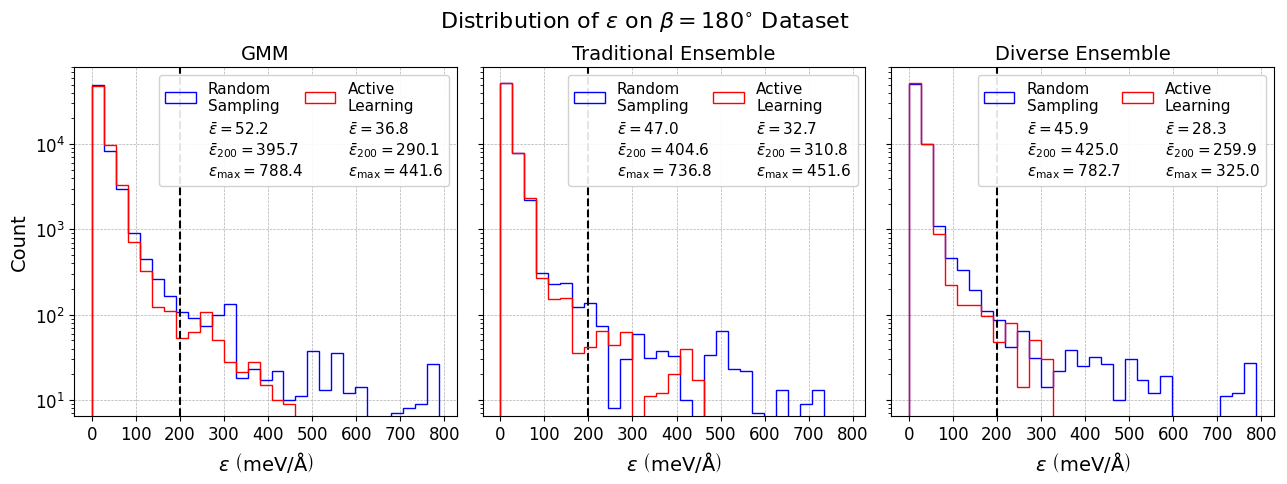}
     \end{subfigure}
    \caption{Distribution of $\epsilon$ of the GMM and ensembles on $D_{\beta=120^{\circ}}$, $D_{\beta=150^{\circ}}$, and $D_{\beta=180^{\circ}}$ for models with $f=32$, trained on $D_{\mathrm{train,50}}^{300}$.}
    \label{fig:active-learning-betas-n50-f32-si}
\end{figure*}

\begin{table*}[!htbp]
\centering
\resizebox{\textwidth}{!}
{\begin{tabular}{c|c|c|c|c|c|c|c}
\hline \hline
 \multicolumn{2}{c|}{\multirow{2}{*}{}} & \multicolumn{3}{c|}{Absolute Improvement ($\mathrm{meV/\AA}$)} & \multicolumn{3}{c}{Percentage Improvement (\%)} \\
\cline{3-8} 
\multicolumn{2}{c|}{} & GMM & Traditional & Diverse & GMM & Traditional & Diverse \\
\hline \hline
\multirow{3}{*}{$D_{\beta=120^{\circ}}$} & $\bar{\epsilon}$ & -0.8 & 12.4 & 5.2 & -1.7 & 27.3 & 11.7 \\ \cline{2-8}
& $\bar{\epsilon}_{200}$ & 85.1 & -3.3 & 31.2 & 28.1 & -1.3 & 11.6 \\ \cline{2-8}
& $\epsilon_{\mathrm{max}}$ & 189 & 223.3 & 262.3 & 33.4 & 32.8 & 40.1 \\ 
\hline \hline
\multirow{3}{*}{$D_{\beta=150^{\circ}}$} & $\bar{\epsilon}$ & 4.8 & 8.0 & 4.7 & 10.9 & 23.0 & 13.2 \\ \cline{2-8}
& $\bar{\epsilon}_{200}$ & 55.2 & 57.8 & 18.9 & 18.6 & 20.6 & 7.0 \\ \cline{2-8}
& $\epsilon_{\mathrm{max}}$ & 157.4 & 95.2 & 129 & 35.0 & 27.3 & 32.5 \\ 
\hline \hline
\multirow{3}{*}{$D_{\beta=180^{\circ}}$} & $\bar{\epsilon}$ & 15.4 & 14.3 & 17.6 & 29.5 & 30.4 & 38.3 \\ \cline{2-8}
& $\bar{\epsilon}_{200}$ & 105.6 & 93.8 & 165.1 & 26.7 & 23.2 & 38.8 \\ \cline{2-8}
& $\epsilon_{\mathrm{max}}$ & 346.8 & 285.2 & 457.7 & 44.0 & 38.7 & 58.5 \\ 
\hline \hline
\end{tabular}}
\caption{Improvement of $\bar{\epsilon}$, $\bar{\epsilon}_{200}$, and $\epsilon_{\mathrm{max}}$ with active learning over random sampling for each of the three methods on $D_{\beta=120^{\circ}}$, $D_{\beta=150^{\circ}}$, and $D_{\beta=180^{\circ}}$ (for models with hidden feature dimension $f=32$ and initially trained on $D_{\mathrm{train,50}}^{300}$).}
\end{table*}

\begin{table*}[!htbp]
\centering
\resizebox{\textwidth}{!}
{\begin{tabular}{c|c|c|c|c|c|c|c}
\hline \hline
 \multicolumn{2}{c|}{\multirow{2}{*}{}} & \multicolumn{3}{c|}{Absolute Improvement ($\mathrm{meV/\AA}$)} & \multicolumn{3}{c}{Percentage Improvement (\%)} \\
\cline{3-8} 
\multicolumn{2}{c|}{} & GMM & Traditional & Diverse & GMM & Traditional & Diverse \\
\hline \hline
\multirow{3}{*}{$D_{\beta=120^{\circ}}$} & $\bar{\epsilon}$ & 5.9 & 6.4 & 4.3 & 15.7 & 20.6 & 13.7 \\ \cline{2-8}
& $\bar{\epsilon}_{200}$ & 26.3 & -15.3 & -7.9 & 9.9 & -5.8 & -2.9 \\ \cline{2-8}
& $\epsilon_{\mathrm{max}}$ & 225.0 & 110.2 & -66.5 & 41.2 & 19.9 & -16.4 \\ 
\hline \hline
\multirow{3}{*}{$D_{\beta=150^{\circ}}$} & $\bar{\epsilon}$ & 5.7 & 3.1 & -0.2 & 16.4 & 14.2 & -0.7 \\ \cline{2-8}
& $\bar{\epsilon}_{200}$ & 16.7 & -- & 16.1 & 6.4 & -- & 6.1 \\ \cline{2-8}
& $\epsilon_{\mathrm{max}}$ & 69.4 & 18.4 & 54.0 & 18.3 & 9.2 & 16.1 \\ 
\hline \hline
\multirow{3}{*}{$D_{\beta=180^{\circ}}$} & $\bar{\epsilon}$ & 19.7 & 10.3 & 7.5 & 39.5 & 34.4 & 18.6 \\ \cline{2-8}
& $\bar{\epsilon}_{200}$ & 133.7 & 25.1 & 40.3 & 32.1 & 8.1 & 10.0 \\ \cline{2-8}
& $\epsilon_{\mathrm{max}}$ & 327.8 & 160.1 & 260.3 & 49.2 & 33.8 & 34.4 \\ 
\hline \hline
\end{tabular}}
\caption{Improvement of $\bar{\epsilon}$, $\bar{\epsilon}_{200}$, and $\epsilon_{\mathrm{max}}$ with active learning over random sampling for each of the three methods on $D_{\beta=120^{\circ}}$, $D_{\beta=150^{\circ}}$, and $D_{\beta=180^{\circ}}$ (for models with hidden feature dimension $f=16$ and initially trained on $D_{\mathrm{train,100}}^{300}$). In the case of $D_{\beta=150^{\circ}}$, the traditional ensemble did not make any errors larger than 200 $\mathrm{meV/\AA}$ with active learning.}
\end{table*}

\begin{table*}[!htbp]
\centering
\resizebox{\textwidth}{!}
{\begin{tabular}{c|c|c|c|c|c|c|c}
\hline \hline
 \multicolumn{2}{c|}{\multirow{2}{*}{}} & \multicolumn{3}{c|}{Absolute Improvement ($\mathrm{meV/\AA}$)} & \multicolumn{3}{c}{Percentage Improvement (\%)} \\
\cline{3-8} 
\multicolumn{2}{c|}{} & GMM & Traditional & Diverse & GMM & Traditional & Diverse \\
\hline \hline
\multirow{3}{*}{$D_{\beta=120^{\circ}}$} & $\bar{\epsilon}$ & 4.2 & 1.1 & -1.4 & 13.0 & 4.1 & -5.0 \\ \cline{2-8}
& $\bar{\epsilon}_{200}$ & -44.9 & 1.3 & 75.1 & -20.7 & 0.5 & 24.8 \\ \cline{2-8}
& $\epsilon_{\mathrm{max}}$ & -215.1 & -1.3 & 226.5 & -94.1 & -0.3 & 42.3 \\ 
\hline \hline
\multirow{3}{*}{$D_{\beta=150^{\circ}}$} & $\bar{\epsilon}$ & 3.4 & 2.5 & 3.8 & 12.6 & 11.5 & 13.2 \\ \cline{2-8}
& $\bar{\epsilon}_{200}$ & -4.5 & -11.1 & 27.6 & -2.0 & -5.3 & 11.1 \\ \cline{2-8}
& $\epsilon_{\mathrm{max}}$ & -0.8 & -17.1 & 74.9 & -0.3 & -7.9 & 22.4 \\ 
\hline \hline
\multirow{3}{*}{$D_{\beta=180^{\circ}}$} & $\bar{\epsilon}$ & 8.5 & 8.2 & 4.0 & 26.8 & 27.9 & 12.4 \\ \cline{2-8}
& $\bar{\epsilon}_{200}$ & 28.8 & 59.5 & 8.9 & 11.1 & 18.7 & 2.6 \\ \cline{2-8}
& $\epsilon_{\mathrm{max}}$ & 62.4 & 141.1 & 47.7 & 18.1 & 29.6 & 9.9 \\ 
\hline \hline
\end{tabular}}
\caption{Improvement of $\bar{\epsilon}$, $\bar{\epsilon}_{200}$, and $\epsilon_{\mathrm{max}}$ with active learning over random sampling for each of the three methods on $D_{\beta=120^{\circ}}$, $D_{\beta=150^{\circ}}$, and $D_{\beta=180^{\circ}}$ (for models with hidden feature dimension $f=32$ and initially trained on $D_{\mathrm{train,100}}^{300}$).}
\end{table*}


\begin{table*}[!htbp]
\centering
\resizebox{\textwidth}{!}
{\begin{tabular}{c|c|c|c|c|c|c|c}
\hline \hline
 \multicolumn{2}{c|}{\multirow{2}{*}{}} & \multicolumn{3}{c|}{Absolute Improvement ($\mathrm{meV/\AA}$)} & \multicolumn{3}{c}{Percentage Improvement (\%)} \\
\cline{3-8} 
\multicolumn{2}{c|}{} & GMM & Traditional & Diverse & GMM & Traditional & Diverse \\
\hline \hline
\multirow{3}{*}{$D_{\mathrm{test}}^{300}$} & $\bar{\epsilon}$ & 0 & 1.1 & 0.4 & 0.0 & 4.0 & 1.2 \\  \cline{2-8}
& $\bar{\epsilon}_{200}$ & -13.2 & -- & 5.7 & -5.8 & -- & 2.5 \\ \cline{2-8}
& $\epsilon_{\mathrm{max}}$ & -53.0 & 62.9 & 1.1 & -21.5 & 25.7 & 0.5 \\ 
\hline \hline
\multirow{3}{*}{$D_{\mathrm{test}}^{600}$} & $\bar{\epsilon}$ & 1.6 & 4.9 & 3.2 & 2.1 & 7.8 & 4.7 \\ \cline{2-8}
& $\bar{\epsilon}_{200}$ & -2.2 & 8.9 & 0.8 & -0.8 & 3.1 & 0.3 \\ \cline{2-8}
& $\epsilon_{\mathrm{max}}$ & -24.8 & 89.7 & 61.4 & -1.9 & 8.5 & 6.4 \\ 
\hline \hline
\multirow{3}{*}{$D_{\mathrm{test}}^{1200}$} & $\bar{\epsilon}$ & 5.4 & 9.3 & 9.1 & 2.8 & 5.3 & 5.4 \\ \cline{2-8}
& $\bar{\epsilon}_{200}$ & -1.5 & 9.2 & 4.2 & -0.4 & 2.4 & 1.1 \\ \cline{2-8}
& $\epsilon_{\mathrm{max}}$ & 59.4 & 161.6 & 169.7 & 2.8 & 7.8 & 8.1 \\ 
\hline \hline
\multirow{3}{*}{$D_{\mathrm{test}}^{\mathrm{mixed}}$} & $\bar{\epsilon}$ & 3.5 & 4.0 & -0.9 & 3.9 & 5.8 & -1.1 \\ \cline{2-8}
& $\bar{\epsilon}_{200}$ & 27.3 & 16.3 & 10.8 & 8.5 & 5.2 & 3.5 \\ \cline{2-8}
& $\epsilon_{\mathrm{max}}$ & 347.4 & 132.5 & 119.7 & 23.3 & 11.5 & 11.4 \\ 
\hline \hline
\end{tabular}}
\caption{Improvement of $\bar{\epsilon}$, $\bar{\epsilon}_{200}$, and $\epsilon_{\mathrm{max}}$ with active learning over random sampling for each of the three methods on $D_{\mathrm{test}}^{300}$, $D_{\mathrm{test}}^{600}$, $D_{\mathrm{test}}^{1200}$, and $D_{\mathrm{test}}^{\mathrm{mixed}}$ (for models with hidden feature dimension $f=16$, initially trained on $D_{\mathrm{train,50}}^{300}$ for the single-temperature test sets and $D_{\mathrm{train,50}}^{\mathrm{mixed}}$ for the mixed-temperature test sets). In the case of $D_{\mathrm{test}}^{300}$, the traditional ensemble did not make any errors larger than 200 $\mathrm{meV/\AA}$ with active learning.}
\end{table*}


\begin{table*}[!htbp]
\centering
\resizebox{\textwidth}{!}
{\begin{tabular}{c|c|c|c|c|c|c|c}
\hline \hline
 \multicolumn{2}{c|}{\multirow{2}{*}{}} & \multicolumn{3}{c|}{Absolute Improvement ($\mathrm{meV/\AA}$)} & \multicolumn{3}{c}{Percentage Improvement (\%)} \\
\cline{3-8} 
\multicolumn{2}{c|}{} & GMM & Traditional & Diverse & GMM & Traditional & Diverse \\
\hline \hline
\multirow{3}{*}{$D_{\mathrm{test}}^{300}$} & $\bar{\epsilon}$ & -1.4 & 0.5 & 0.8 & -4.3 & 1.8 & 2.6 \\ \cline{2-8}
& $\bar{\epsilon}_{200}$ & -2.6 & 38.5 & -29.6 & -1.1 & 15.9 & -13.5 \\ \cline{2-8}
& $\epsilon_{\mathrm{max}}$ & 19.7 & 63.8 & -21.5 & 7.2 & 23.8 & -9.4 \\ 
\hline \hline
\multirow{3}{*}{$D_{\mathrm{test}}^{600}$} & $\bar{\epsilon}$ & -1.0 & 2.9 & 3.2 & -1.5 & 4.9 & 5.2 \\ \cline{2-8}
& $\bar{\epsilon}_{200}$ & 0.6 & -0.3 & 4.6 & 0.2 & 0.1 & 1.7 \\ \cline{2-8}
& $\epsilon_{\mathrm{max}}$ & -122.0 & -169.9 & -96.3 & -14.5 & -20.3 & -14.6 \\ 
\hline \hline
\multirow{3}{*}{$D_{\mathrm{test}}^{1200}$} & $\bar{\epsilon}$ & -2.4 & 10.6 & 8.9 & -1.5 & 6.7 & 5.8 \\ \cline{2-8}
& $\bar{\epsilon}_{200}$ & -7.5 & 5.6 & 10.9 & -2.1 & 1.5 & 3.1 \\ \cline{2-8}
& $\epsilon_{\mathrm{max}}$ & -152.6 & 127.3 & 254.3 & -8.0 & 6.2 & 12.9 \\ 
\hline \hline
\multirow{3}{*}{$D_{\mathrm{test}}^{\mathrm{mixed}}$} & $\bar{\epsilon}$ & 1.6 & 3.4 & 4.7 & 2.0 & 5.2 & 5.9 \\ \cline{2-8}
& $\bar{\epsilon}_{200}$ & 9.5 & 21.5 & 14.4 & 3.2 & 7.0 & 4.7 \\ \cline{2-8}
& $\epsilon_{\mathrm{max}}$ & 40.5 & 194.1 & -221.5 & 3.8 & 18.6 & -22.8 \\ 
\hline \hline
\end{tabular}}
\caption{Improvement of $\bar{\epsilon}$, $\bar{\epsilon}_{200}$, and $\epsilon_{\mathrm{max}}$ with active learning over random sampling for each of the three methods on $D_{\mathrm{test}}^{300}$, $D_{\mathrm{test}}^{600}$, $D_{\mathrm{test}}^{1200}$, and $D_{\mathrm{test}}^{\mathrm{mixed}}$ (for models with hidden feature dimension $f=32$, initially trained on $D_{\mathrm{train,50}}^{300}$ for the single-temperature test sets and $D_{\mathrm{train,50}}^{\mathrm{mixed}}$ for the mixed-temperature test sets).}
\end{table*}


\begin{table*}[!htbp]
\centering
\resizebox{\textwidth}{!}
{\begin{tabular}{c|c|c|c|c|c|c|c}
\hline \hline
 \multicolumn{2}{c|}{\multirow{2}{*}{}} & \multicolumn{3}{c|}{Absolute Improvement ($\mathrm{meV/\AA}$)} & \multicolumn{3}{c}{Percentage Improvement (\%)} \\
\cline{3-8} 
\multicolumn{2}{c|}{} & GMM & Traditional & Diverse & GMM & Traditional & Diverse \\
\hline \hline
\multirow{3}{*}{$D_{\mathrm{test}}^{300}$} & $\bar{\epsilon}$ & 2.0 & 0.7 & -0.1 & 7.9 & 4.1 & -0.4 \\ \cline{2-8}
& $\bar{\epsilon}_{200}$ & -- & -- & -- & -- & -- & -- \\ \cline{2-8}
& $\epsilon_{\mathrm{max}}$ & 31.5 & -0.6 & 27.4 & 15.4 & -0.4 & 16.6 \\ 
\hline \hline
\multirow{3}{*}{$D_{\mathrm{test}}^{600}$} & $\bar{\epsilon}$ & 6.2 & 4.0 & 2.5 & 11.0 & 9.7 & 5.1 \\ \cline{2-8}
& $\bar{\epsilon}_{200}$ & 14.5 & 9.5 & 9.7 & 5.1 & 3.5 & 3.5 \\ \cline{2-8}
& $\epsilon_{\mathrm{max}}$ & 40.5 & 204.7 & 174.7 & 3.9 & 24.4 & 20.6 \\ 
\hline \hline
\multirow{3}{*}{$D_{\mathrm{test}}^{1200}$} & $\bar{\epsilon}$ & 15.8 & 13.8 & 7.8 & 10.7 & 11.5 & 6.2 \\ \cline{2-8}
& $\bar{\epsilon}_{200}$ & 8.7 & 19.1 & 9.9 & 2.4 & 5.4 & 2.9 \\ \cline{2-8}
& $\epsilon_{\mathrm{max}}$ & 293.7 & 114.3 & 338.8 & 12.9 & 6.6 & 18.0 \\ 
\hline \hline
\multirow{3}{*}{$D_{\mathrm{test}}^{\mathrm{mixed}}$} & $\bar{\epsilon}$ & 2.7 & 2.3 & 2.0 & 4.3 & 4.8 & 3.2 \\ \cline{2-8}
& $\bar{\epsilon}_{200}$ & 18.0 & 15.0 & 13.9 & 6.3 & 5.1 & 4.8 \\ \cline{2-8}
& $\epsilon_{\mathrm{max}}$ & 115.2 & 268.8 & 239.8 & 14.5 & 29.8 & 27.0 \\ 
\hline \hline
\end{tabular}}
\caption{Improvement of $\bar{\epsilon}$, $\bar{\epsilon}_{200}$, and $\epsilon_{\mathrm{max}}$ with active learning over random sampling for each of the three methods on $D_{\mathrm{test}}^{300}$, $D_{\mathrm{test}}^{600}$, $D_{\mathrm{test}}^{1200}$, and $D_{\mathrm{test}}^{\mathrm{mixed}}$ (for models with hidden feature dimension $f=16$, initially trained on $D_{\mathrm{train,100}}^{300}$ for the single-temperature test sets and $D_{\mathrm{train,100}}^{\mathrm{mixed}}$ for the mixed-temperature test sets). In the case of $D_{\mathrm{test}}^{300}$, none of the methods made any errors larger than 200 $\mathrm{meV/\AA}$ with active learning.}
\end{table*}


\begin{table*}[!htbp]
\centering
\resizebox{\textwidth}{!}
{\begin{tabular}{c|c|c|c|c|c|c|c}
\hline \hline
 \multicolumn{2}{c|}{\multirow{2}{*}{}} & \multicolumn{3}{c|}{Absolute Improvement ($\mathrm{meV/\AA}$)} & \multicolumn{3}{c}{Percentage Improvement (\%)} \\
\cline{3-8} 
\multicolumn{2}{c|}{} & GMM & Traditional & Diverse & GMM & Traditional & Diverse \\
\hline \hline
\multirow{3}{*}{$D_{\mathrm{test}}^{300}$} & $\bar{\epsilon}$ & 0.1 & 0.1 & -0.1 & 0.7 & 0.8 & -0.4 \\ \cline{2-8}
& $\bar{\epsilon}_{200}$ & -- & -- & -- & -- & -- & -- \\ \cline{2-8}
& $\epsilon_{\mathrm{max}}$ & 25.7 & -0.8 & 5.9 & 13.7 & -0.6 & 3.5 \\ 
\hline \hline
\multirow{3}{*}{$D_{\mathrm{test}}^{600}$} & $\bar{\epsilon}$ & 3.4 & 2.1 & 1.6 & 7.3 & 5.3 & 3.5 \\ \cline{2-8}
& $\bar{\epsilon}_{200}$ & 10.8 & -2.3 & 6.4 & 3.9 & -0.8 & 2.2 \\ \cline{2-8}
& $\epsilon_{\mathrm{max}}$ & 249.1 & -57.3 & 0.0 & 32.1 & -8.2 & 0.0 \\ 
\hline \hline
\multirow{3}{*}{$D_{\mathrm{test}}^{1200}$} & $\bar{\epsilon}$ & 17.0 & 9.2 & 7.2 & 13.7 & 8.3 & 6.0 \\ \cline{2-8}
& $\bar{\epsilon}_{200}$ & 35.1 & 11.2 & 6.8 & 9.9 & 3.3 & 6.8 \\ \cline{2-8}
& $\epsilon_{\mathrm{max}}$ & -36.7 & 36.6 & 126.0 & -2.0 & 2.0 & 6.1 \\ 
\hline \hline
\multirow{3}{*}{$D_{\mathrm{test}}^{\mathrm{mixed}}$} & $\bar{\epsilon}$ & 3.6 & 5.9 & 3.7 & 6.4 & 12.1 & 6.0 \\ \cline{2-8}
& $\bar{\epsilon}_{200}$ & 12.5 & 21.9 & 5.2 & 4.3 & 7.2 & 1.8 \\ \cline{2-8}
& $\epsilon_{\mathrm{max}}$ & 235.5 & 215.8 & 310.9 & 22.6 & 23.3 & 25.4 \\ 
\hline \hline
\end{tabular}}
\caption{Improvement of $\bar{\epsilon}$, $\bar{\epsilon}_{200}$, and $\epsilon_{\mathrm{max}}$ with active learning over random sampling for each of the three methods on $D_{\mathrm{test}}^{300}$, $D_{\mathrm{test}}^{600}$, $D_{\mathrm{test}}^{1200}$, and $D_{\mathrm{test}}^{\mathrm{mixed}}$ (for models with hidden feature dimension $f=32$, initially trained on $D_{\mathrm{train,100}}^{300}$ for the single-temperature test sets and $D_{\mathrm{train,100}}^{\mathrm{mixed}}$ for the mixed-temperature test sets). In the case of $D_{\mathrm{test}}^{300}$, none of the methods made any errors larger than 200 $\mathrm{meV/\AA}$ with active learning.}
\end{table*}

\end{document}